\documentclass{article}

\usepackage{arxiv}

\usepackage[utf8]{inputenc} 
\usepackage[T1]{fontenc}    
\usepackage{hyperref}       
\usepackage{url}            
\usepackage{booktabs}       
\usepackage{amsfonts}       
\usepackage{nicefrac}       
\usepackage{microtype}      
\usepackage{lipsum}
\usepackage{fancyhdr}       
\usepackage{graphicx}       
\graphicspath{{media/}}     

\usepackage{lineno}

\usepackage{titlesec}

\title{Unraveling the Molecular Magic: AI Insights on the Formation of Extraordinarily Stretchable Hydrogels
}

\author{
  Shahriar Hojjati Emmami \\
  Department of Biomedical Engineering\\
  Amirkabir University of Technology \\
   \And
  Ali Pilehvar Meibody \\
  Department of Applied Science and Technology\\
  Polytechnic University of Turin\\
   \And
  Lobat Tayebi \\
  Department of Developmental \\Sciences\\
  Marquette University \\
   \And
  Mohammadamin Tavakoli \\
  Department of Computing and \\Mathematical Sciences\\
  California Institute of Technology \\
   \And
  Pierre Baldi \\
  Department of Computer Science\\
  University of California, Irvine \\
}

\begin{document}
\maketitle


\begin{abstract}
The deliberate manipulation of ammonium persulfate, methylenebisacrylamide, dimethyleacrylamide, and polyethylene oxide concentrations resulted in the development of a hydrogel with an exceptional stretchability, capable of extending up to 260 times its original length. This study aims to elucidate the molecular architecture underlying this unique phenomenon by exploring potential reaction mechanisms, facilitated by an artificial intelligence prediction system. Artificial intelligence predictor introduces a novel approach to interlinking two polymers, involving the formation of networks interconnected with linear chains following random chain scission. This novel configuration leads to the emergence of a distinct type of hydrogel, herein referred to as a "Span Network." Additionally, Fourier-transform infrared spectroscopy (FTIR) is used to investigate functional groups that may be implicated in the proposed mechanism, with ester formation confirmed among numerous hydroxyl end groups obtained from chain scission of PEO and carboxyl groups formed on hydrogel networks.
\end{abstract}

\keywords{Stretchable, Hydrogel, Reaction Prediction}

\section{Introduction}\label{sec1}
\begin{figure}[!]
  \centering
  \includegraphics[width=\linewidth]{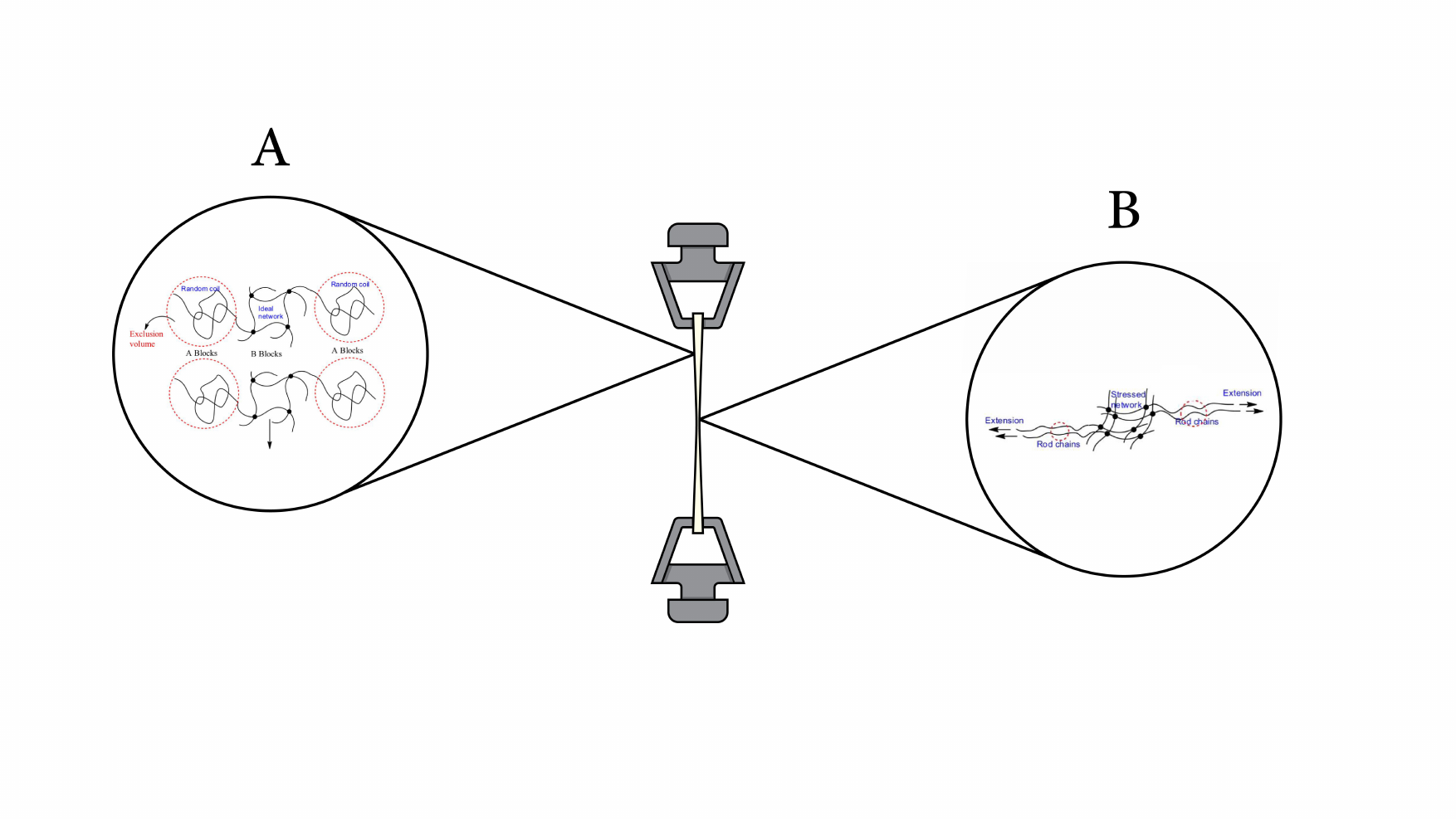}
  \caption{Structural Representation of Span Network Hydrogels: The hydrogel structure illustrates the interconnection of Dimethylacrylamide chains through methylene bisacrylamide crosslinkers, forming networks. These networks are further linked by polyethylene oxide (PEO). In the pre-stretch state, the PEO exhibits a coiled conformation, constituting a stable structure. Under tension, the networks align parallel to each other. This redistribution of stress ensures that pressure is uniformly distributed throughout the hydrogel, allowing for the unrestricted movement of PEO.}
  \label{fig:fig1}
\end{figure}

In the +100 years since the term was coined, hydrogels have been defined in a variety of ways. The most widely accepted definition characterizes hydrogels as water-swollen, crosslinked polymeric networks arising from a straightforward reaction involving one or more monomers \cite{liu2015highly}. These hydrogels owe their remarkable water-absorption capabilities to hydrophilic functional groups attached to the polymeric backbone, while the crosslinks between network chains confer resistance to dissolution \cite{ahmed2015hydrogel}. Being "soft and wet" materials, hydrogels exhibit distinct properties, including biocompatibility, responsiveness to diverse stimuli, low surface friction, and environmental compatibility \cite{gong2006friction}. Consequently, hydrogels have attracted substantial attention as innovative functional materials in the medical and industrial sectors, with applications including 3D printing for soft tissues \cite{bakarich2014three}, drug delivery vehicles \cite{shirakura2014hydrogel}, and actuators and sensors \cite{dong2006adaptive}, among others.
Traditionally, hydrogels have been categorized into two primary groups: homo-polymeric hydrogels and co-polymeric hydrogels. However, these conventional hydrogels suffer from inherent drawbacks, including low mechanical strength and limited stretchability under high stress, strain, impact loads, and cyclic loads, thereby restricting their application ranges. This limitation has spurred the emergence of a novel class of hydrogels known as interpenetrating polymeric hydrogels (IPN). In this approach, additional networks help the hydrogel have better performance. The underlying concept of IPN is the use of two networks to form the hydrogel instead of relying on a single network. In the double network which is a special case of IPN, one network is intentionally sacrificed under tension and stress, leading to significantly enhanced mechanical strength and relatively superior stretchability. An alternative approach involves manipulating crosslinkers to create double-network hydrogels \cite{gong2003double}, consisting of covalent and non-covalent networks that enable reversible rather than permanent damage. These mechanisms of energy dissipation during loading are further complemented by other strategies, including nanocomposite hydrogels \cite{zhu2006novel}, slide-ring hydrogels \cite{okumura2001polyrotaxane}, triblock copolymer hydrogels \cite{henderson2010ionically}, hydrophobic-modified hydrogels \cite{li2012hydrophobically}, tetra-PEG gels \cite{sakai2008design}, macromolecular microsphere composite (MMC) hydrogels \cite{huang2007novel}, ionic associations/hydrogen bonds \cite{long2018salt,shi2018stretchable}, and supramolecular structures \cite{zhang2019supramolecular,dutta2018highly,wang2018highly}. Supplementary Section 1 explores various IPN classifications.
Despite numerous advantages and ongoing improvements in their mechanical properties, until recently, the literature has reported hydrogels of only limited stretchability \cite{liu2015highly,gong2003double,bin2014extremely,cai2017extremely}. The highest reported stretchability was approximately 100 times the original length \cite{jeon2016extremely}, though even the gold standard, rubber (not a hydrogel), can reach up to 50 times its initial length \cite{goff2016soft}. However, recent studies have demonstrated significant advances in this property. For instance, an organogel has exhibited a remarkable 210 times increase in its original length \cite{zhang2018extremely}, and in another example, a polymer network has been stretched to 130 times its initial length \cite{zhang2019superstretchable}.
The fundamental mechanism behind stretching involves transmitting pressure to the hydrogel, resulting in the elongation of its backbone. Polymers possess an exclusion volume around their chains, and when multiple chains are present, they become entangled with each other. Due to the presence of such entanglement clusters, the pressure acts predominantly on these entanglements, eventually leading to their rupture. By strategically incorporating polymer chains between two networks where there are no other chains within the exclusion volume, pressure applied to the hydrogel allows the unrestricted movement of chains at the atomic level. This results in erratic motion and energy dissipation through collisions, creating a fascinating connection between microscopic physics and the macroscopic world \cite{buchanan2015dissipate}.  
Prominent scientists, including Erwin Schrödinger in his 1944 work "What is Life?" \cite{schrodinger1944life} and Nobel Laureate Richard Feynman in his 1959 lecture "There is plenty of room at the bottom," \cite{feynman1960there} have highlighted the significance of understanding the behavior of molecules and the potential to manipulate materials on a nanoscale. To achieve this goal, our proposed approach will create multiple networks comprising N, N-Dimethylacrylamide as our monomer and N, N-Methylenebisacrylamide as the crosslinker, followed by the incorporation of a single linear chain (polyethylene oxide) into the networks. This PEO can attach all networks together. When pressure is applied to these networks, free PEO chains, unencumbered by neighboring chains, can move freely and stretch to an unprecedented extent, reaching up to 260 times their original length—a world record \cite{emami2021ultra}. This allows us to harness the linear chain’s exceptional stretchability (Fig.  \ref{fig:fig1}). The structure of the resulting network, shaped by varying monomer and crosslinking agent concentrations, governs the hydrogel’s chemical and physical properties \cite{garcia2019n}. Manipulating these initial concentrations enables the creation of this unique structure; we term this new hydrogel classification a “span network.” Mechanical tests provide further insights into these hydrogels’ extraordinary properties.
Given the span networks’ mechanistic complexity and product instability, it is crucial to understand each reactant’s mechanistic-level reactions. As it helps to better explain and understand the exact reactions and behaviors at play.

However, manual navigation through the vast space of possible mechanistic reactions involved in hydrogel synthesis is not feasible. Therefore, to explore this space computational techniques must be deployed with a fundamental tradeoff between speed and accuracy.
Computationally intensive methods based on quantum mechanics (QM), such as Coupled Cluster (CC) \cite{knowles1993coupled}, yield accurate results but are computationally prohibitive for large systems due to their exponential scaling with system size. Density Functional Theory (DFT) \cite{parr1979local}, a less computationally demanding QM-based method, faces several challenges when it comes to simulating radical species \cite{cohen2012challenges}, which are common in hydrogel synthesis processes. On the other extreme hand of the computational spectrum, hand-curated rule-based systems are generally fast but require curating a set of rules that typically cover only a limited range of chemistry \cite{baldinoelectron09,chentutor08}.

Between these two extremes, AI-based reaction predictors seem to strike an optimal compromise offering both fast inference and reasonable accuracy and generalization capabilities, provided sufficient data to train them is available \cite{baldi2021call,tavakoli2023rmechdb, tavakoli2024pmechdb}. There is a handful of published AI-based reaction prediction systems \cite{fooshee2018deep, coley2017prediction, irwin2022chemformer, tavakoli2024ai,tavakoli2022quantum}. For hydrogel synthesis, a good predictor must generalize well to radical chemistry, as most of the mechanistic reactions in hydrogel synthesis involve radical species.  Furthermore, it must be capable of predicting all possible mechanistic pathways and byproducts throughout the entire hydrogel synthesis.
In addition to AI-based reaction predictions, Fourier-transform infrared spectroscopy (FTIR) is employed to investigate the presence of functional groups implicated in the proposed structure.

\section{Result and Discussion}\label{sec2}
\subsection{Investigation of Possible Mechanisms}\label{subsec2}
\begin{figure}[!]
  \centering
  \includegraphics[width=\linewidth]{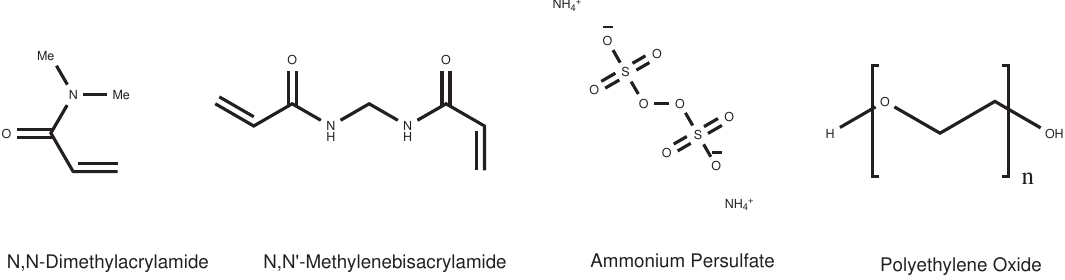}
  \caption{The Initial reactants: N, N-Dimethylacrylamide as the monomer, N, N’-Methylenebisacrylamide as a crosslinker, Ammonium Persulfate as initiator, and Polyethylene Oxide as polymer.}
  \label{fig:fig2}
\end{figure}

The desired molecular arrangement, encompassing network segments and linear chains, employed a modified free radical polymerization approach, making specific adjustments to the initiator/crosslinker concentration (see Fig. \ref{fig:fig2}). This process combined four distinct initial components into a single reaction vessel, with the objective of incorporating linear polyethylene oxide into the network composed of N, N-Dimethylacrylamide (with N, N'-Methylene Bisacrylamide as the crosslinker), initiated by ammonium persulfate.
Before beginning the investigation, certain assumptions were made. For computational efficiency and to accommodate limitations in the predictor system regarding input atom number, we treated two monomer units as a dimer, akin to a polymer unit. This simplification is acceptable because the number of units in the growing chain from dimer and above does not affect the outcome of reactions based solely on the molecule size \cite{kayik2014stereoselective,krajnc2001kinetic}. For instance, Coote et al. demonstrated that in free radical polymerization of acrylonitrile and vinyl chloride, using trimers or longer chains had an insignificant impact on the rate constants \cite{izgorodina2006accurate}.
The system incorporated various critical parameters, including reactants, the depth to which the reaction proceeds, selectivity parameters determining the preferred structure among possible reactions at each depth, and context parameters specifying elements reintroduced into the system at each depth. In the initial phase, we examined the effect of adding monomer as context (Supplementary Section 2.1) and the impact of water as a reactant or context parameter (Supplementary Section 2.2). The findings indicated that adding monomers as context is crucial for mimicking real-world conditions and facilitating chain growth. Moreover, no significant disparities were observed between using water as a reactant or context parameter, leading us to select water as our reactant.
Our initial investigation focused on ammonium persulfate decomposition, the pivotal first stage of reactions and initiation steps. The predictor system delineated 12 of the most plausible reaction pathways for ammonium persulfate decomposition, resulting in 13 distinct products across all reactions (see Fig. \ref{fig:fig3}). These reactions are discussed comprehensively in Supplementary Section 3.1. Notably, it is well-documented in the literature \cite{herrera2022role,herrera2023new}, and empirically that all products can be attained, but only specific products are responsible for initiating polymerization and facilitating growth. According to the literature, sulfate radical anion (SO4·-) \cite{kamagate2018activation}, hydroxyl radical \cite{shi2018stretchable,khan2020effect,bashir2021flexible} and bisulfate \cite{oun2017effect} play pivotal roles in activating the monomer and converting it into a monomer radical for polymerization growth.
We ran these initiator radicals individually with each monomer, crosslinker, and polymer (Supplementary Section 3.2). Due to their high reactivity, these initial reactions primarily involved radical-radical interactions and rearrangements, with limited interactions with reactants. This situation incurred computational costs. To address this issue, we adopted three approaches.
In the first approach, we ran all reactants (monomer, crosslinker, and polymer) with ammonium persulfate (Supplementary Section 3.3.1). We categorized all reactions into four groups, each associated with one of the four radical initiators predicted by ammonium persulfate. Group A was initiated by the decomposition of ammonium persulfate (APS) into sulfur trioxide, sulfate radical anion (SO4·-), and oxygen radical anion. These species subsequently combined to form peroxymonosulfate. Notably, this peroxymonosulfate initiator deviates from real-world behavior, as it initiates reactions by attacking the methyl group of monomers, a phenomenon unlikely to occur in reality. Group B closely mirrored the initial decomposition of ammonium persulfate (APS) observed in Group A. However, in subsequent reactions, a novel occurrence unfolded as the sulfate radical anion (SO4·-) and oxygen radical anion combined to form a rarely reported SO5 structure. This SO5 species initiated the reaction by attacking a monomer’s double bond, initiating polymerization growth. Group C deviated from the previous groups in its reaction sequence. In the initial reaction, the decomposition of ammonium persulfate (APS) yielded two sulfate radical anions (SO4·-), leading to the formation of sulfur trioxide and oxygen anion radicals. Notably, the oxygen anion radical served as an activator, attacking the double bond and generating a radical monomer. Group D was initiated by the decomposition of ammonium persulfate (APS) to produce two sulfate radical anions (SO4·-). In contrast to previous mechanisms, sulfate radical anion directly attacked the monomer. However, in the third reaction, a novel occurrence unfolded as sulfate radical anion on the monomer interacted with oxygen radicals on one end and SO3 on the other.
\begin{figure}[!]
  \centering
  \includegraphics[width=\linewidth]{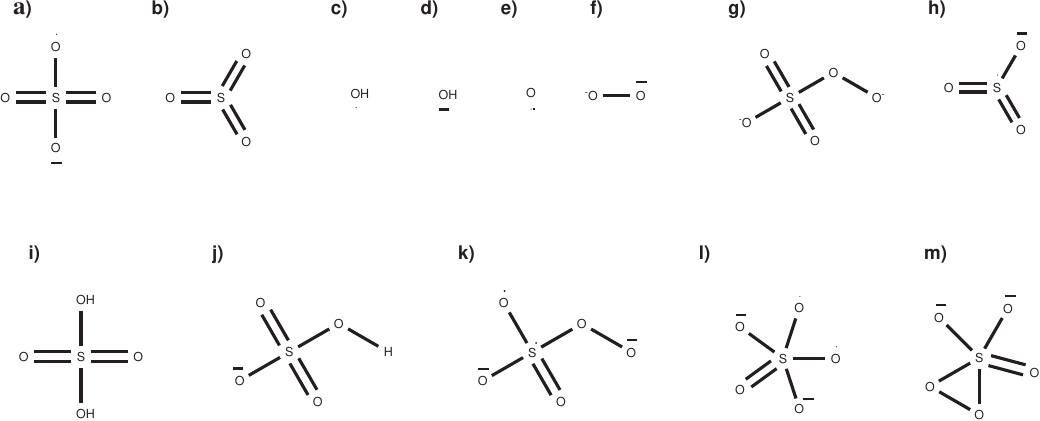}
  \caption{Products Attainable from Ammonium Persulfate (APS) Decomposition: a) Sulfate Radical Anion (SO4·-); b) Sulfur Trioxide; c) Hydroxyl Radical; d) Hydroxyl Ion; e) Oxygen Radical Anion; f) Superoxide Anion; g) Peroxymonosulfate; h) Sulfite Radical Anion; i) Sulfuric Acid; j) Bisulfate; k) Dianion Bisulfate; l) SO5 Anion Diradical; m) SO5 Dianion.}
  \label{fig:fig3}
\end{figure}
In accordance with findings from the literature \cite{bednarz2015free}, we adopted the approach designated as "Group C" for our primary investigation (Supplementary Section 3.3.2). This approach used N, N-Dimethyleacrylamide as the monomer,  N, N'-Methylene Bisacrylamide as the crosslinker, water as a conditioning factor, and polyethylene oxide (PEO) as the polymer. The investigation extended to a depth of 4 with a selectivity factor of 2. Duplicate structures were eliminated, and the context parameter was set to "monomer" to closely emulate real-world conditions, resulting in the identification of 11 distinct structures.
Among these 11 structures, 7 exhibited a reaction pathway in which the sulfate radical anion (SO4·-) decomposed into sulfur trioxide and oxygen radical anion, subsequently initiating an attack on the monomer (See Fig. \ref{fig:fig4}A). In the remaining 5 cases, the sulfate radical anion directly attacked the reactant components, leading to the incorporation of ether groups into the polymer backbone, a scenario inconsistent with real-world behavior. While these 7 structures share similarities, the order of initiator attack varies, with some structures involving the initial attack on the monomer followed by an attack on the crosslinker, and vice versa. In actuality, the initiator simultaneously engages all components, a phenomenon that becomes apparent only at deeper reaction depths due to the high reactivity of monomers and sulfate radical anion (SO4·-).
To address these complexities, we pursued an alternative approach. We investigated the involvement of hydroxyl radical, another initiator radical that is reported extensively in the literature (Supplementary Section 4.1). As illustrated in Fig. \ref{fig:fig4}B, the hydroxyl radical initiated an attack on the monomer, leading to the formation of a monomer radical and subsequent connection of monomers and crosslinkers. This confirmed that either the sulfate radical anion (SO4·-) or the hydroxyl radical, or potentially other radical anions, are exclusively responsible for initiating attacks on monomers and crosslinkers, with the specific radical initiator being of secondary importance. In other words, the specific radical initiator that attacks the reactant is not important because whether sulfate radical anion (SO4.-) or hydroxyl radical attacks the monomer and crosslinkers, it is more important that these radicals attack all reactants and condition them for the next steps in growth polymerization in the same way.
In all cases involving the sulfate radical anion (SO4·-), PEO did not participate in the reactions. However, in select runs involving the hydroxyl radical, PEO lost its hydrogen and had a radical on its end chain. This is detailed in Supplementary Fig. 62 and exemplified in Fig. \ref{fig:fig4}C1. According to the literature \cite{lee2020persulfate}, the sulfate radical anion (SO4·-) may initially take on this role, after which the hydroxyl radical is able to engage. It is worth noting that PEO involvement occurs at deeper reaction depths due to the high reactivity of monomers and the sulfate radical anion (SO4·-).
Subsequently, PEO, bearing a radical on its end group, engages in reactions with the 3D hydrogel network, facilitating the interconnection of networks (Fig. \ref{fig:fig4}D). In this manner, PEO becomes incorporated between networks, allowing for free movement when pressure is applied to the hydrogel. This represents the first proposed final structure.
In reality, the possibility of obtaining hydrogen from the PEO end group is implausible, given its random coil structure and the inherent difficulties in accessibility. As mentioned, we opted to represent PEO with a two-unit segment of ethylene oxide, though perhaps using a longer segment would better align with the system predictor's capabilities. In Supplementary Section 4.2, we investigated PEO’s size. These runs predicted chain scission, as depicted in Fig. \ref{fig:fig4}C2 \cite{emami2002peroxide}.
Further investigations explored additional potential components, such as superoxide anion, and their interactions with water and our final structure. These explorations led to the production of the C=O radical, as illustrated in Fig. \ref{fig:fig4}E (comprehensive details are available in Supplementary Section 5). References in the literature indicate that C=O can be formed from HOO and other species, such as HOO- (Supplementary Fig. 70) \cite{gomez2003kinetics} and that the presence of hydroxyl radical in the products enables the C=O radical and hydroxyl radical to produce a carboxyl group. The system predictor is trained first on radical species, and the final step may involve the carboxyl group. Considering the potential occurrence of chain scission, short chains bearing OH groups may react with this carboxyl, resulting in ester formation via stepwise growth (Fig. \ref{fig:fig4}f). This represents the second proposed final structure.
In summary, either the sulfate radical anion (SO4·-) or the hydroxyl radical plays a key role in initiating attacks on monomers and crosslinkers, with their specific identities being of secondary importance. These radicals drive the polymerization and network formation process. In one mechanism, PEO may acquire hydrogen from the end group, attaching itself to networks through radical interactions facilitated by backbiting \cite{reber2007thermodynamic}. In the second mechanism, PEO chain scission may occur, allowing for the formation of ester bonds through reactions involving carboxyl groups. In this manner, PEO effectively connects all networks.

\begin{figure}[!]
  \centering
  \includegraphics[width=0.95\linewidth]{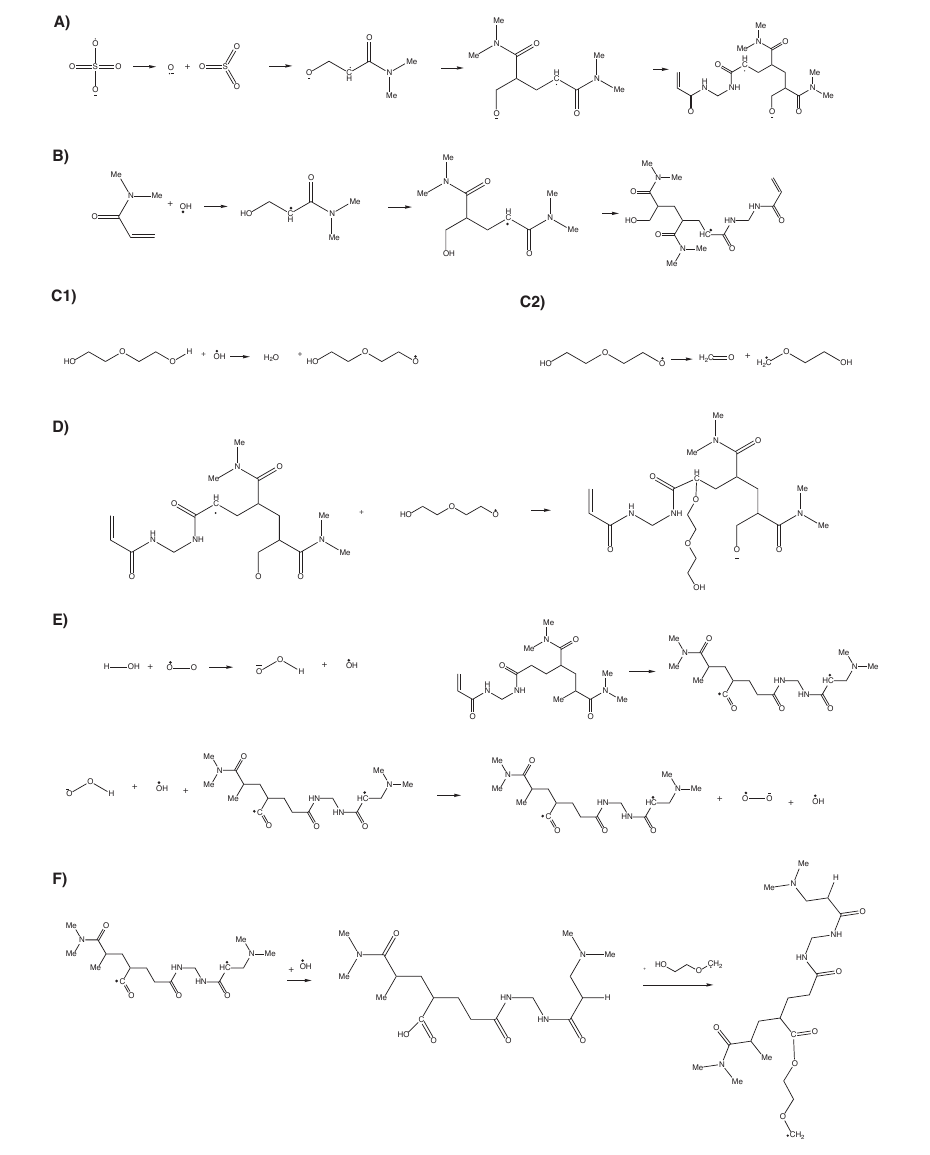}
  \caption{Reaction Mechanisms and Proposed PEO Involvement: A) Gelation process initiated by sulfate radical anion. B) Gelation process initiated by hydroxyl radical (OH·). C) Two proposed mechanisms for participating in polyethylene oxide (PEO): 1. Acquisition of hydrogen from PEO end groups. 2. Chain scission involving PEO. D) First Mechanism: Radical PEO attaches to the hydrogel network. E) Interaction between water and superoxide ion. F) Second Mechanism: Radical carbonyl group and hydroxyl group lead to the emergence of a carboxyl and ester.}
  \label{fig:fig4}
\end{figure}

\subsection{FTIR and Characterization}\label{subsec2}
\begin{figure}[!]
  \centering
  \includegraphics[width=0.9\linewidth]{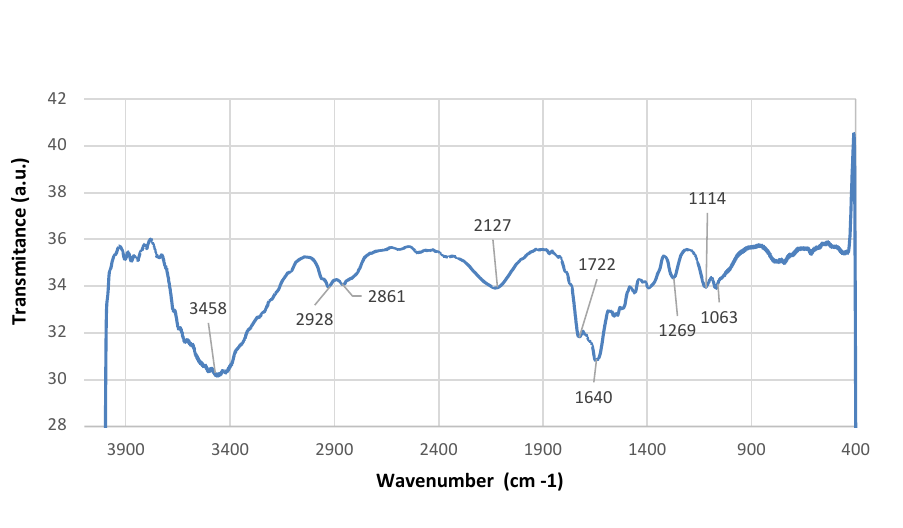}
  \caption{FTIR spectrum of drawn hydrogel left for several months to be air-dried.}
  \label{fig:fig5}
\end{figure}

As shown in Fig. \ref{fig:fig5}, peaks at 1063 and 1147 cm-1 presumably point to the C-O stretch of primary alcohols, numerous PEO hydroxyl end groups after chain scission, and aliphatic ethers (PEO backbone), respectively. Absorption at 1269 is consistent with 1240-1340 cm-1 absorption of amides. The band in 1640 cm-1 pertains to alkenes, which may also reveal PEO chain scission and vinyl ether formation. The 1722 peak is due to a carbonyl group and confirms ester and carboxylic acid formation. However, it appears to indicate ester formation since there is a peak in 1147 for aliphatic ethers. Absorption at 2127 shows ketene formation and confirms PEO degradation and ketene formation. Peaks at 2861 and 2928 illustrate -CH2- in backbone and carboxylic acid, respectively. Broad absorption at 3458 is due to numerous OH end groups in PEO after degradation, which indicates an increase in chain scission (this relationship demonstrates the second structure \cite{socrates2004infrared,larkin2017infrared}).
The possibility of these mechanisms depends crucially on the initiator’s concentration. In the event that the initiator is added to a level exceeding the critical concentration, all carbonyl groups may undergo oxidation and convert to carboxylic acid, leading to the formation of an brittle crosslinking network. Consequently, the hydrogel structure would be unable to withstand mechanical stress and might break easily. Conversely, adding insufficient initiator concentration may result in a few carbonyl groups being oxidized to carboxylic acid, leading to the presence of a few linear chains within the network. As a result, the hydrogel would exhibit poor stretchability and might break upon initial stretching. When the critical concentration of the initiator is added, the carbonyl chains are broken at an appropriate rate, and PEO molecules can be incorporated into the network. As a result, a network of N, N-Dimethylacrylamide, and a linear PEO chain can be formed, enabling the PEO chain to stretch freely when stress is applied.
In the polymer field, Flory's pioneering work suggested a random coil theory \cite{flory1953principles}, which dictates macromolecules’ intrinsic properties, including end-to-end distance, gyration radius, exclusion volume, and other mechanical and soluble properties \cite{flory1969statistical,sperling2005introduction}. In conventional polymers, the stretching of a single polymer chain is hindered by its exclusion volume, which converts stress into polymer chains, resulting in breakage at the entanglement point. This limitation impedes the development of ultra-stretchable polymers through entanglement alone. However, by deliberately generating a linear chain that is devoid of any nearby chains and has ample exclusion volume to facilitate easy stretching, the polymer can be stretched until all the chains are converted into the linear configuration. Such behavior at the atomic scale holds promise for producing hydrogels with extraordinary properties. To this end, we intend to create a series of networks comprising a single linear chain between them to realize the desired level of stretchability.
Consequently, in typical PEO, chain entanglement occurs due to neighboring chains’ exclusion volume effect, resulting in the polymer being broken under stress at the entanglement point. Incorporating linear PEO chains between hydrogel networks allows PEO chains to stretch more easily and increases the hydrogel network’s flexibility.

\subsection{Stretchability and Properties}
\label{subsec2}
\begin{figure}[!]
  \centering
  \includegraphics[width=0.9\linewidth]{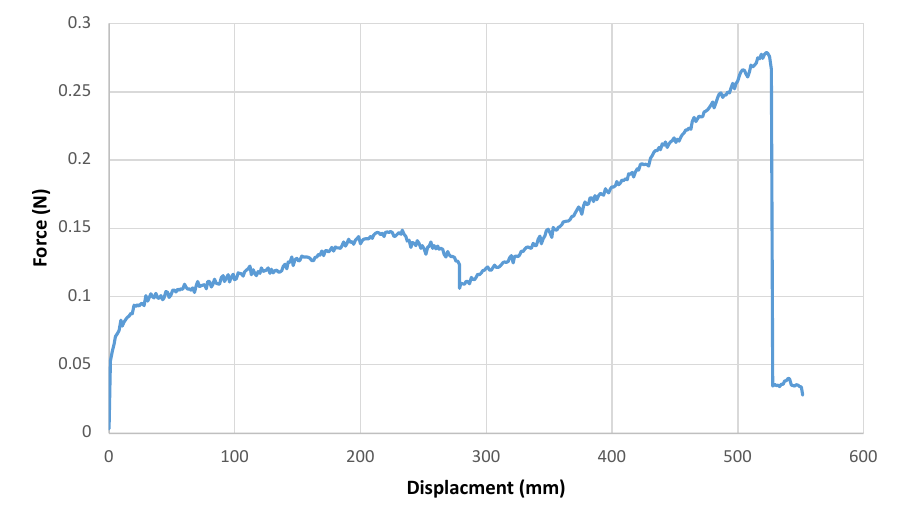}
  \caption{Tensile curve, resembles to a great extent typical elastomers’ axial extension curve shape.}
  \label{fig:fig6}
\end{figure}

To visualize our approach’s unique capabilities, we have conducted an experimental study to create a highly stretchable construct. The monomer was polymerized at properly tuned low concentrations of initiator/crosslinker in the presence of linear polymer chains at their excluded volume dilution to avoid either extremely soft or brittle premature breaking (see Synthesis Method for details). A sample was drawn while the monomer was wrapped around the spatula from the hood surface to the laboratory floor; the sample showed extraordinary extensibility potential, to a magnitude that is nearly invisible (hypothetically, at their limit of extensibility a few fully elongated single chains are difficult to see by the naked eye) and left there for few weeks. Random coil retraction to the original disordered equilibrium state confirmed the polymer chains’ entropy transition: from order to disorder (Supplementary Fig. 72 b and c, respectively). The associated movies capture the unorthodox behaviors in the structure proposed above (see supplemental movies for details). The orchestrated synchronized concentrations between initiator/crosslinker, linear polymer, and monomer at extreme dilution provide a narrow range that facilitates molecular-level capabilities to a degree that imposes eccentric behaviors. For out-of-tune concentrations (either below or beyond values provided in the supplementary synthesis method), gradual thinning upon extension and collapsing demonstrates these effects at different angles, enabling one to grasp different aspects of molecular-level arrangement competence and its profound effect on macroscopic behaviors. Videos arranged sequentially illustrate samples of extensibility, low extensibility, and non-extensibility (for deviation from a narrow range of concentration implemented in the synthesis method). Retraction at a slow pace confirms thermodynamically driven phenomena. 
A tensile test (Fig. \ref{fig:fig6}) showed a 26000\% increase in length compared to the initial 2 mm gauge size. 
Strength at the breaking point of 9 KPa was obtained from the original sample diameter of 3.1 mm and force at a break of 0.27 N. Other experiments, including dynamic mechanical analysis (DMA), in shear and axial modes and stress relaxation tests, are provided in Supplementary Section 6. 
Stress relaxation was implemented at 150 mm displacement for 20 hours.  An incremental force increase to 0.21 N showed a sharp contrast to other solid materials, with stress decay under constant strain \cite{chaudhuri2016hydrogels,bauer2017hydrogel}. The drying effect may justify the rise in stress compared to typical materials, but its effect is not profound past a short period for an extremely thin-diameter hydrogel. A stress surge from almost 0.07 N to 0.21 N displayed random coil retraction to the original entropic disordered state. This behavior is especially useful for applications that need a material to stay intact under constant strain. 

\section{Conclusion}\label{sec3}
In conclusion, our study introduces a novel approach to initiation and crosslinking in free radical polymerization, specifically designed to induce deliberate chain scission through strategically positioned chemical groups within polyethylene oxide samples.
Our research provides a robust theoretical foundation, rooted in condensed-matter physics, supporting the emerging understanding that exploring molecular-level phenomena yields significant implications at the macroscopic scale. The spatial arrangement of macromolecules and polymer chains, transitioning from a compact to a fully extended form, exemplifies this macro-level phenomenon. Leveraging artificial intelligence prediction systems, we have shed light on the mechanisms underlying this phenomenon, enabling the transformation of a flexible macromolecule from a compact state to full extension while circumventing the exclusion volume around linking chains.
The fundamental structure of span network hydrogels involves interconnections among N, N-dimethylacrylamide networks, linked by methylene bisacrylamide crosslinkers, with these networks further connected through polyethylene oxide (PEO) chains featuring ester group connections. This intricate molecular structure comprises ideal networks interconnected with linear chains within a system composed of rod chain-network segments, purposefully designed to prevent interference and entanglement with other molecules. The observed partial retraction of elongated chains over time, driven by entropic processes during extensional testing and reflected in stress relaxation diagrams, corroborates our previously proposed theory, demonstrating the return of rod chains to their original state of random coil entropic disorder.

\section{Methods}
\subsection{Synthesis Method}\label{subsec4}
N, N’-Dimethyl acrylamide (Aldrich 99\%) inhibitor was removed by an aluminum oxide column and purged with nitrogen gas for at least 10 minutes. Polymerization (400µL monomer) and crosslinking were performed in a water bath at 50°C in the presence of ammonium persulfate (Sigma 98\%) as an initiator and N, N’-Methylenebisacrylamide (Sigma 99\%), as a crosslinking agent (both at 30µL). Polyethylene oxide (2,000,000 Aldrich) was added to the reaction vessel based on the theoretical text calculations provided previously. The reaction container was left for one hour in a water bath under constant stirring. After polymerization and crosslinking, water was added to the sample and remained for at least one hour before removing hydrogel from the reaction vessel with a spatula.
\subsection{Mechanical Tensile Experiments}\label{subsec4}
A tensile experiment (using a Shimadzu AGS-X 5Kn test frame) was performed on hydrogel to determine ultimate strength at break. Hydrogel was mounted between the device’s upper and lower jaws and adjusted to maintain an upright position. Grips were tightened to avoid slippage at high extensions, but not to the extent that they cut the sample from overexertion of shear force. The experiment was run at a 2 mm gauge length, at 80 mm/s head speed, and the number of collected points was adjusted to 1 point per second, and the diameter of the original sample was 3.1 mm. The experiment was paused at 272 mm extension, as it was necessary to tighten the jaws again prior to resuming tensile force. 
\subsection{Fourier Infrared Spectroscopy (FTIR)  }\label{subsec4}
Infrared measurement was carried out on a Bomen (Canada) FTIR spectrophotometer. The hydrogel was drawn on a tensile machine until it reached the breaking point and left for several months to completely remove any entrapped water inside the drawn fiber (dried gel). Dried gel was broken into little pieces, mixed with potassium bromide (KBR), and compressed into a pellet shape.

\subsection{Radical Reaction Predictor}\label{subsec4}
To predict the entire pathways of elementary reaction steps involved in the synthesis of the hydrogel, we employ the mechanistic reaction predictor \cite{tavakoli2024ai} that is specifically designed for radical chemistry. More details on the parameters, models, and pathway search specifications can be found in the Supporting Information.

\section*{Acknowledgment}
The research on ultra-stretchable hydrogel was mainly done by S.H.Emmami at Marquette University and Michigan State University.
The deep learning models and the reaction predictors were developed by M.Tavakoli and P.Baldi at the University of California, Irvine.
The authors would like to express their appreciation to the Marquette University Dental School for providing computational sources.







\bibliographystyle{unsrt}  
\bibliography{main}  

\newpage
\section*{Supporting Information}
\setcounter{section}{0}
\renewcommand{\thesection}{S\arabic{section}}

\section{Classification of Hydrogels}\label{sec1}
\begin{figure}[!]
  \centering
  \includegraphics[width=\linewidth]{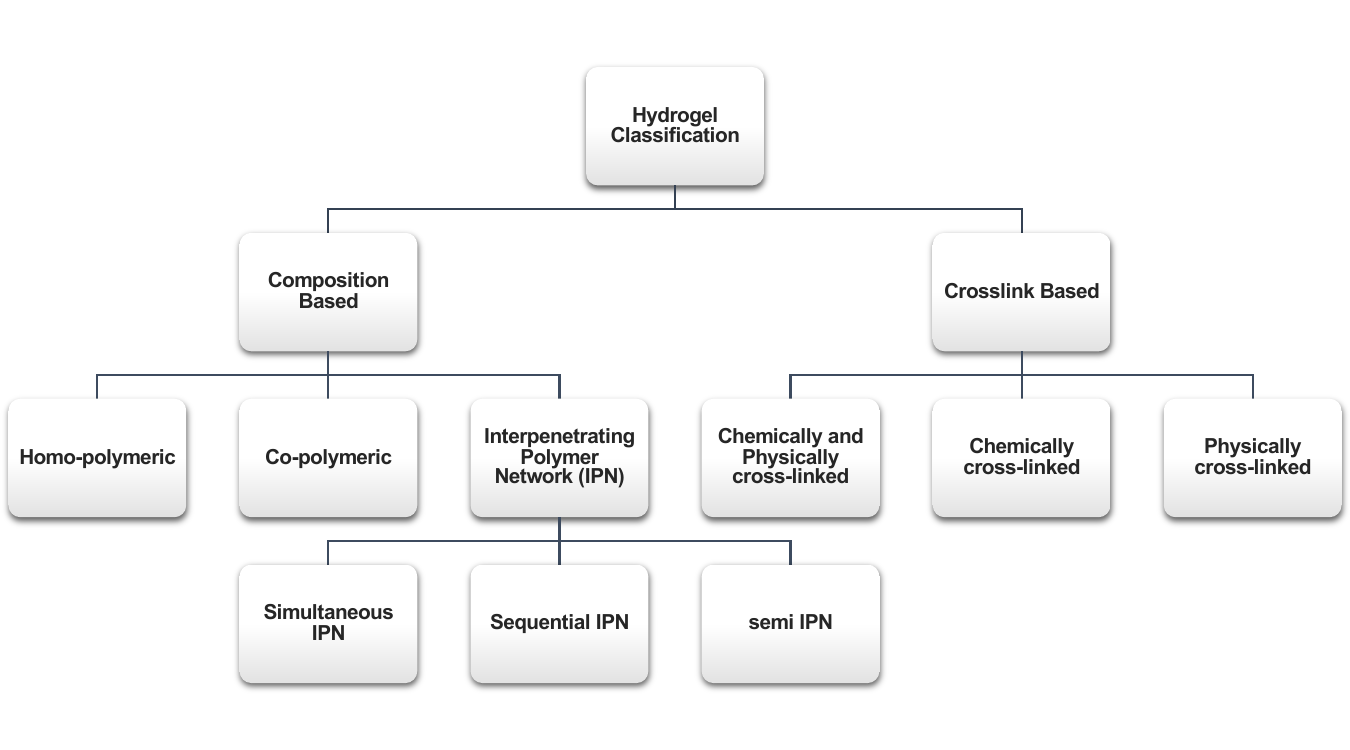}
  \caption{Hydrogel classification based on composition and crosslink nature.}
  \label{fig:Supplementary Fig 1.}
\end{figure}

An alternative conceptualization of hydrogels underscores their capacity for absorbing water and retaining a substantial portion thereof within their structure, without undergoing dissolution in aqueous environments\cite{sun2012highly}. The primary limitation of hydrogels derived from a single polymer resides in their suboptimal mechanical performance
and extensibility. For instance, when an alginate hydrogel is stretched to approximately 1.2 times its original length, it is susceptible to rupture\cite{sun2012highly}. Hydrogel classification ( see Fig. \ref{fig:Supplementary Fig 1.} based on specific attributes has been made feasible by introducing interpenetrating polymer network (IPN) hydrogels and other strategies aimed at augmenting their mechanical characteristics. Hydrogels, as the name suggests, are composed of a singular monomer unit forming a network, and these networks can be structured in various configurations. Consequently, hydrogels can be dichotomized into two categories predicated on the chemical or physical nature of their crosslink junctions. Chemically crosslinked networks possess enduring junctions, whereas physical networks feature transient junctions stemming from phenomena such as polymer chain entanglements or physical interactions, including ionic interactions, hydrogen bonds, or hydrophobic interactions\cite{atala2018principles}. Hydrogels can be further classified into three categories based on their polymeric composition\cite{zhao2013degradable}. Their preparation method yields several notable classes of hydrogels: (a) Homopolymeric hydrogels, constituted by polymer networks derived from a single monomer species, serve as the fundamental structural unit of any polymer network\cite{iizawa2007synthesis}. Depending on the monomer’s inherent characteristics and the polymerization technique employed, homopolymers may exhibit a crosslinked skeletal structure. (b) Co-polymeric hydrogels encompass two or more distinct monomer species, at least one of which is hydrophilic, arranged in a random, block, or alternating configuration along the polymer network chain\cite{yang2002colon}. (c) Interpenetrating Polymeric Hydrogel (IPN), a pivotal category of hydrogels, involves two distinct crosslinked synthetic and/or natural polymer components coexisting within a network structure\cite{maolin2000swelling}. This approaches used for have multiple function instead of using one network with one property.

In semi-IPN hydrogels, one component is crosslinked, while the other remains non-crosslinked. In hydrogel synthesis, the entrapment of a linear polymer in a matrix can yield a semi-interpenetrating network (IPN) hydrogel, which subsequently can be transformed into a fully IPN hydrogel by selectively crosslinking the linear polymer chains\cite{dragan2012macroporous}. IPN represents a versatile polymer system amenable to various classifications\cite{dragan2014design}. Depending on the synthesis chemistry employed, IPN hydrogels can be categorized as (a) Simultaneous IPN, in which both networks’ precursors are mixed and independently synthesized through non-interfering pathways such as chain and stepwise polymerization\cite{myung2008progress,sperling1994interpenetrating,wang2013enhanced}. (b) Sequential IPN, in which a single-network hydrogel swells within a solution that contains a mixture of monomers, initiators, and activators, either with or without a crosslinker. (c) The presence of crosslinker results in a full IPN, while its absence leads to a network featuring linear polymers embedded within the primary network (semi-IPN)\cite{myung2008progress,sperling1994interpenetrating,chivukula2006synthesis}.  

Despite their numerous advantages and improvements in mechanical properties, hydrogels are recognized for their limited stretchability and mechanical strength, constraining their application range. When an IPN network is prepared, it can possess two chemical crosslinks, one of which acts as a sacrificial bond, or two independent crosslinks, one of which is physical and capable of reversibility. This article introduces novel methodologies for fabricating ultra-stretchable hydrogels incorporating linear polymer chains into polymeric networks. To attain heightened mechanical performance and stretchability, our objective is to harness the attributes of linear chains within a polymeric network, thereby achieving an unprecedented level of stretchability.

\section{Radical Reaction Predictor}

For all experiments, we use a specific version of the radical reaction predictor introduced in \cite{tavakoli2024ai}. We use the contrastive learning model presented in \cite{tavakoli2024ai} with an atomic descriptor of length one. This model is trained on the entire dataset of RMechDB \cite{tavakoli2023rmechdb}. 
Even though the RMechDB dataset is dominated by atmospheric data, the reaction predictor is highly generalizable to other radical domains such as polymerization. 

We follow the exact procedure of pathway search described in \cite{tavakoli2024ai}, with the use of different starting materials, branching factors (also referred to as selection factors), depth, and reaction contexts. The selection of these parameters is described in the following sections.

\section{Pre-setting Before Run}\label{sec2}
Before starting the experimental procedures as elucidated in the methodology section, it is imperative to establish certain requisite settings. These settings, crucial to the experiment’s accurate execution, pertain predominantly to the reactants, serving as the fundamental core of the pre-setting parameters. The first important parameter is the reactants, which must be adjusted before being run. Subsequently, these reactants have the potential to engage in chemical reactions with one another. However, it becomes imperative to ascertain the depth to which these reactions are pursued. The question arises: how profound an exploration of the reaction depths should be undertaken? Moreover, within each depth, the reactants can undergo multifarious reactions, prompting the need to discern and prioritize the most probable products—a facet encapsulated within the ambit of selectivity. It is also essential to underscore the context parameters’ significance. The term “context” denotes the specific composition that must be incorporated into each reaction or the depth to which such a composition should extend. The final data format is presented in SMILE notation, characterized by distinct paragraphs delineating the possible products. Each paragraph signifies the distinct number of reactions required to yield these products. Certain paragraphs may exhibit duplications, which have been eliminated to enhance clarity and comprehensibility. In light of the above considerations, it becomes evident that the pre-setting parameters, encompassing reactants, depth, selectivity at various depths, and context, play an indispensable role in shaping the experimentation course. The ensuing sections explore two pivotal questions that merit elucidation: The first query asks whether the monomer should serve exclusively as a reactant or whether it can also be reintroduced into the reaction as part of the context, thereby mirroring real-world scenarios. The second inquiry addresses the role of water in the experimental framework— specifically, whether it should be considered as a reactant or, conversely, as part of the contextual parameters.

\subsection{Effect of Adding Monomer as Context}\label{subsec2}
\begin{figure}[!]
  \centering
  \includegraphics[width=0.8\linewidth]{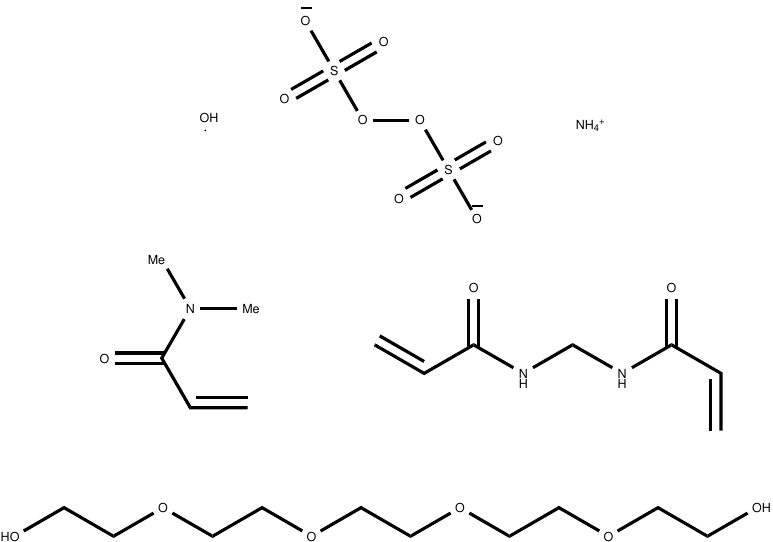}
  \caption{The initial reactants, comprising Hydroxyl Radical, Ammonium Persulfate, N, N- Dimethylacrylamide, N, N’-Methylenebisacrylamide, and a limited quantity of Polyethylene Oxide (constrained to five units due to system limitations). The study encompasses three iterative depths, evaluating all predicted products and highlighting the eight most promising reactions for further consideration.}
  \label{fig:Supplementary Fig. 2}
\end{figure}

We conducted two experimental runs employing N, N-dimethyleneacrylamide, N, N’- Methylenebisacrylamide, and ammonium persulfate as our primary reactants, each explored to a depth of three levels with a selectivity factor of eight, as illustrated in Fig. \ref{fig:Supplementary Fig. 2}. Both runs share a commonality in their initial step, where a monomer serves as the first reactant. However, the key distinction between these two runs lies in their context parameter. In the first run, the monomer is not added back, whereas in the second run, the monomer is added back to the next two levels. In practical, real-world experimentation, a system comprises a multitude of monomers. In such a scenario, radical attacks are not limited to a single monomer; rather, they propagate the chain by incorporating other monomers. However, in our experimental context, the absence of an additional monomer leads to a unique situation where only the initial monomer undergoes activation, as depicted in the accompanying figure. Due to the absence of additional monomers, this monomer rearranges itself to maximize the use of its radical electrons.

\subsubsection{Monomer as Reactant}\label{subsubsec2}
\begin{figure}[!]
  \centering
  \includegraphics[width=\linewidth]{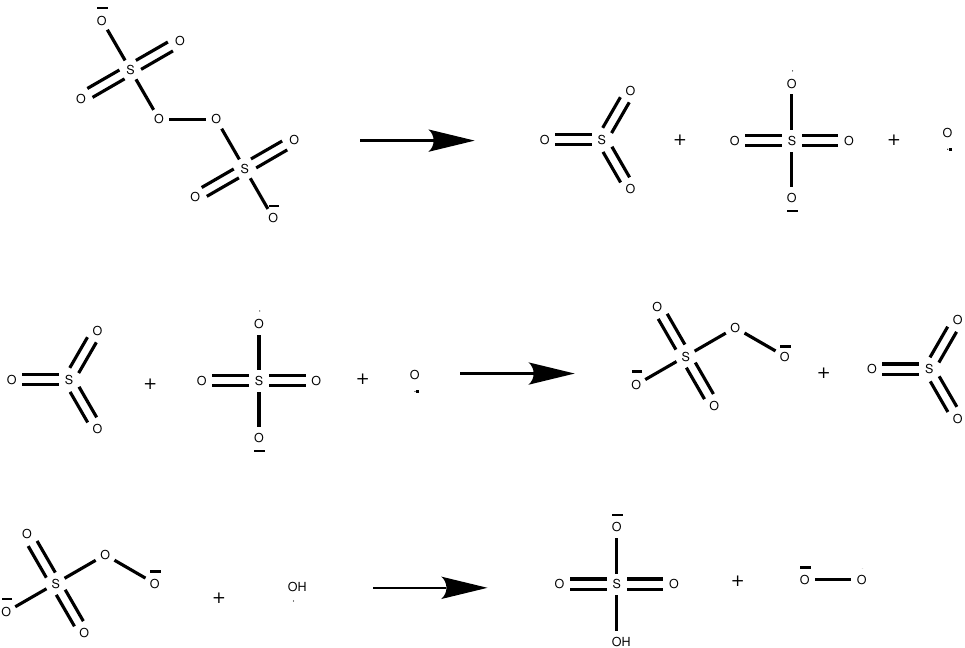}
  \caption{In cases like the one illustrated in this prediction pathway, the initial decomposition of Ammonium Persulfate leads to the formation of reactive intermediates. Due to their high reactivity, these intermediates subsequently engage with other species, ultimately giving rise to stable initiator radicals. Such pathways are potentially applicable across a broad spectrum of systems.}
  \label{fig:Supplementary Fig. 3}
\end{figure}

\begin{figure}[!]
  \centering
  \includegraphics[width=\linewidth]{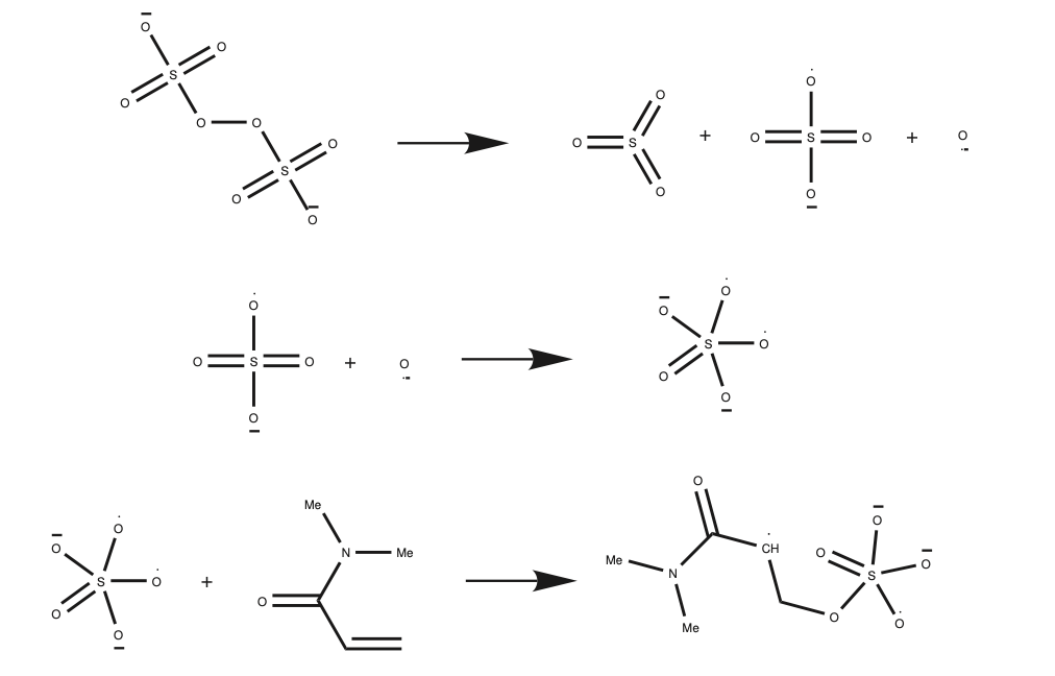}
  \caption{In an alternate scenario, Ammonium Persulfate may undergo decomposition, resulting in the generation of radicals. After a rearrangement process leading to the formation of a stable radical species, the desired radical may initiate a reaction with a monomer. This activation process readies the monomer for subsequent steps in the reaction sequence.}
  \label{fig:Supplementary Fig. 4}
\end{figure}

\begin{figure}[!]
  \centering
  \includegraphics[width=\linewidth]{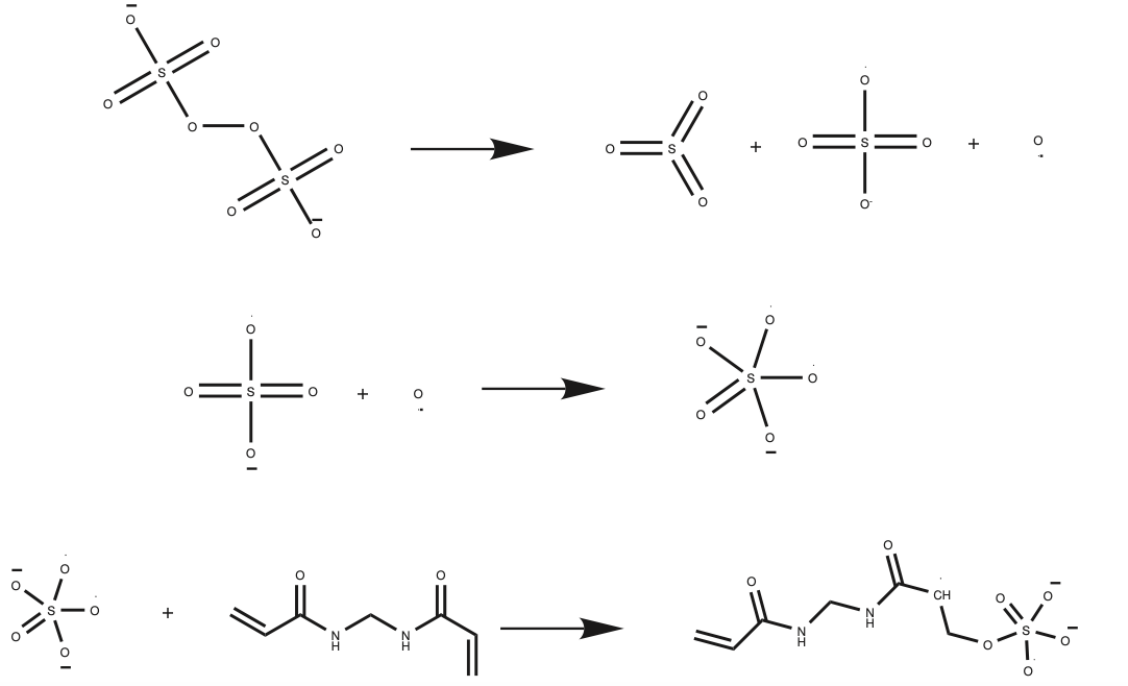}
  \caption{According to predictions, within the spectrum of the most likely reactions, stable radicals originating from the decomposition of Ammonium Persulfate possess the capacity to activate the crosslinker, priming it for the gelation process.}
  \label{fig:Supplementary Fig. 5}
\end{figure}

\begin{figure}[!]
  \centering
  \includegraphics[width=\linewidth]{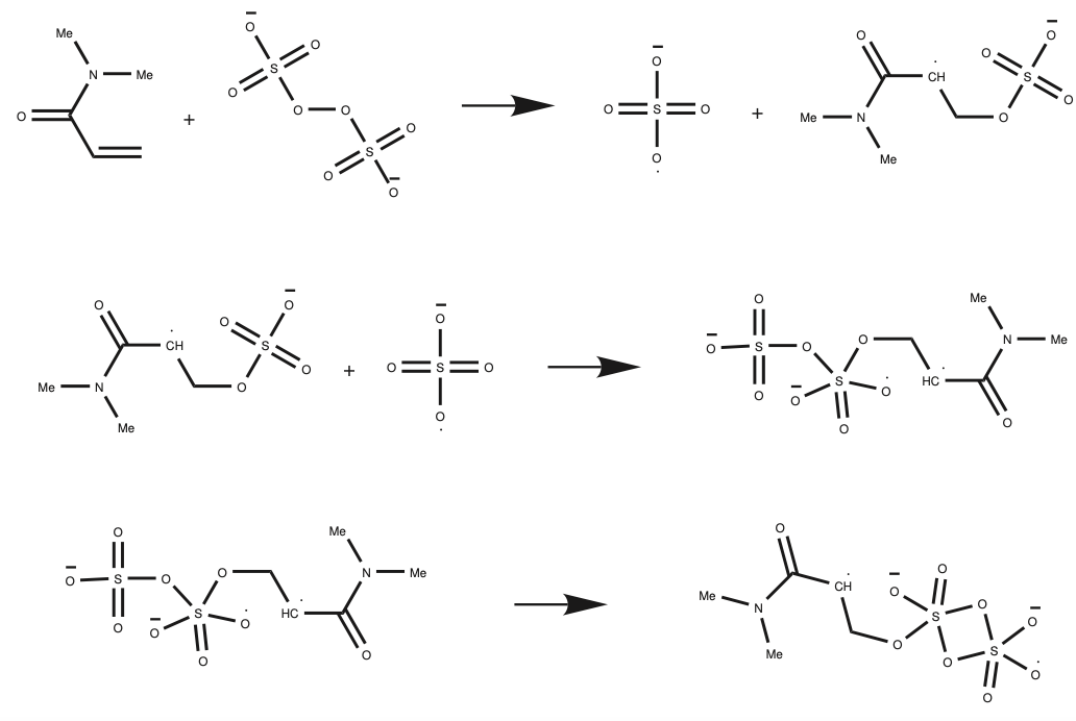}
  \caption{In some cases, the reaction begins with the initiator’s direct attack on the monomer, while in contrast, other scenarios involve an initial rearrangement step leading to a more stable product before monomer interaction. Notably, when no additional monomer is present in the context, these reactions may self-rearrange until reaching their final stages.}
  \label{fig:Supplementary Fig. 6}
\end{figure}

In the first run, we conducted experiments using the aforementioned set of reactants without adding the monomer back at each depth. Among the numerous reactions performed, we have selected and investigated the most significant ones, as elaborated in Figures \ref{fig:Supplementary Fig. 3}-\ref{fig:Supplementary Fig. 6}.

\subsubsection{Monomer as Reactant and Context}\label{subsubsec2}
\begin{figure}[!]
  \centering
  \includegraphics[width=0.8\linewidth]{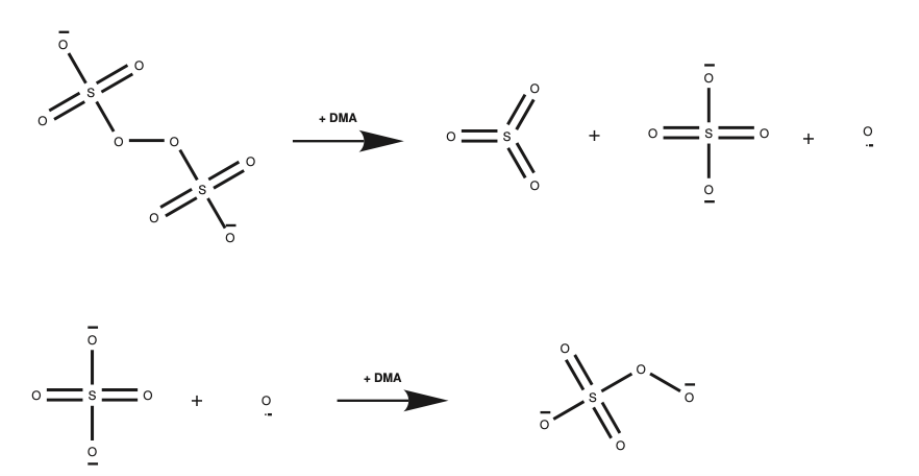}
  \caption{Highlighting a consistent observation across all cases, the addition of monomer does not interfere with or alter the decomposition of Ammonium Persulfate (APS). This consistent behavior is a favorable aspect of the reaction process.}
  \label{fig:Supplementary Fig. 7}
\end{figure}

\begin{figure}[!]
  \centering
  \includegraphics[width=\linewidth]{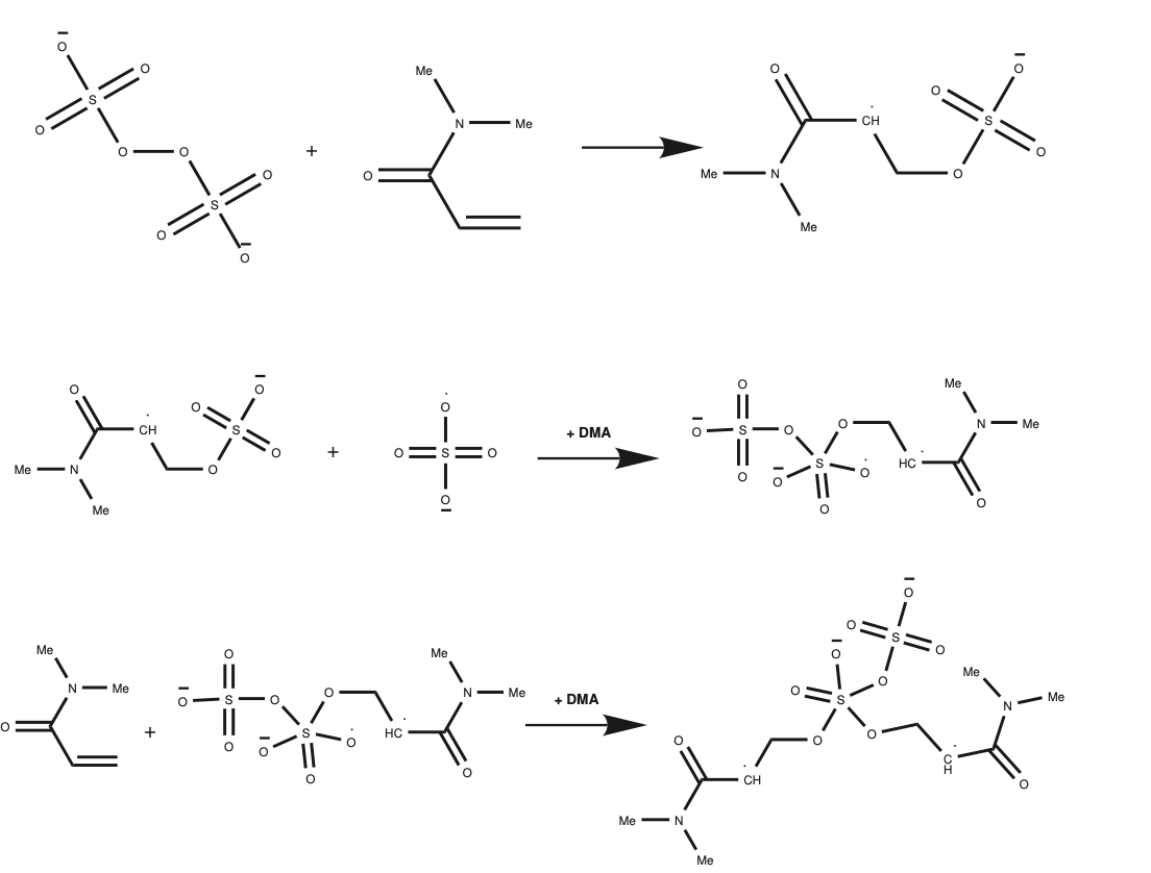}
  \caption{In scenarios following the activation of a monomer, the presence of an adjacent monomer serves as a contextual element, creating an opportunity for chain growth through dimer formation. In these reactions, it is postulated that the dimer, composed of two monomers, symbolizes the inception of a polymer chain.}
  \label{fig:Supplementary Fig. 8}
\end{figure}

\begin{figure}[!]
  \centering
  \includegraphics[width=0.9\linewidth]{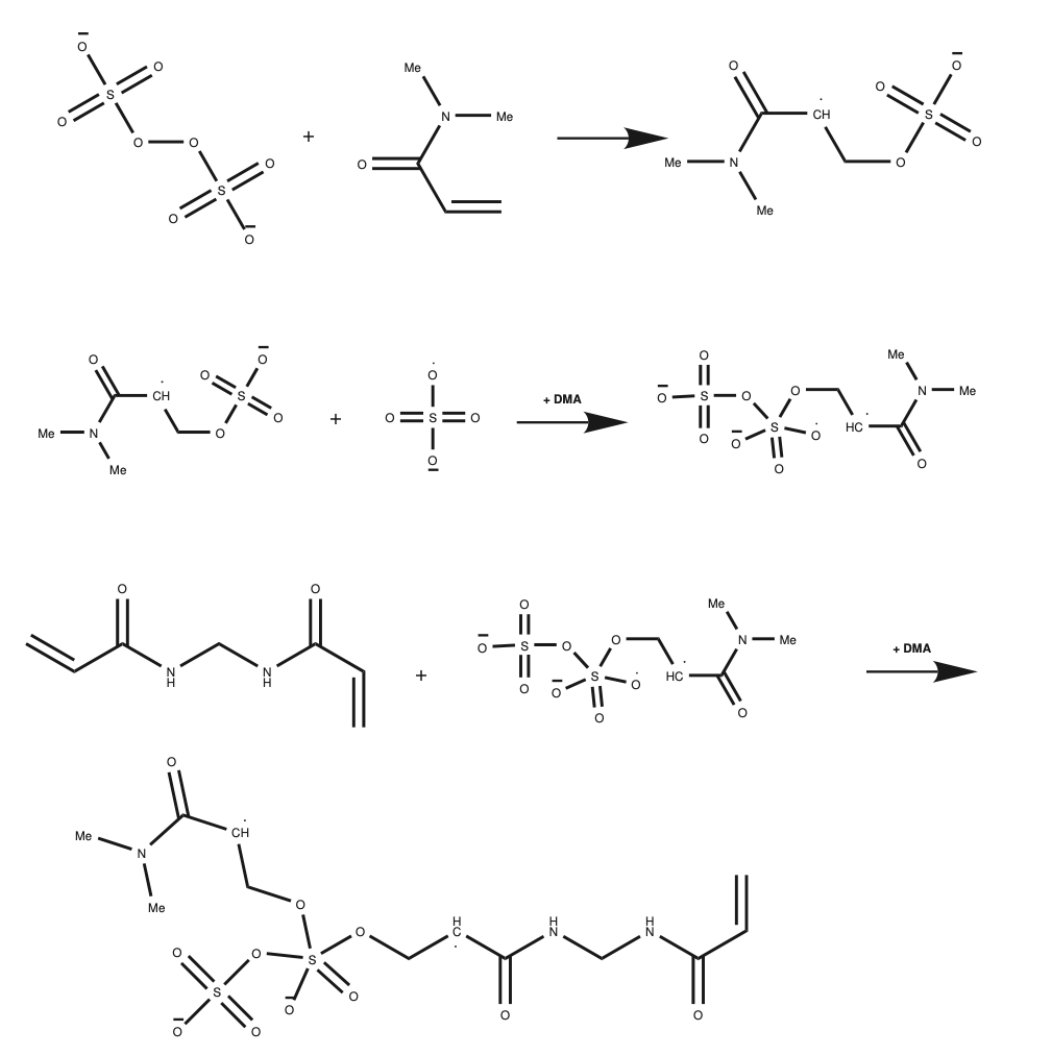}
  \caption{A key observation demonstrates that even in the presence of another monomer, the initially activated monomer retains the capacity to engage with a crosslinker, facilitating their linkage and setting the stage for subsequent reactions.}
  \label{fig:Supplementary Fig. 9}
\end{figure}

As established in the preceding section, it is evident that following the initiation step, in which the initiator radical interacts with the monomer, the chain’s subsequent growth depends on re-adding the monomer to the system. This phenomenon is commonly referred to as the system’s “context setting.” In the present scenario, all experimental parameters remain consistent with those elucidated in the previous section, except for one crucial distinction—monomer reintegration as a contextual element. Under these conditions, it becomes apparent that the activation of one monomer can facilitate the addition of another monomer, thereby providing the necessary substrate for chain growth, as depicted below. Among the myriad reactions predicted by the system, we have selected specific cases for detailed examination, as illustrated in Figures \ref{fig:Supplementary Fig. 7}-\ref{fig:Supplementary Fig. 9}. These selected cases are intended to offer a deeper understanding of the system’s functionality and the impact of certain factors on presetting parameters. Consequently, to emulate real-world conditions more faithfully, we have opted to utilize the monomer as both a reactant and a context parameter.

\subsection{Water Condition}\label{subsec2}
\begin{figure}[!]
  \centering
  \includegraphics[width=\linewidth]{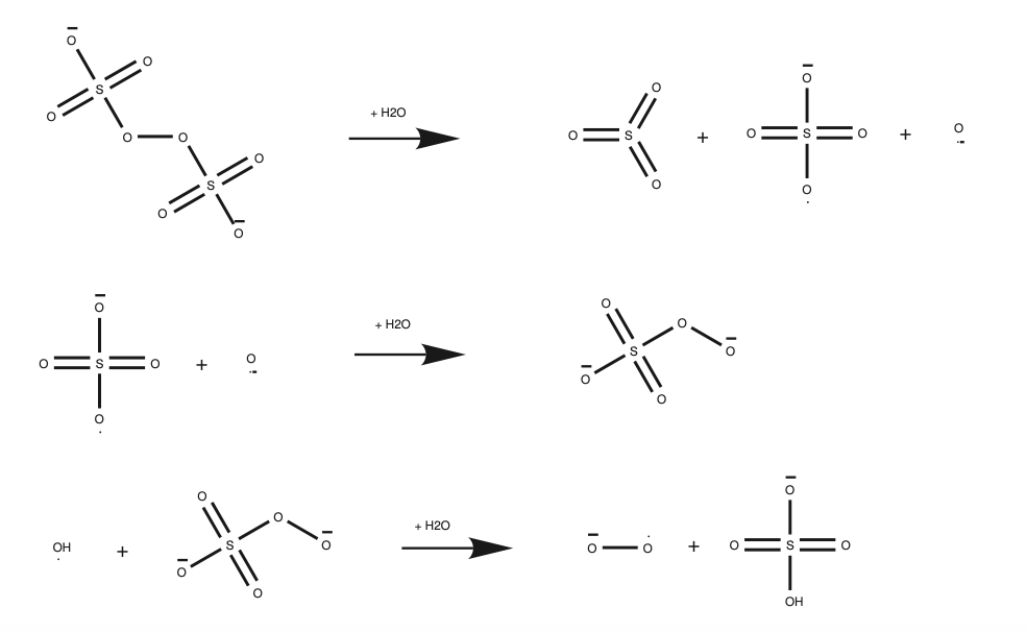}
  \caption{In this scenario, water serves as a contextual component rather than a reactant. The process begins with the decomposition of APS, followed by interactions with monomers or other reactants. Importantly, throughout these reactions, water does not exert any influence on monomers, initiators, or initiator radicals.}
  \label{fig:Supplementary Fig. 10}
\end{figure}

\begin{figure}[!]
  \centering
  \includegraphics[width=\linewidth]{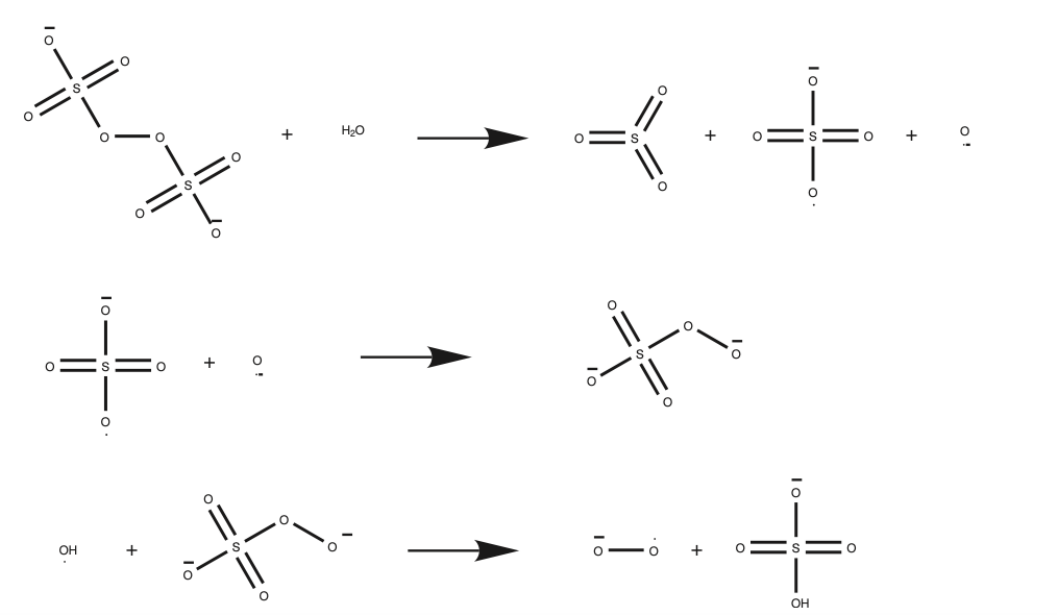}
  \caption{In this scenario, water serves as a reactant. The study explores the influence of water, revealing that its presence has no discernible effect on monomers or APS decomposition.}
  \label{fig:Supplementary Fig. 11}
\end{figure}

\begin{figure}[!]
  \centering
  \includegraphics[width=\linewidth]{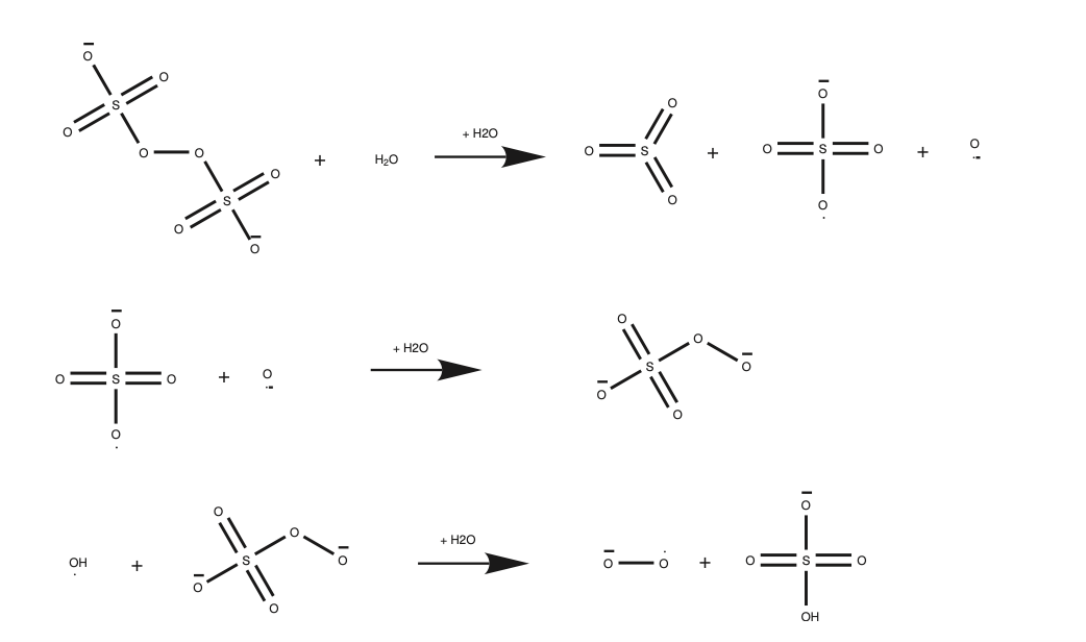}
  \caption{In this case, water serves dual roles as both a reactant and a contextual component. Importantly, it is observed that whether a reaction involves water as a reactant or a contextual component, there is no discernible distinction in the outcomes. Additionally, in specific instances where initiator radicals seek to obtain hydrogen (H) from H2O, it is evident that they can do so equivalently from either the contextual or reactant water, with no distinguishable differences in behavior.}
  \label{fig:Supplementary Fig. 12}
\end{figure}

We conducted a total of three experimental runs, each encompassing three levels of depth and exhibiting a selectivity factor of 8. The reactants employed remained consistent with those presented in Fig. \ref{fig:Supplementary Fig. 2}.In the first run, the reactants remained unchanged; notably, water was not included as part of the reactant composition, but rather, it was used as a contextual element, as illustrated in Fig. \ref{fig:Supplementary Fig. 10}. The second run introduced water as a reactant without subsequent reintegration, as depicted in Fig. \ref{fig:Supplementary Fig. 11}. The final run also incorporated water as a reactant, but it was re-added as a contextual component, as visualized in Fig. \ref{fig:Supplementary Fig. 12}. Building on the insights gained from the preceding section, which considered the results of reintroducing monomers in all cases, we have thoroughly analyzed these three scenarios and their corresponding reactions. This meticulous examination found that there exist no significant discernible differences among these cases. Consequently, we have decided to include water as a reactant in the main and final runs, omitting the re-addition of water as a reactant for the sake of computational efficiency.


\section{Main Run}\label{sec3}
Following the determination of the pre-run setting parameters, it has been unequivocally established that the context parameter bears significant importance. In this regard, it has become evident that the monomer must be added back as a context parameter at each depth, thus playing a crucial role in the system setup. Furthermore, it has been ascertained that water, when used as a contextual element, does not exert any discernible effects on the predictor outcomes. For the primary run, conducted with these established parameters, our initial inquiry focuses on the initiation step: specifically, the decomposition of APS (ammonium persulfate). Within this context, our primary objective is to explore the various reactions that the modeling system can predict. This entails selecting the most plausible radicals produced during the decomposition of ammonium persulfate in conjunction with our primary reactants: namely, the monomer and crosslinkers. Such an approach enables us to deduce the most probable final molecular structures. It is important that our investigation of the decomposition process incorporates two scenarios: one in the presence of water within the reactants and another without water. This comparative analysis underscores the significance of water’s inclusion in these reactions, shedding light on its potential impact.
\subsection{Decomposition of APS}\label{subsec3}
\subsubsection{Decomposition of APS with Water}\label{subsubsec3}
\begin{figure}[!]
  \centering
  \includegraphics[width=\linewidth]{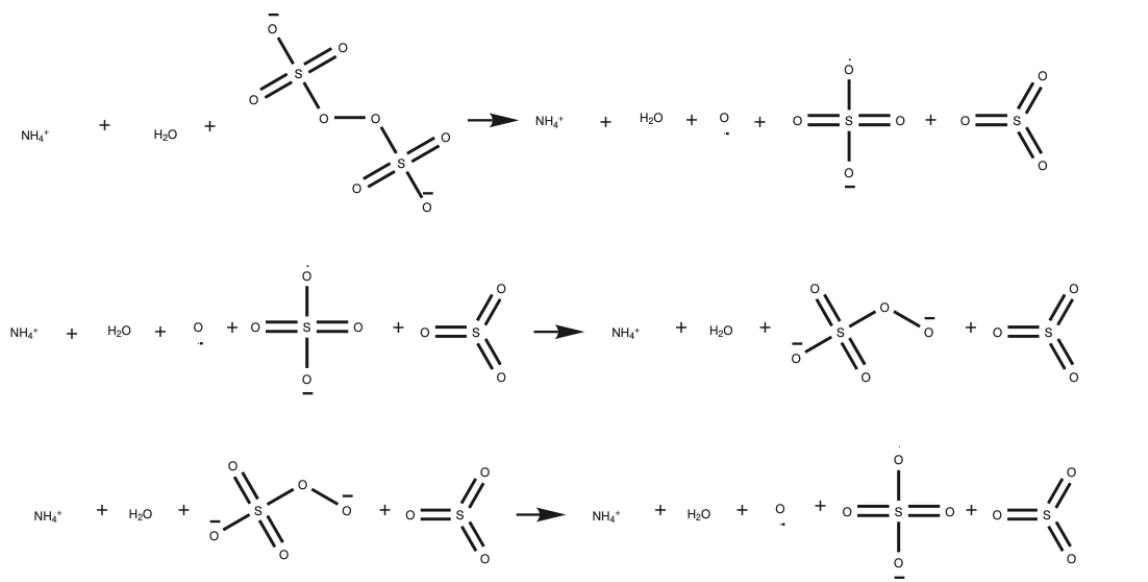}
  \caption{This figure provides valuable mechanistic insights into ammonium persulfate (APS) decomposition. Initially, APS decomposes into the sulfate radical anion (SO4·-) along with sulfur trioxide and the oxygen radical anion, which may represent an unstable form of oxygen. The figure shows that at this mechanistic level, in a subsequent reaction, the oxygen radical anion has the potential to re-engage with sulfate radical anion (SO4·-) to form Peroxymonosulfate. Upon further examination at subsequent depths, this cycle continues, with the transformation oscillating between the formation of sulfate radical anion (SO4·-) and Peroxymonosulfate. This mechanism highlights a dynamic tradeoff between these two radical species.}
  \label{fig:Supplementary Fig. 13}
\end{figure}

\begin{figure}[!]
  \centering
  \includegraphics[width=\linewidth]{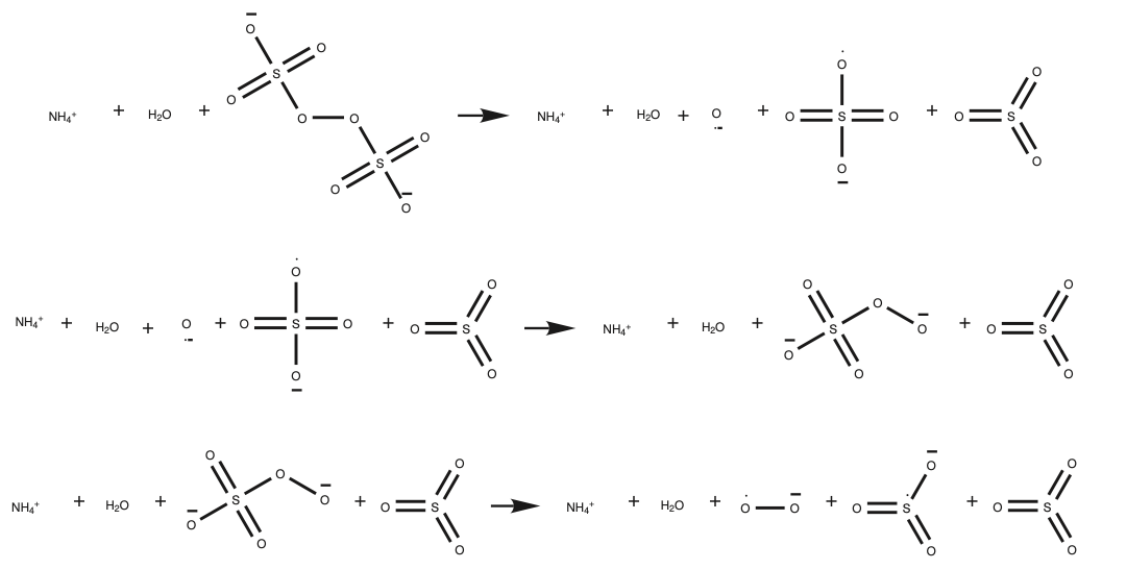}
  \caption{This diagram mirrors the decomposition mechanism observed in Fig. \ref{fig:Supplementary Fig. 13}, involving two primary reactions. However, in this case, the final step reveals that Peroxymonosulfate can undergo further decomposition, forming the sulfite radical and the superoxide anion.}
  \label{fig:Supplementary Fig. 14}
\end{figure}

\begin{figure}[!]
  \centering
  \includegraphics[width=\linewidth]{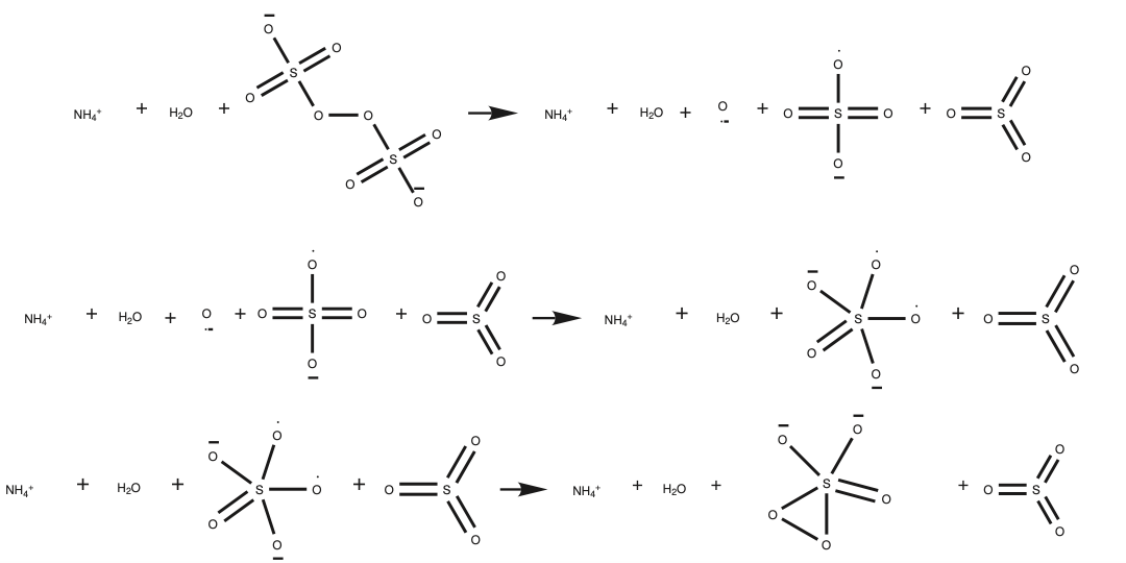}
  \caption{A reactive pathway in which the sulfate radical anion (SO4·-) engages with the oxygen radical anion (O·-) in the second reaction. This interaction leads to the formation of a rare species, SO5, which subsequently undergoes further structural transformation into even more distinctive configurations.}
  \label{fig:Supplementary Fig. 15}
\end{figure}

\begin{figure}[!]
  \centering
  \includegraphics[width=\linewidth]{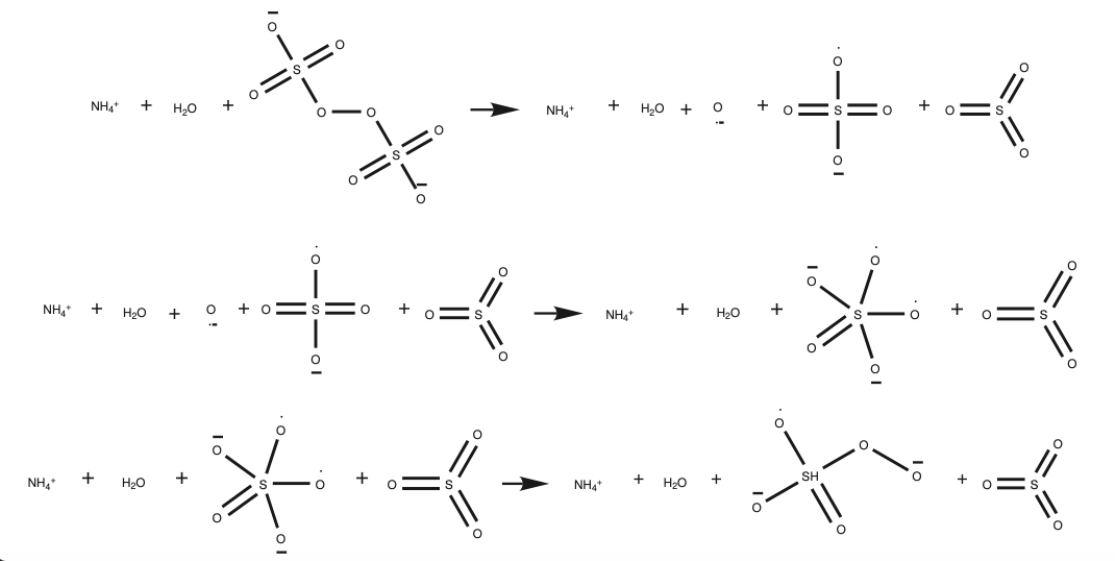}
  \caption{In this decomposition pathway, as the progression advances toward the formation of the rare SO5 species, predictions may encounter challenges when dealing with the inherently unstable structure of SO5.}
  \label{fig:Supplementary Fig. 16}
\end{figure}

\begin{figure}[!]
  \centering
  \includegraphics[width=\linewidth]{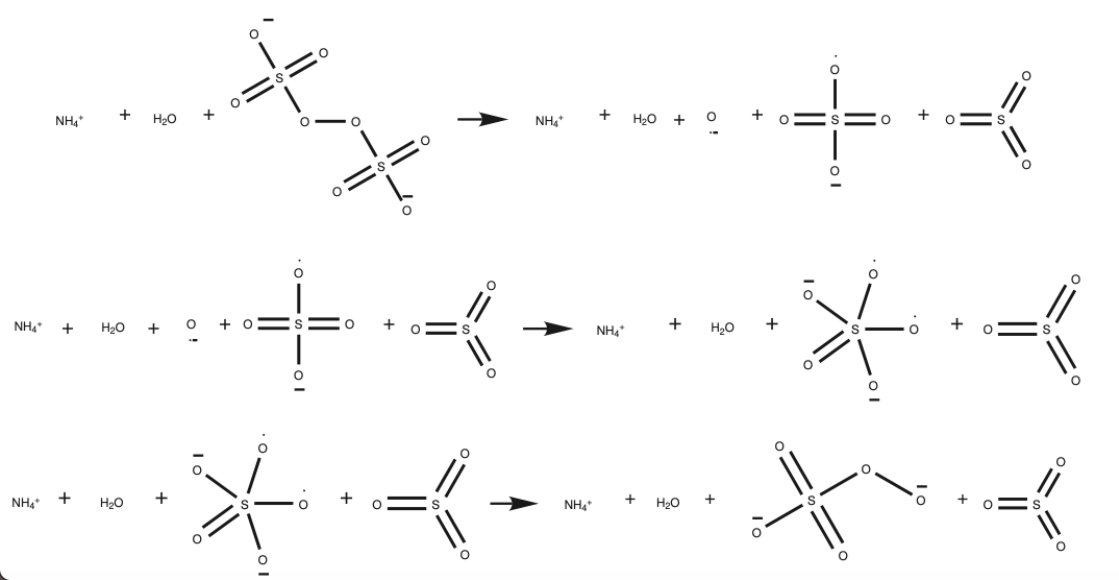}
  \caption{In certain instances, it is possible for the rare and unstable SO5 species to undergo a conversion process, returning to the Peroxymonosulfate state.}
  \label{fig:Supplementary Fig. 17}
\end{figure}

\begin{figure}[!]
  \centering
  \includegraphics[width=\linewidth]{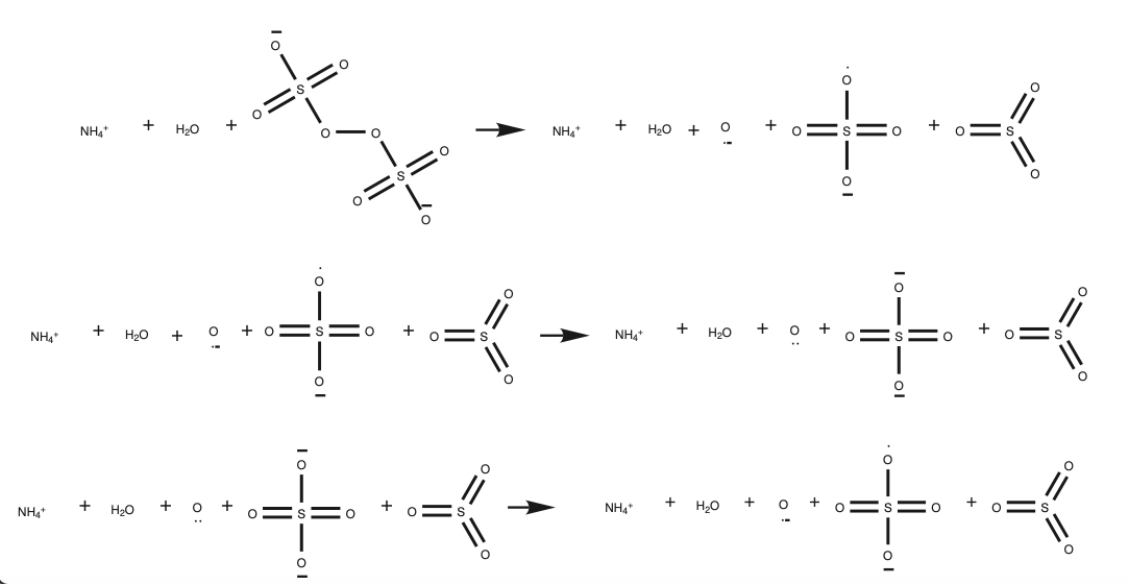}
  \caption{A transformation pathway in which the sulfate radical anion undergoes conversion to the sulfate dianion, eventually combining with oxygen(-2) to generate a sulfate radical anion (SO4·-) again.}
  \label{fig:Supplementary Fig. 18}
\end{figure}

\begin{figure}[!]
  \centering
  \includegraphics[width=\linewidth]{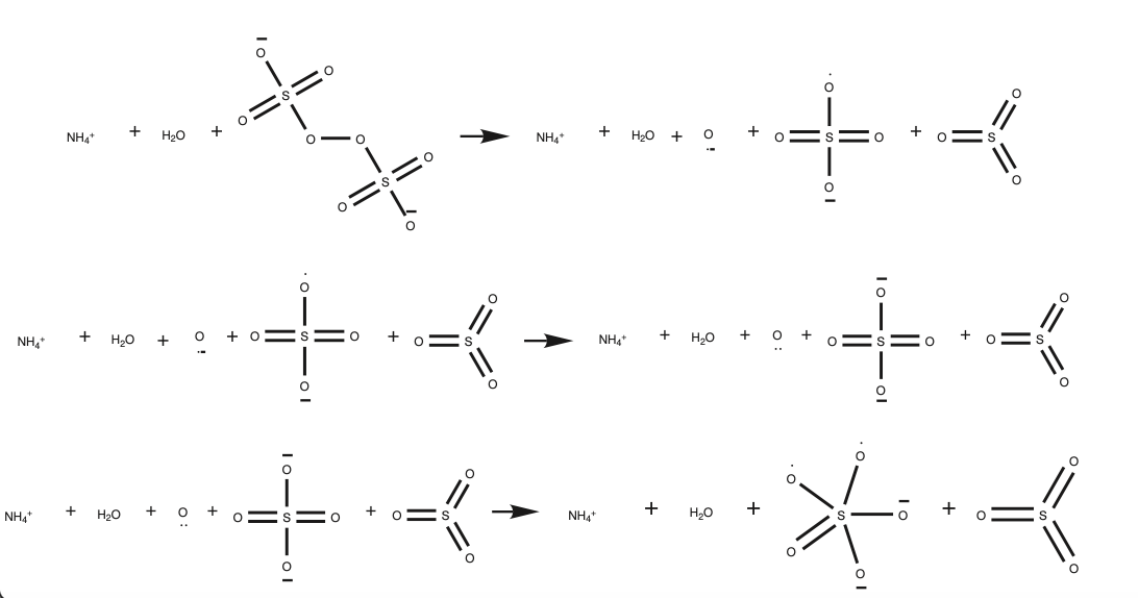}
  \caption{Occasionally, upon further progression beyond the standard steps, the prevalent sulfate radical anion (SO4·-) is found to convert into the rare SO5 species, involving an additional, nonessential step in the reaction pathway.}
  \label{fig:Supplementary Fig. 19}
\end{figure}

\begin{figure}[!]
  \centering
  \includegraphics[width=\linewidth]{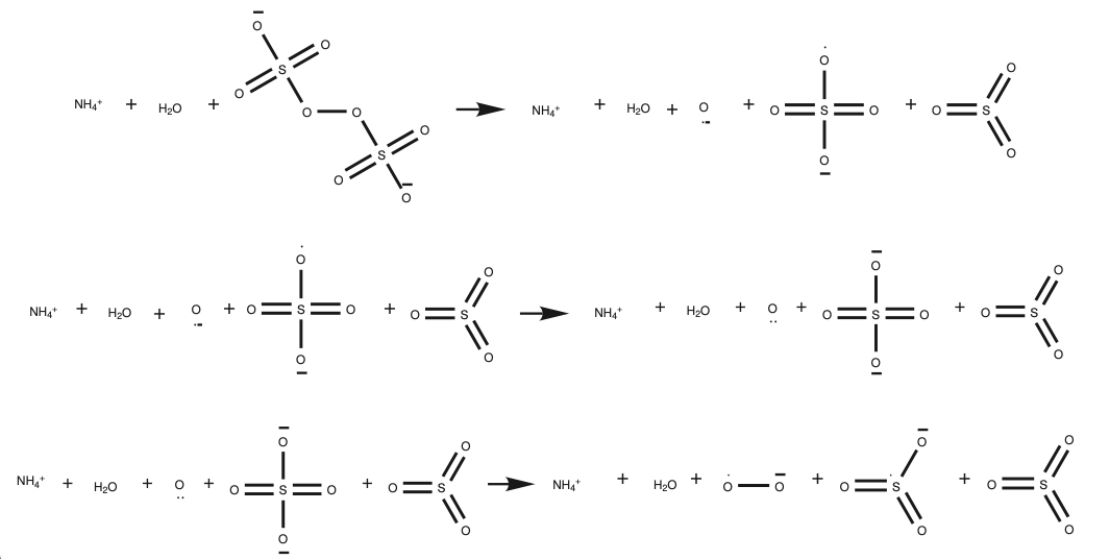}
  \caption{Produced SO3 anion radical and oxygen and peroxide.}
  \label{fig:Supplementary Fig. 20}
\end{figure}

\begin{figure}[!]
  \centering
  \includegraphics[width=\linewidth]{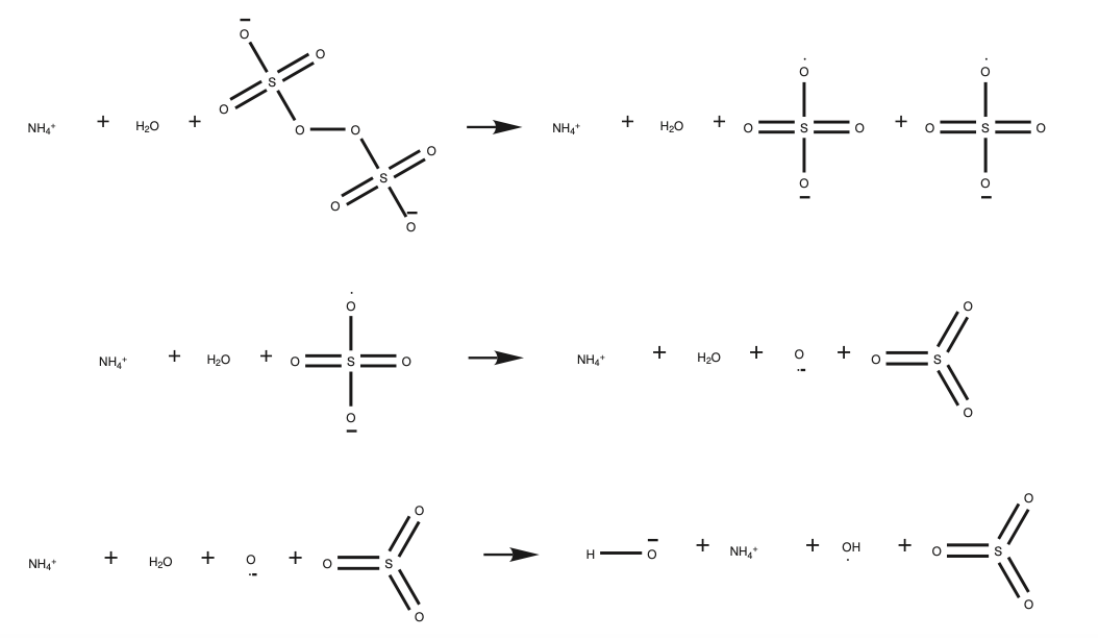}
  \caption{A distinct decomposition pathway for ammonium persulfate, unlike the preceding reactions. In this scenario, ammonium persulfate decomposes into two sulfate radical anions (SO4·-), emphasizing that the initial reaction yields sulfate radical anion (SO4·-). Of notable significance is water’s role in this decomposition process. Here, the oxygen radical anion can interact with water, forming both hydroxyl radicals and hydroxide ions in the final stages.}
  \label{fig:Supplementary Fig. 21}
\end{figure}

\begin{figure}[!]
  \centering
  \includegraphics[width=\linewidth]{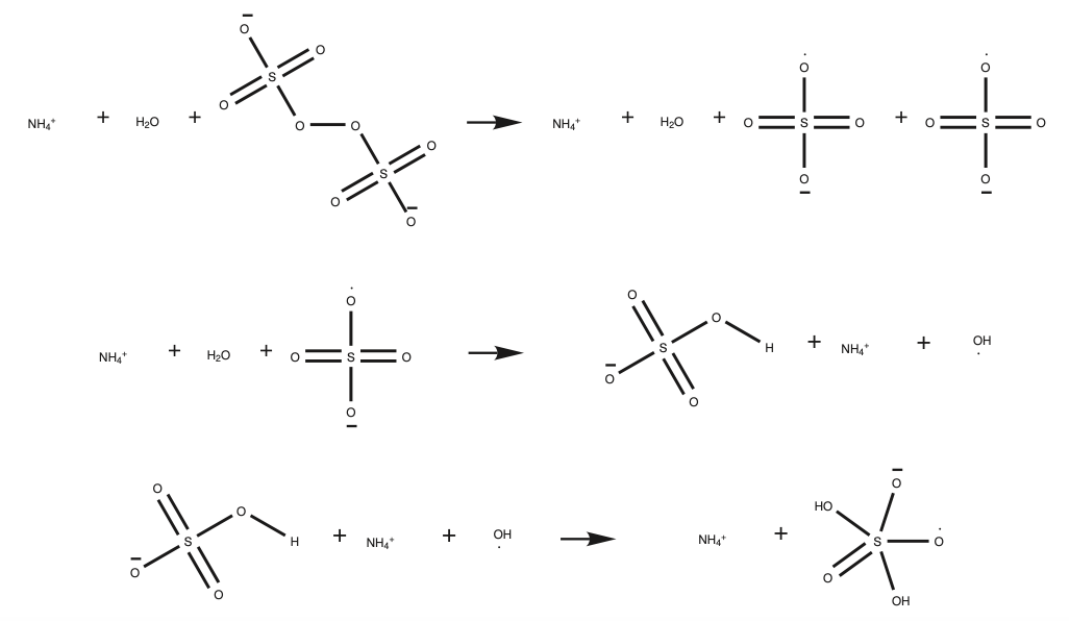}
  \caption{A reaction pathway where water and sulfate radical anion (SO4·-) react to produce hydroxyl radicals and hydrogen sulfate ions. The final stages of this sequence give rise to a further decomposition event, generating a rare h2so5 anion radical.}
  \label{fig:Supplementary Fig. 22}
\end{figure}

\begin{figure}[!]
  \centering
  \includegraphics[width=\linewidth]{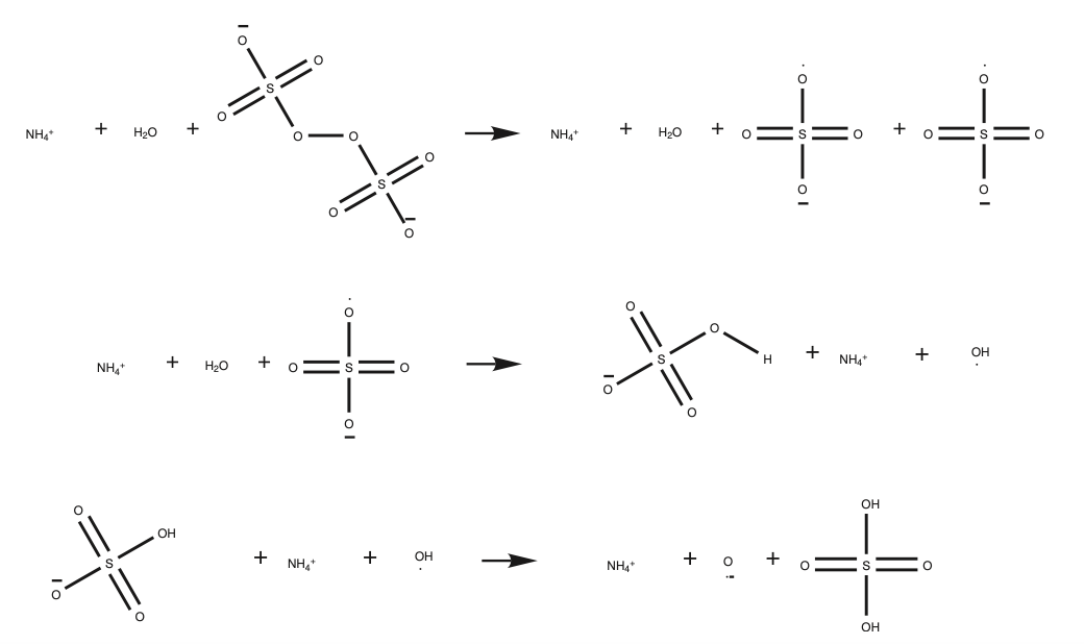}
  \caption{As the reaction sequence continues, hydrogen sulfate ions interact with hydroxyl radicals to produce sulfuric acid. This accumulation of sulfuric acid can subsequently influence the surrounding water’s acidity, potentially affecting various chemical reactions.}
  \label{fig:Supplementary Fig. 23}
\end{figure}

\begin{figure}[!]
  \centering
  \includegraphics[width=\linewidth]{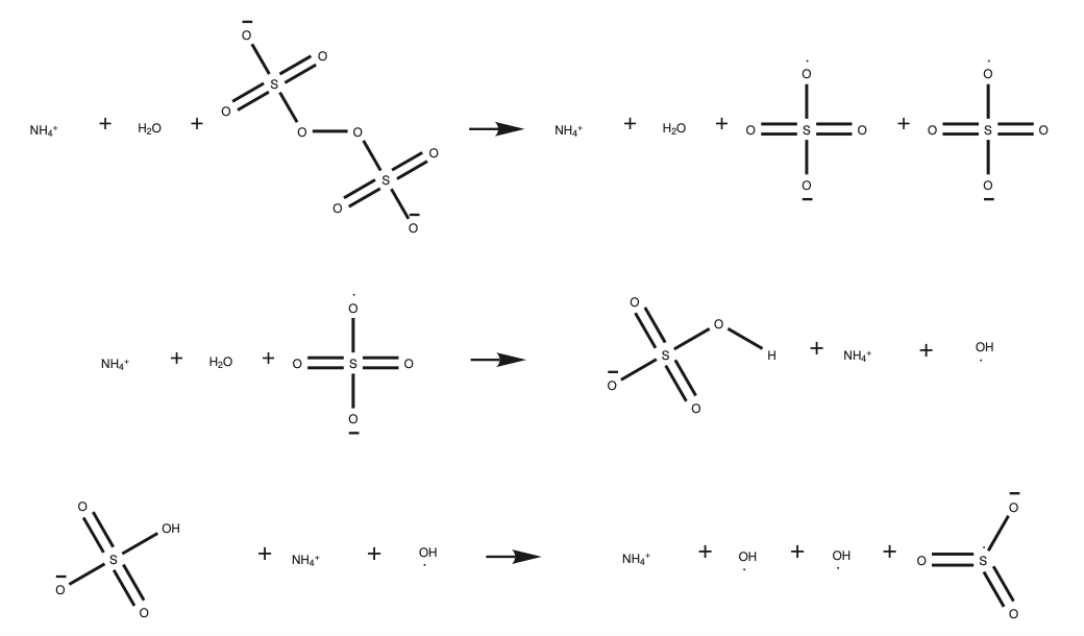}
  \caption{Under certain conditions, the interaction between hydrogen sulfate ions and hydroxyl radicals can lead to the production of double hydroxyl radicals and sulfate anion radicals. This observation underscores the potential significance of hydroxyl radicals’ presence, consistent with findings reported in the existing literature.}
  \label{fig:Supplementary Fig. 24}
\end{figure}

The decomposition of APS, as documented in the existing literature, remains somewhat opaque, with certain products such as sulfate radical anion\cite{borisov2015kinetic,naim2013chemical}, hydroxyl radical\cite{rodrigues2014superabsorbent}, hydroxyl ion\cite{liegeois1972influence}, Peroxymonosulfate\cite{herrera2022role}, bisulfate\cite{johnson2008persulfate,oun2017effect} and superoxide anion\cite{do2013persulfate} being reported. In certain instances, various products are documented across different sources, yet it remains elusive precisely how this initiator’s activation of monomer molecules initiates and catalyzes growth. Our first step, which constitutes the initiation phase, employed an artificial intelligence predictor to predict APS decomposition, as visually depicted in Figures \ref{fig:Supplementary Fig. 13}-\ref{fig:Supplementary Fig. 24}. The experimental parameters used ammonium persulfate and water, and the  experimentation extended to a depth of three levels. Notably, selectivity factors of $(2, 2, 3)$ were applied in each respective depth. It is pertinent that water was not added back as a contextual component. The adjustment of selectivity values aimed to prioritize the most probable outcomes, subsequently eliminating any duplicated structures. Ultimately, this process identified the 12 most prominent products. In accordance with findings in the existing literature, all products are documented. However, it has been reported that compounds such as sulfate radical anion (SO4·-) and hydroxyl radical play pivotal roles in the subsequent growth stage\cite{rodrigues2014superabsorbent,mahdavinia2004modified,khan2020effect,bashir2021flexible}. All the data pertaining to both this section and the subsequent section (Supplementary Section 3.1.2) are comprehensively documented . Please be aware that the order of reactions and products presented here does not imply ranking; rather, it signifies that these represent the most plausible outcomes based on the applied selectivity and depth parameters, considering all conceivable interactions and reactions among the reactants.


\subsubsection{Decomposition of APS Without Water (Water as Context)}\label{subsubsec3}
\begin{figure}[!]
  \centering
  \includegraphics[width=\linewidth]{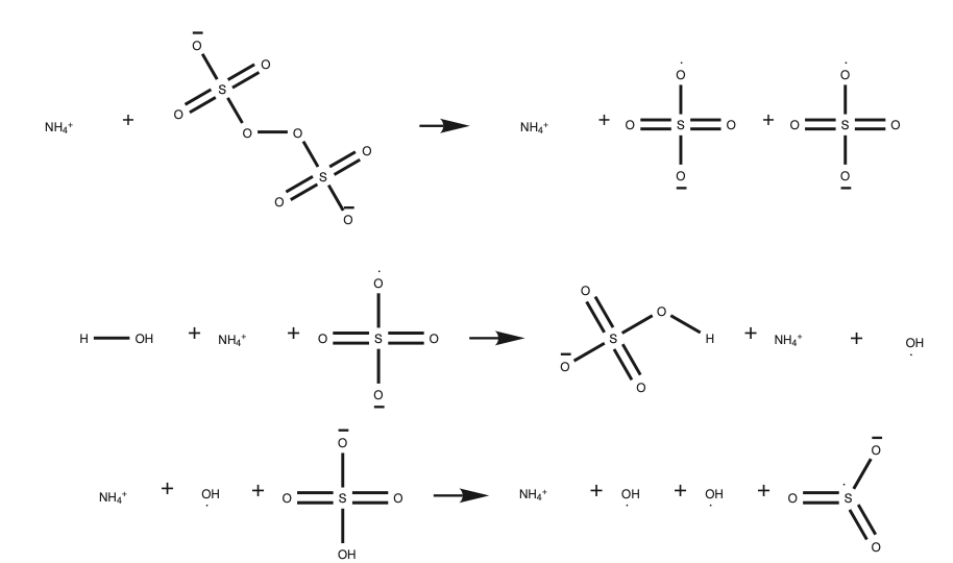}
  \caption{This figure demonstrates that whether water is considered a reactant or a contextual component, the desired product can obtain hydrogen from water in both scenarios. In this case, the sulfate radical anion (SO4·-) acquires hydrogen from water, which is added either as a reactant or context. This observation suggests that while it is essential for water to be present as a reactant due to the importance of these hydrogen-transfer reactions, its inclusion as context yields comparable results. This approach is adopted to optimize computational efficiency, as the choice between reactant and context does not significantly affect the outcome.}
  \label{fig:Supplementary Fig. 25}
\end{figure}

We further explored the decomposition of ammonium persulfate, this time in the absence of water, and subsequently reintroduced water as a contextual element. This exploration was undertaken to elucidate water’s potential impact, a departure from our previous investigation in Supplementary Section 3.1.1. The experimental parameters were held constant, mirroring those outlined in Supplementary Section 3.1.1, with the exception that the reactants did not include water. Instead, water was added back exclusively as a contextual component. A representative sample of the 12 most probable products is presented in Fig. \ref{fig:Supplementary Fig. 25}. Our findings lead us to conclude that the presence of water plays a significant role, potentially exerting an influence in certain scenarios, perhaps in relation to the presence of hydroxyl radicals and hydroxide ions.
\subsection{Run with All Decomposition Products}\label{subsec3}
\begin{figure}[!]
  \centering
  \includegraphics[width=0.8\linewidth]{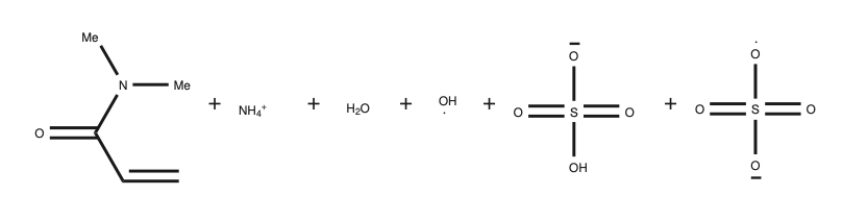}
  \caption{The primary reactants employed in the first run include N, N- Dimethylacrylamide, ammonium, hydroxyl radical, bisulfate, and sulfate radical anion.}
  \label{fig:Supplementary Fig. 26}
\end{figure}

\begin{figure}[!]
  \centering
  \includegraphics[width=0.6\linewidth]{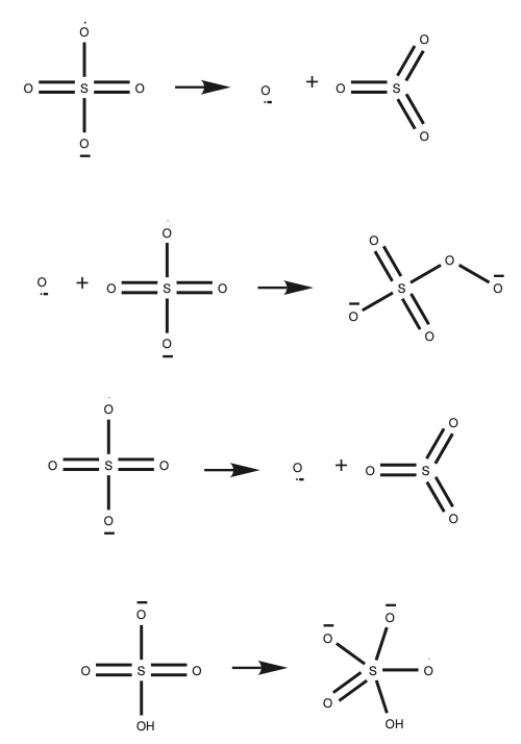}
  \caption{In a significant aspect of the reaction dynamics, the sulfate radical anion (SO4·-) undergoes rearrangement, yielding an unstable oxygen radical anion. This highly reactive species subsequently interacts with other sulfate radical anions (SO4·-), ultimately forming peroxymonosulfate. Notably, this sequence of reactions is primarily consumed by the rearrangement of initiator radicals due to their high reactivity and accessibility to other molecules. Consequently, these initiator radicals have only limited opportunities to engage with the monomers present in the reactant, leading to an emphasis on rearrangement.}
  \label{fig:Supplementary Fig. 27}
\end{figure}

\begin{figure}[!]
  \centering
  \includegraphics[width=0.85\linewidth]{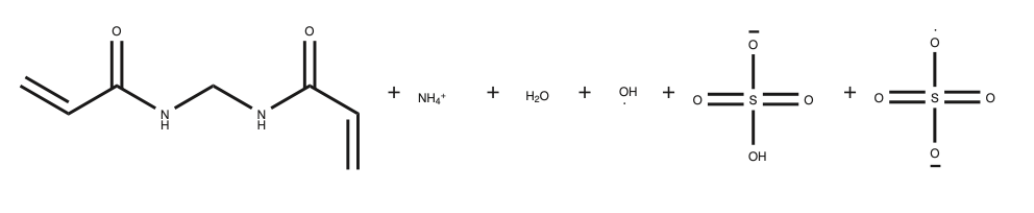}
  \caption{The reactants employed in the second run consist of N, N’- Methylenebisacrylamide, ammonium, hydroxyl radical, bisulfate, and sulfate radical anion.}
  \label{fig:Supplementary Fig. 28}
\end{figure}

\begin{figure}[!]
  \centering
  \includegraphics[width=0.85\linewidth]{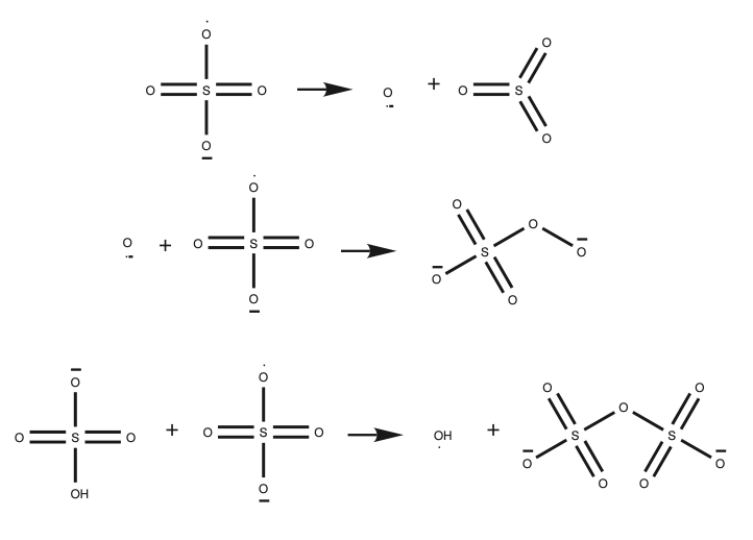}
  \caption{In a pattern similar to previous reactions, certain steps in the decomposition of Ammonium Persulfate (APS) recur. These recurring steps, along with reactions between some intermediates, may lead to APS regeneration. Consequently, this impedes the opportunity for crosslinker attack, potentially hindering the mimicry of real-world reactions.}
  \label{fig:Supplementary Fig. 29}
\end{figure}

\begin{figure}[!]
  \centering
  \includegraphics[width=\linewidth]{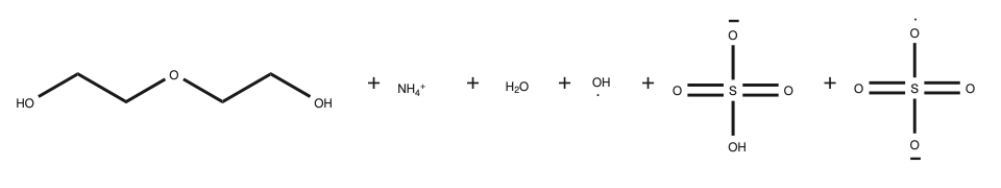}
  \caption{The third run uses the reactants Polyethylene Oxide (PEO), ammonium, hydroxyl radical, bisulfate, and sulfate radical anion. Notably, for computational efficiency, only two units of PEO are considered to emphasize distinctions between monomers and polymers.}
  \label{fig:Supplementary Fig. 30}
\end{figure}

\begin{figure}[!]
  \centering
  \includegraphics[width=0.85\linewidth]{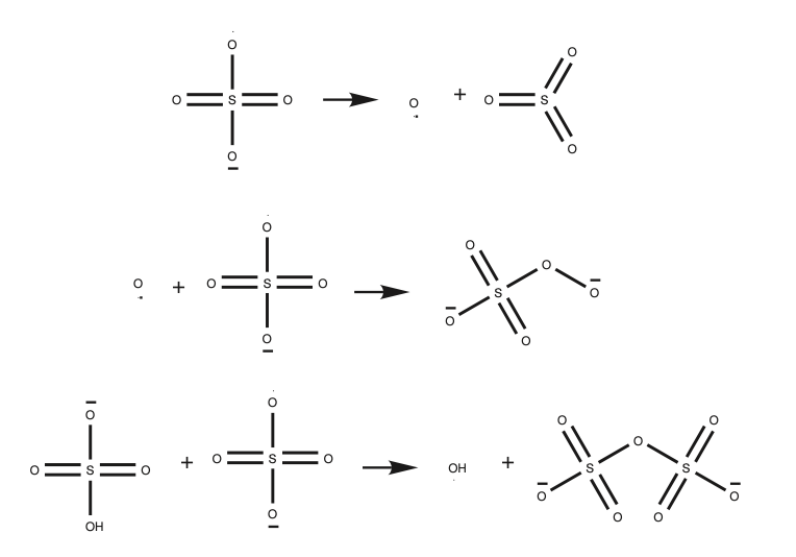}
  \caption{A recurring sequence of crosslinker related reactions leads to a constrained opportunity for Polyethylene Oxide (PEO) to actively participate in the reactions.}
  \label{fig:Supplementary Fig. 31}
\end{figure}

Building on the findings from the previous sections (Supplementary Sections 3.1.1 and 3.1.2), which predicted various products such as sulfate radical anion (SO4·-), bisulfate, sulfuric acid, hydroxyl radical, hydroxyl, sulfur trioxide, stable and unstable peroxymonosulfate structures, sulfate dianion, peroxide, superoxide anion, and bisulfate, it is noteworthy that in real-world scenarios, all these decompositions can potentially coexist for a very brief period. Among these species, sulfate radical anion (SO4·-), hydroxyl radical, and bisulfate are considered to play pivotal roles in subsequent reactions and steps, as documented in the literature\cite{oun2017effect,mahdavinia2004modified,bashir2021flexible}. Hence, in this section, we performed three benchmark experiments. All runs used ammonium, water, hydroxyl radical, bisulfate, and sulfate radical anion (SO4·-) as reactants. Each benchmark entails different primary reactants: In the first run, our monomer (N, N-Dimethylacrylamide) served as the primary reactant, as illustrated in Fig. \ref{fig:Supplementary Fig. 26}. The second run employed our crosslinker as the primary reactant, as depicted in Fig. \ref{fig:Supplementary Fig. 28}. In the third run, our polymer (polyethylene oxide) was the primary reactant, as showcased in Fig. \ref{fig:Supplementary Fig. 30}. The first run explored these processes to a depth of 4 levels, selecting only 2 reactions at each depth. An example of this is presented in Fig. \ref{fig:Supplementary Fig. 27}. In the second and third runs, we investigated to a depth of 3 levels. In the former, we selected 2 paths from all possible paths, as shown in Fig. \ref{fig:Supplementary Fig. 29}, and in the latter, we chose 3 paths at each depth, as demonstrated in  Fig. \ref{fig:Supplementary Fig. 31}. All runs included the initiator radicals as contextual elements and reintroduced them at each depth. It is important that in all cases, it was observed that the initiator radicals’ high reactivity rendered these settings impractical. Due to their propensity for rapid rearrangement, interaction, and mutual reaction, the initiator radicals failed to initiate reactions with the monomer or other components. While it was anticipated that activation of the monomer or crosslinkers would occur at the second or third depth, the initiator radicals’ reactivity necessitated a more prolonged period, potentially extending to the sixth or seventh depth for their rearrangement. Such prolonged rearrangement would substantially increase computational costs, rendering these settings infeasible.

\begin{figure}[!]
  \centering
  \includegraphics[width=0.8\linewidth]{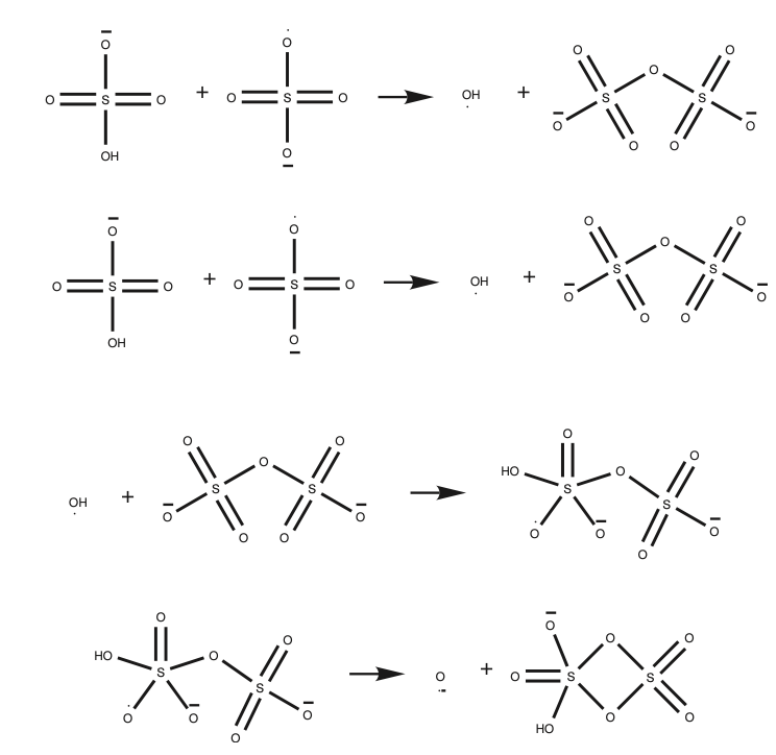}
  \caption{Having other initiator radicals as contextual components can lead to heightened complexity in certain instances across all three runs, necessitating additional computational steps and depths to observe monomer growth. As selectivity becomes a pivotal factor, complexity can escalate exponentially, increasing the number of potential final structures. This increase in complexity may pose challenges in terms of analysis and other aspects of the study.}
  \label{fig:Supplementary Fig. 32}
\end{figure}

\subsection{Run with Desired Initiators}\label{subsec3}
Due to the complexities highlighted in the preceding section, we have encountered difficulties in incorporating all potential initiator radicals into our reactions. To address this challenge, we have devised two distinct strategies. The first approach employs ammonium persulfate as the initial reactant. This strategy is grounded in selecting the most probable initiator decomposition products based on the initial reaction’s selectivity. While this approach permits us to explore a broad spectrum of potential mechanisms, it has two notable drawbacks. First, the limited number of steps allocated for initiator decomposition—typically two or three—can substantially escalate computational costs. Second, it generates an abundance of products, rendering subsequent analysis a formidable task. The multitude of products complicates our ability to discern which ones are more likely to occur in the real-world context. The second approach was adopted to mitigate the challenges faced in the first approach. In this method, we investigate all the products from the first approach, selecting one of them as the primary reactant. This strategy effectively reduces the sample space, enabling us to obtain a specific number of products with decreasing selectivity at each depth. This streamlined approach aids in determining the final molecular structures with greater clarity and precision.
\subsubsection{Run with Ammonium Persulfate}\label{subsubsec3}
\begin{figure}[!]
  \centering
  \includegraphics[width=0.7\linewidth]{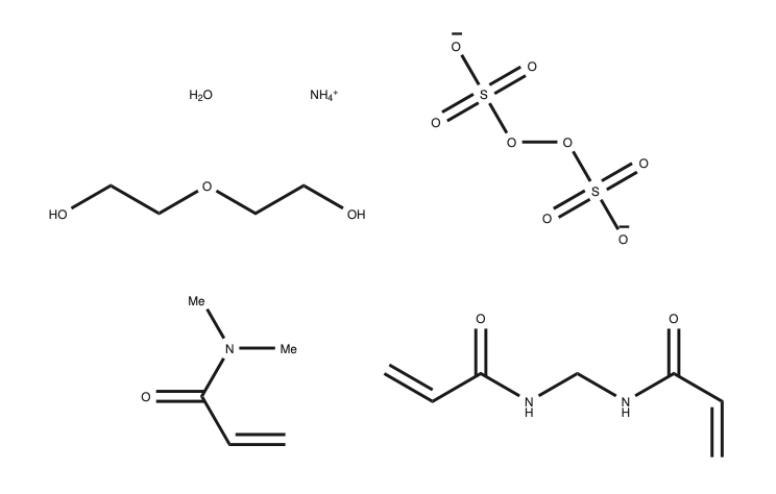}
  \caption{The initial reactants common to all reactions include water, utilized as a conditioning agent, Ammonium Persulfate as the initiator, Polyethylene Oxide (PEO) as polymer, N, N-Dimethylacrylamide as a monomer, and N, N’- Methylenebisacrylamide as the crosslinker.}
  \label{fig:Supplementary Fig. 33}
\end{figure}

\begin{figure}[!]
  \centering
  \includegraphics[width=0.9\linewidth]{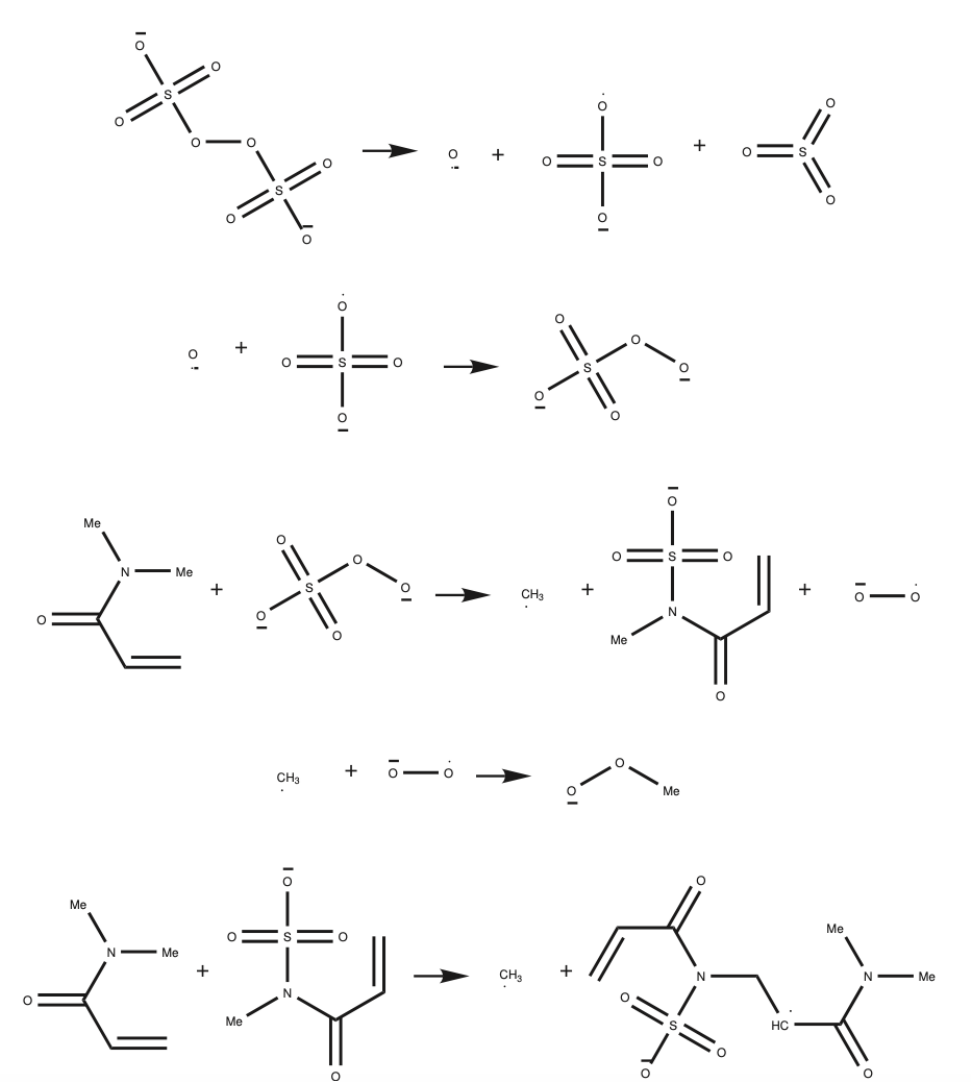}
  \caption{In all mechanisms within Group A, these reactions commence with the decomposition of Ammonium Persulfate (APS) into Sulfur trioxide, sulfate radical anion (SO4·-), and oxygen radical anion. These species subsequently combine to form Peroxymonosulfate. Notably, this peroxymonosulfate initiator deviates from real-world behavior, as it initiates reactions by attacking the methyl group of monomers, a phenomenon unlikely to occur in reality.}
  \label{fig:Supplementary Fig. 34}
\end{figure}

\begin{figure}[!]
  \centering
  \includegraphics[width=0.9\linewidth]{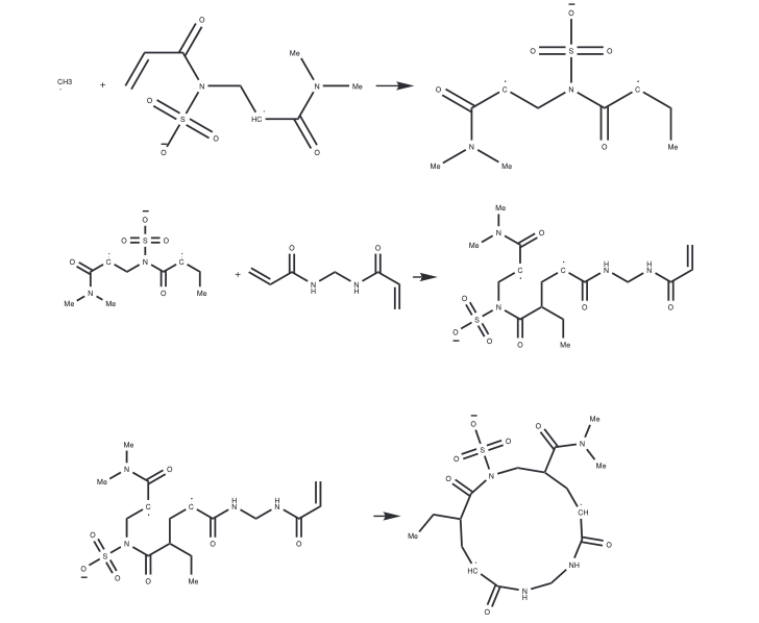}
  \caption{In the final stages of the previous example’s reactions, two monomers come together, allowing the crosslinker to establish connections. Notably, in the absence of adding more monomer units to the system, a unique situation arises where two radicals reside within a single structure, potentially suggesting insight into ring-like termination mechanisms, although such events may be uncommon in practice.}
  \label{fig:Supplementary Fig. 35}
\end{figure}

\begin{figure}[!]
  \centering
  \includegraphics[width=\linewidth]{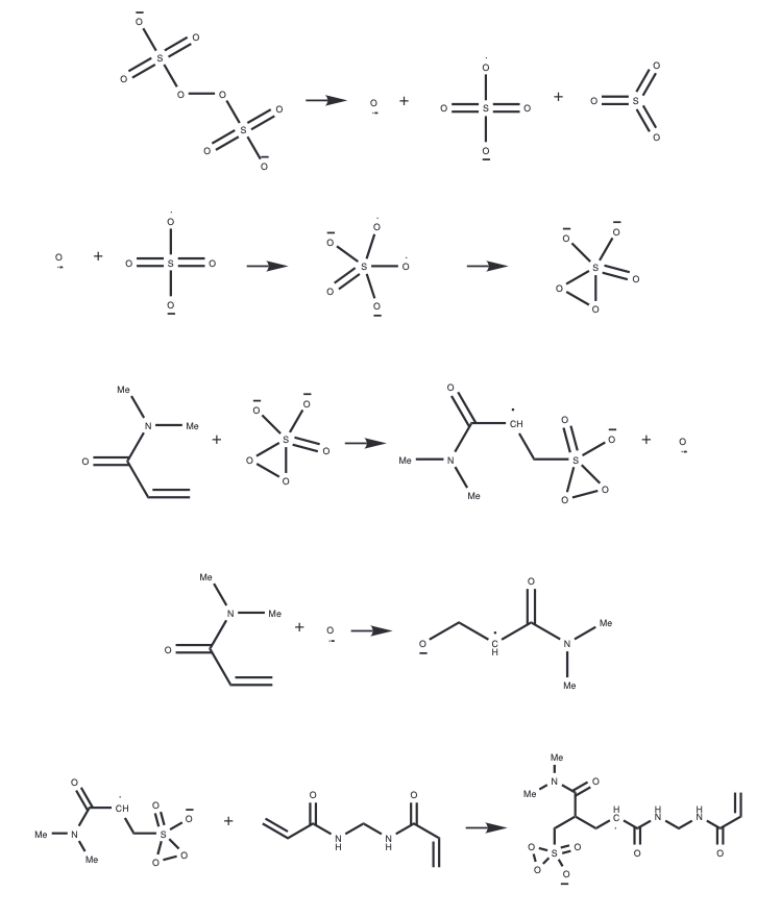}
  \caption{In an example from Group B, the initial decomposition of ammonium persulfate (APS) mirrors that of Group A. However, the subsequent reaction gives rise to a distinctive occurrence, as the sulfate radical anion (SO4·-) and oxygen radical anion combine to form a rarely reported SO5 structure. This SO5 species initiates the reaction by attacking the monomer’s double bond head, setting in motion polymerization growth. Notably, this mechanism shows cases an unconventional pathway where the SO5 structure is instrumental in initiating reactions and can even engage with crosslinkers as the process advances.}
  \label{fig:Supplementary Fig. 36}
\end{figure}

\begin{figure}[!]
  \centering
  \includegraphics[width=\linewidth]{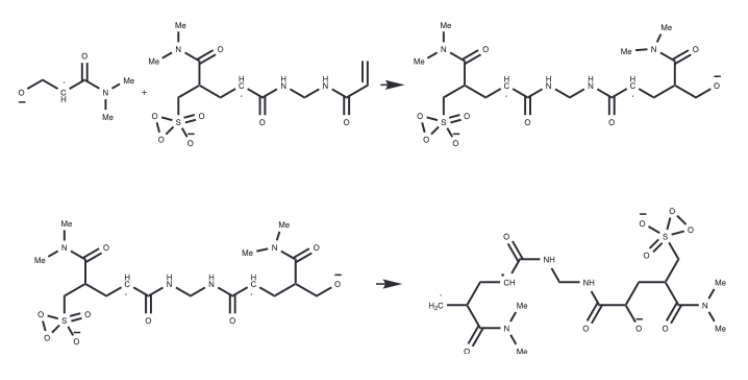}
  \caption{The final two reactions from the previous Group B example produce a unique scenario where two monomers, one activated by SO5 and the other by Oxygen, interact with a crosslinker, resulting in the formation of a semi-gel structure. Importantly, this semi-gel structure contains two radicals, offering the potential for further growth. In the absence of additional monomers, the mechanism's final step rearranges radicals, optimizing their access to other components within the system.}
  \label{fig:Supplementary Fig. 37}
\end{figure}

\begin{figure}[!]
  \centering
  \includegraphics[width=0.9\linewidth]{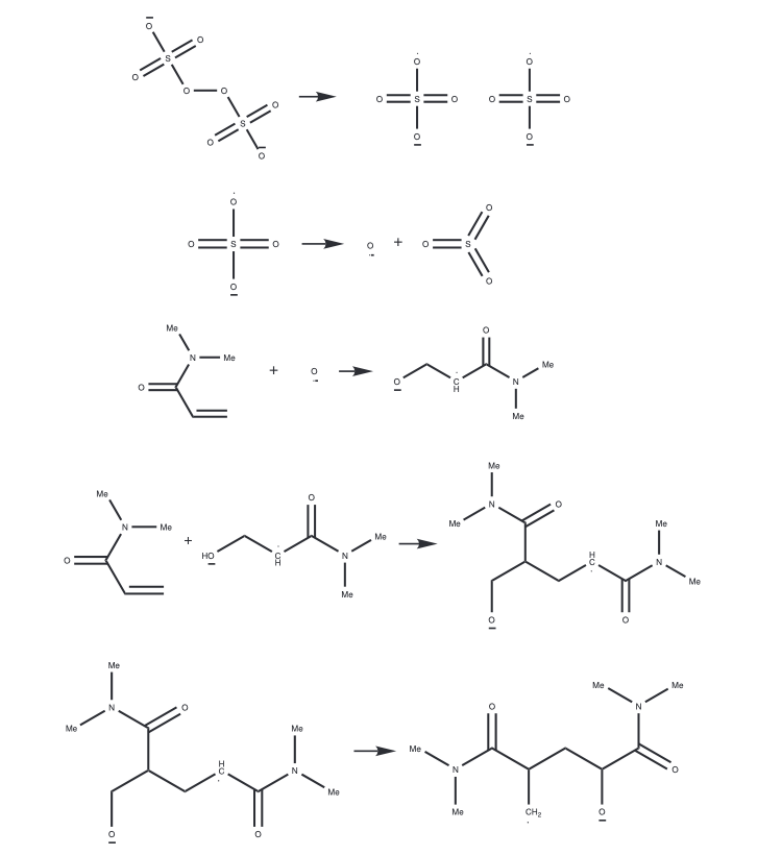}
  \caption{This example from Group C differs from the previous groups in its reaction sequence. In the initial reaction, the decomposition of ammonium persulfate (APS) yields two sulfate radical anions (SO4·-), leading to the formation of Sulfur trioxide and oxygen anion radicals. Notably, the oxygen anion radical serves as an activator, attacking the double bond and generating a radical monomer. This initiates the connection with another monomer, initiating the growth step. Importantly, the resulting dimer represents a head-to-tail configuration. In the real world, simultaneous conversion of radicals occurs among the two monomers and the crosslinker, allowing them to connect. However, due to computational limitations restricting one reaction per depth, this system emulates the real-world process by having the radical monomer interact with another monomer or crosslinker. Remarkably, the dimer’s presence signifies the potential for polymerization to take place, reinforcing this mechanism’s feasibility.}
  \label{fig:Supplementary Fig. 38}
\end{figure}

\begin{figure}[!]
  \centering
  \includegraphics[width=\linewidth]{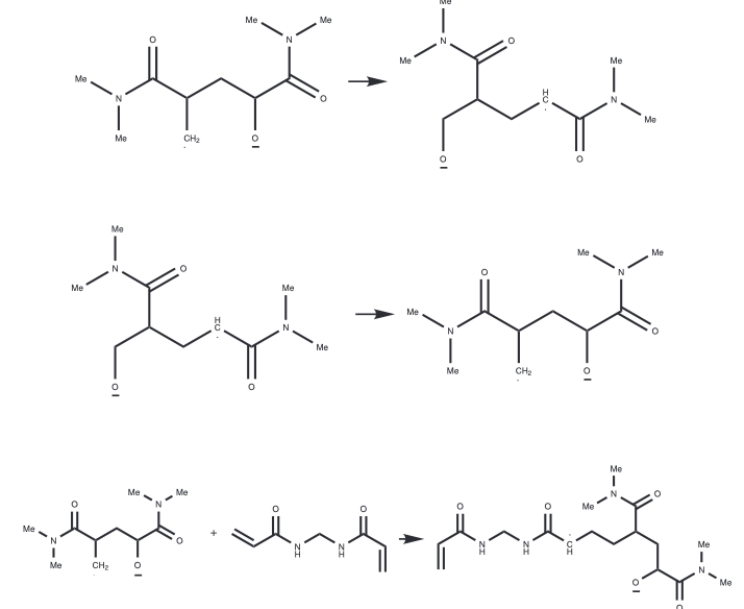}
  \caption{Concluding the previous example from Group C, this example focuses on the last three reactions. Following the rearrangement of radicals within the dimer, the crosslinker attaches to this dimer structure. Notably, the crosslinker exhibits structural symmetry, allowing initiators to potentially attach to both ends, either head or tail. This arrangement suggests the possibility of a symmetric structure forming on the other side of the crosslinkers, ultimately leading to gelation within the system.}
  \label{fig:Supplementary Fig. 39}
\end{figure}

\begin{figure}[!]
  \centering
  \includegraphics[width=\linewidth]{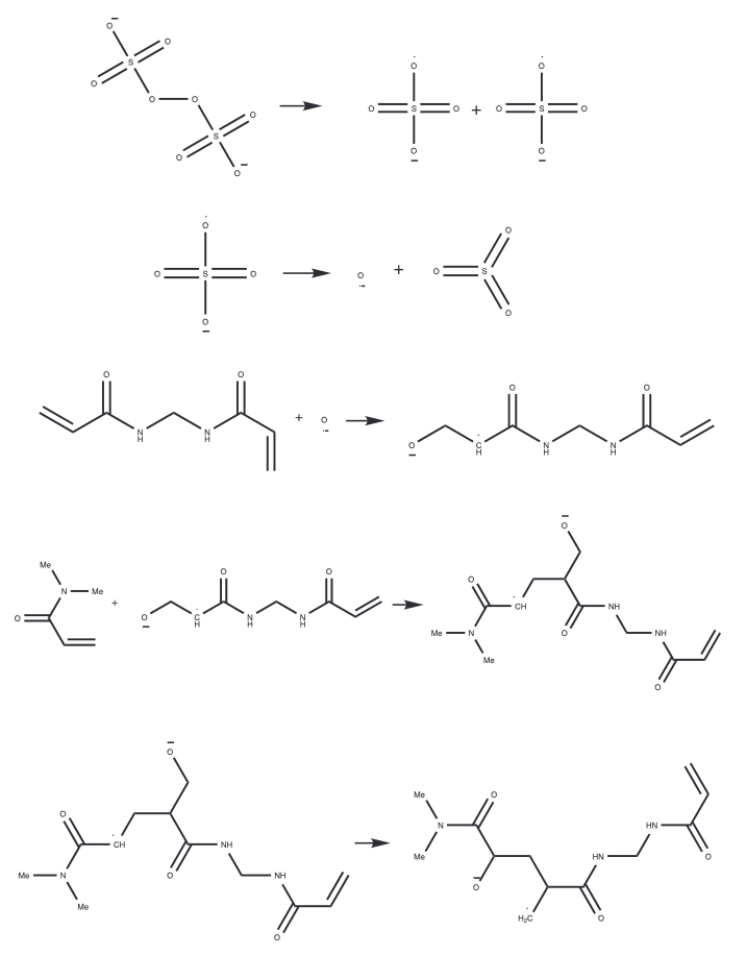}
  \caption{This third example from Group C undergoes the same decomposition reactions. In this scenario, the oxygen anion radical initiates the reaction by attacking the crosslinker, subsequently attaching to a monomer. This observation aligns with the observation discussed earlier: that initiators can simultaneously interact with other components. As a result, the other side of the crosslinker can connect to a different chain, culminating in the formation of a gel-like structure comprising two chains interconnected via the crosslinker.}
  \label{fig:Supplementary Fig. 40}
\end{figure}

\begin{figure}[!]
  \centering
  \includegraphics[width=0.8\linewidth]{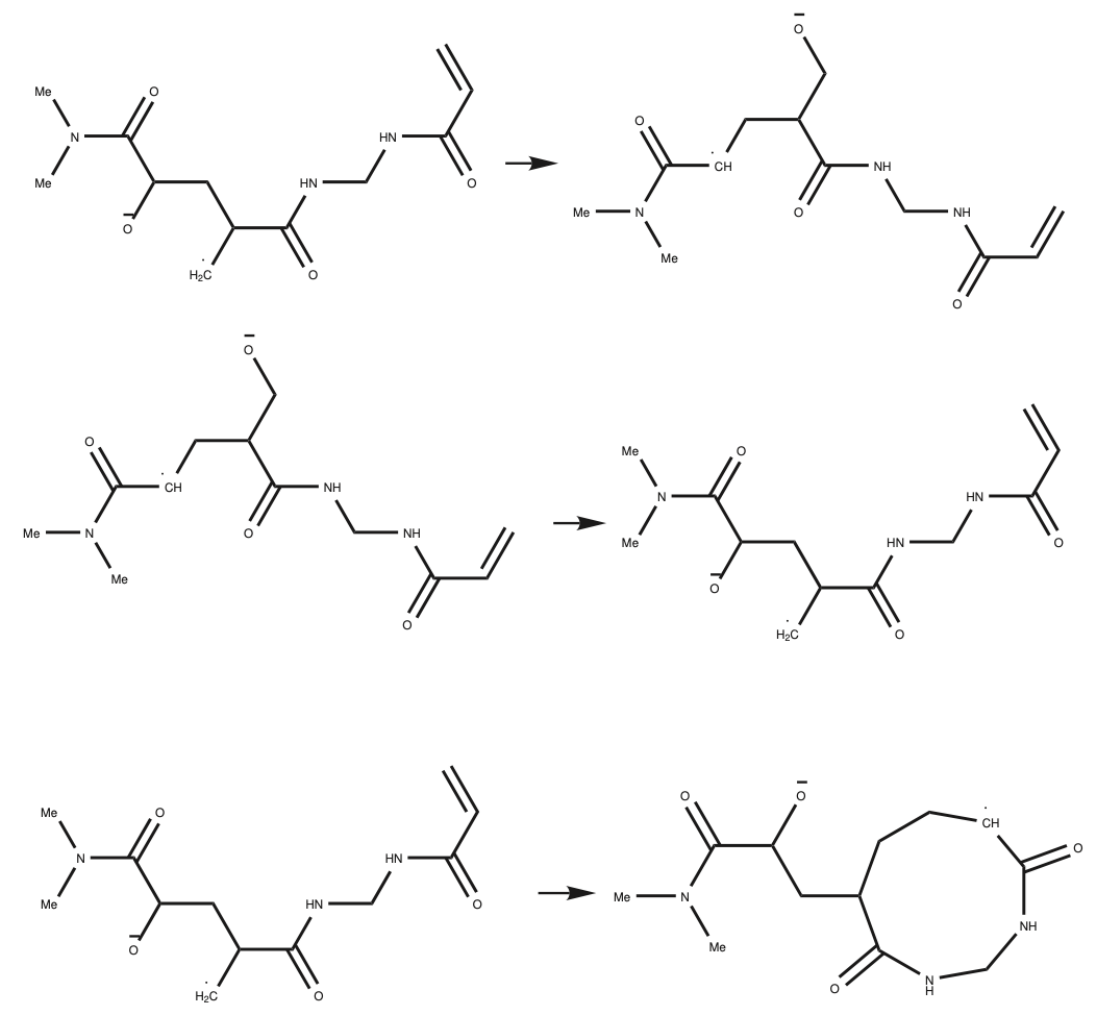}
  \caption{This example depicts the concluding three reactions from the previous example in Group C. The final step shows that in addition to the initiator attacking the tail of the crosslinker, an alternative mechanism becomes apparent. This mechanism transforms a radical from the head to the tail of the crosslinker structure, providing an opportunity for connection to another chain. However, due to the absence of added monomers in the final step, the radical can revert to the other side of the crosslinker, potentially leading to the formation of a ring structure, which is considered a feasible outcome.}
  \label{fig:Supplementary Fig. 41}
\end{figure}

\begin{figure}[!]
  \centering
  \includegraphics[width=0.9\linewidth]{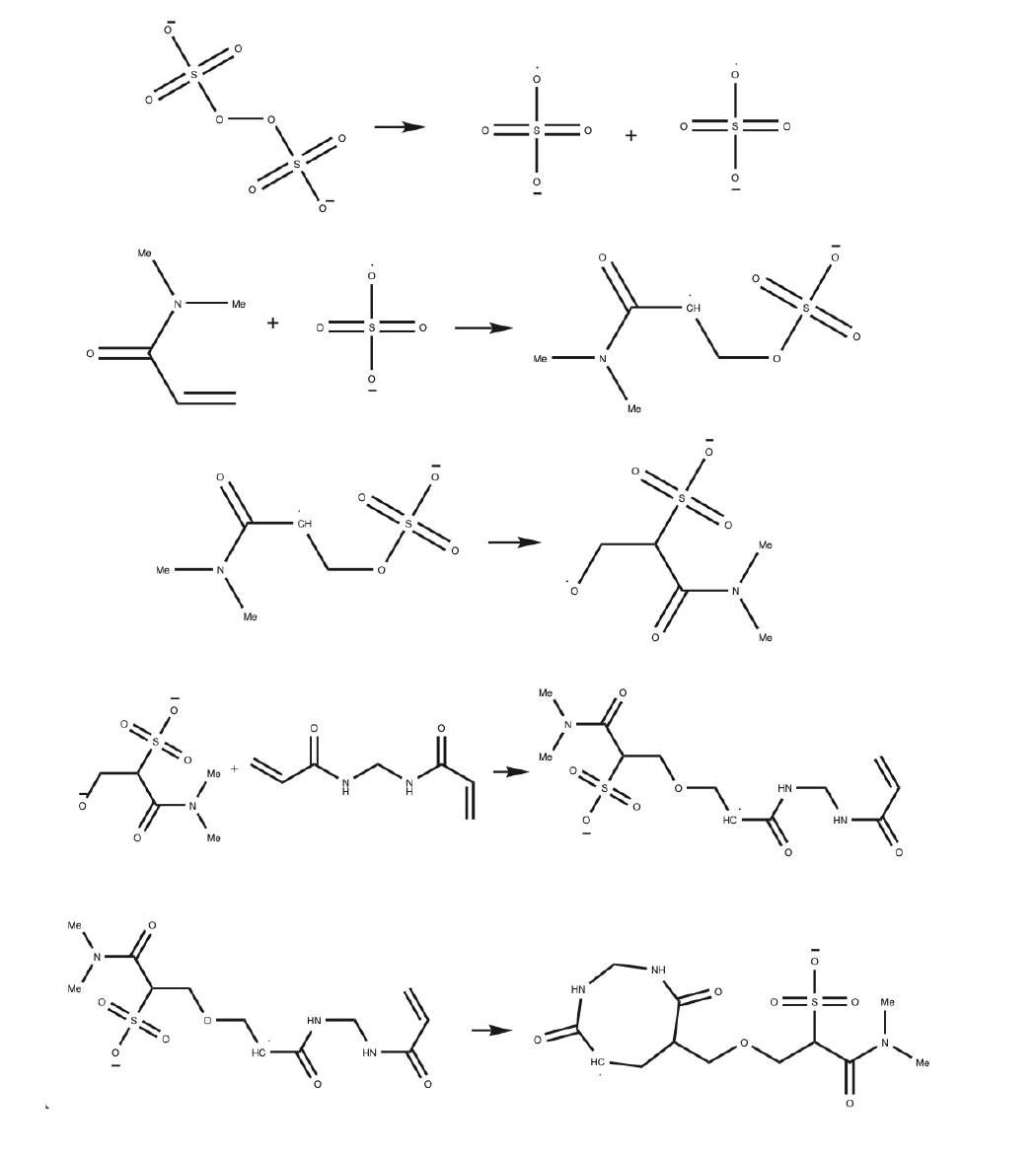}
  \caption{In this example from Group D, the initial reaction decomposes ammonium persulfate (APS) to produce two sulfate radical anions (SO4·-). Notably, in contrast to previous mechanisms, sulfate radical anion (SO4·-) directly attacks the monomer. However, in the third reaction, a distinct occurrence unfolds as sulfate radical anion (SO4·-) on the monomer interacts with two radicals on one end and SO3 on the other. This unique connection between two monomers results in the incorporation of oxygen (O) into the main chain as an ether group. Importantly, incorporation of this ether group, as indicated, does not align with the findings from FTIR (Fourier-transform infrared) spectroscopy results, suggesting a deviation from real-world behavior.}
  \label{fig:Supplementary Fig. 42}
\end{figure}

\begin{figure}[!]
  \centering
  \includegraphics[width=0.9\linewidth]{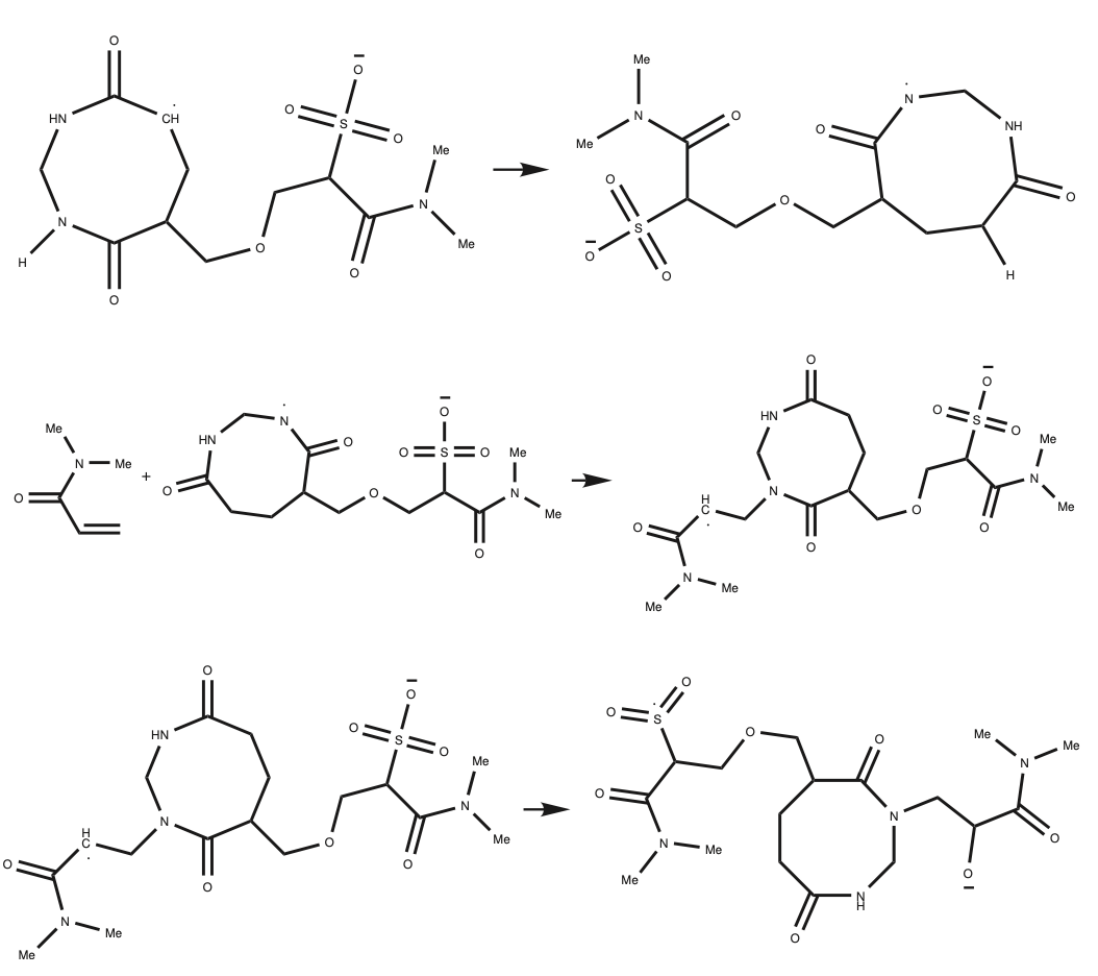}
  \caption{Here, the final three reactions within the previous example from Group D demonstrate an attempt to connect two monomers using a ring crosslinker. However, due to the presence of an ether group within the crosslinker, this connection proves infeasible, deviating from the expected outcome.}
  \label{fig:Supplementary Fig. 43}
\end{figure}

Our first approach restricted our reactants to N, N-Dimethylacrylamide, N, N’- Methylenebisacrylamide, ammonium persulfate, polyethylene oxide (represented by a two-unit polymer segment), and water. Our contextual parameter, N, N-Dimethylacrylamide, served as the monomer, and it remained in use until the seventh depth, as depicted in Fig. \ref{fig:Supplementary Fig. 33}. We conducted investigations up to the eighth depth, applying a selectivity factor of 2 at each depth. Notably, at the final depth, to observe which reactions might occur in the absence of monomer, we intentionally did not reintroduce monomer, thus shedding light on potential termination mechanisms. In this approach, the initial two depths emphasize APS decomposition. We have categorized all 188 possible mechanisms into four distinct groups based on the activation centers, denoted as A, B, C, and D. The investigation of these four groups is documented in Figures \ref{fig:Supplementary Fig. 34} and \ref{fig:Supplementary Fig. 35}, which provide an example for Group A (Figures \ref{fig:Supplementary Fig. 36} and \ref{fig:Supplementary Fig. 37}), illustrating one example from Group B (Figures \ref{fig:Supplementary Fig. 38}-\ref{fig:Supplementary Fig. 41}), showcasing two examples from Group C, and finally, presenting an example from Group D (Figures \ref{fig:Supplementary Fig. 42} and \ref{fig:Supplementary Fig. 43}).

\begin{figure}[!]
  \centering
  \includegraphics[width=0.8\linewidth]{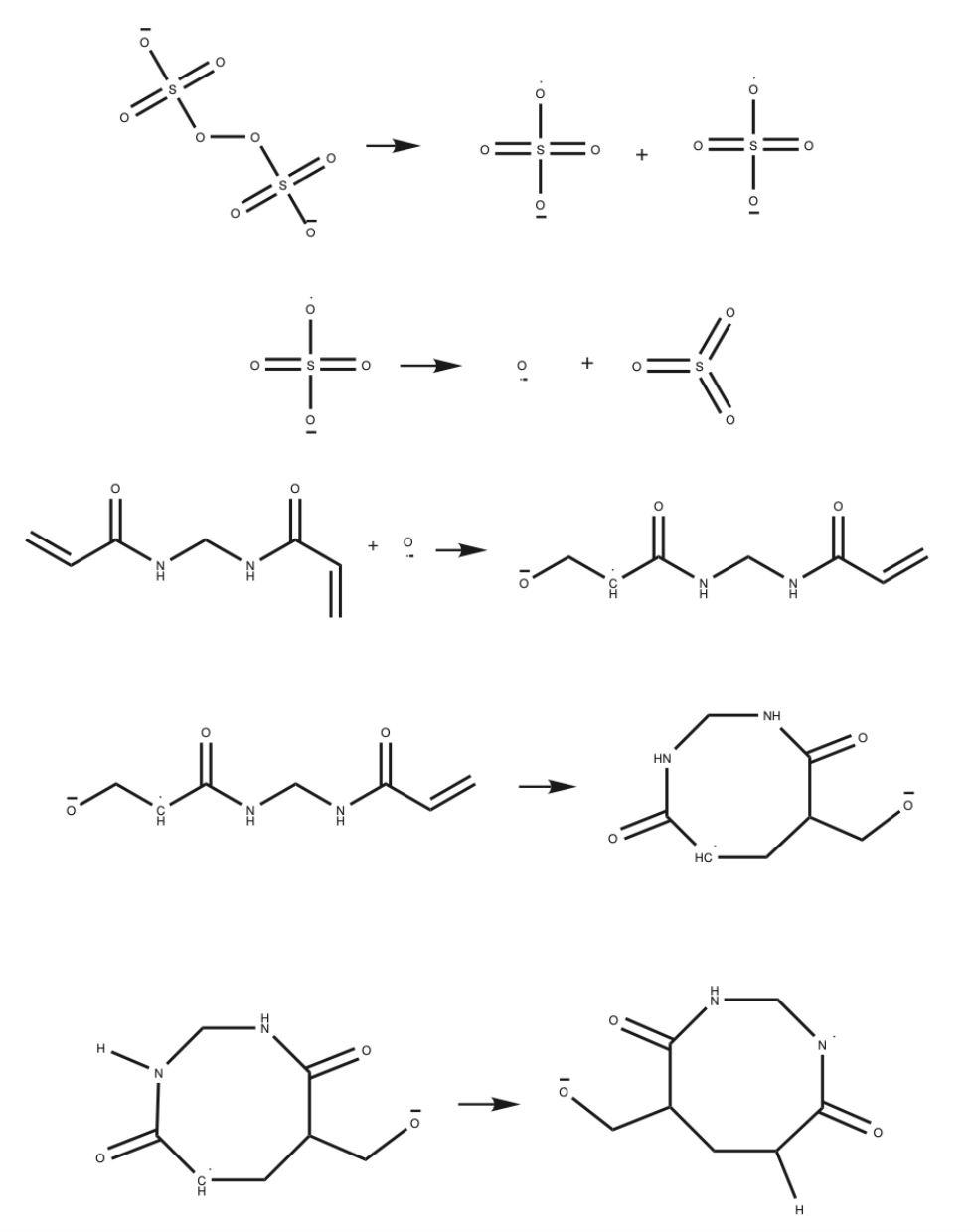}
  \caption{In a remarkable case from Group C, an intriguing sequence of reactions unfolds. Initially, the oxygen radical anion attacks the crosslinker, leading to subsequent interactions between the crosslinker and a monomer. Notably, this unique scenario results in the formation of a ring structure, an outcome that adds an interesting dimension to the reaction mechanisms.}
  \label{fig:Supplementary Fig. 44}
\end{figure}
\begin{figure}[!]
  \centering
  \includegraphics[width=\linewidth]{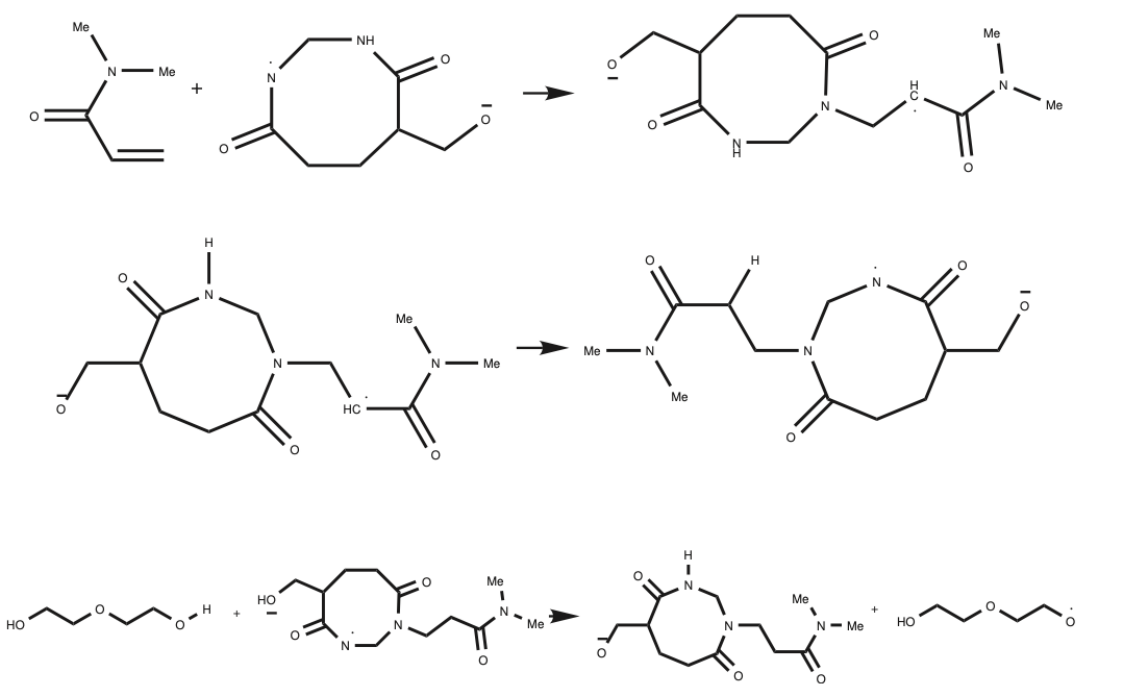}
  \caption{The final three reactions of the previously observed rare mechanism demonstrate that after attacking the monomer, the radical on the head of the monomer may transfer to one of the crosslinker’s nitrogen atoms. This nitrogen atom, being unstable, can acquire a hydrogen (H) atom from the end group of Polyethylene Oxide (PEO). It’s noteworthy that among the 188 possible mechanisms investigated in this run, PEO rarely participated in the reactions. This reaction highlights a rare instance in which PEO becomes involved by providing a hydrogen atom for activation. Across all groups, radicals are located primarily within the main chain or crosslinker, subsequently enabling interactions with PEO.}
  \label{fig:Supplementary Fig. 45}
\end{figure}
\subsubsection{Most Possible Products of the Main Run}\label{subsubsec3}
\begin{figure}[!]
  \centering
  \includegraphics[width=0.55\linewidth]{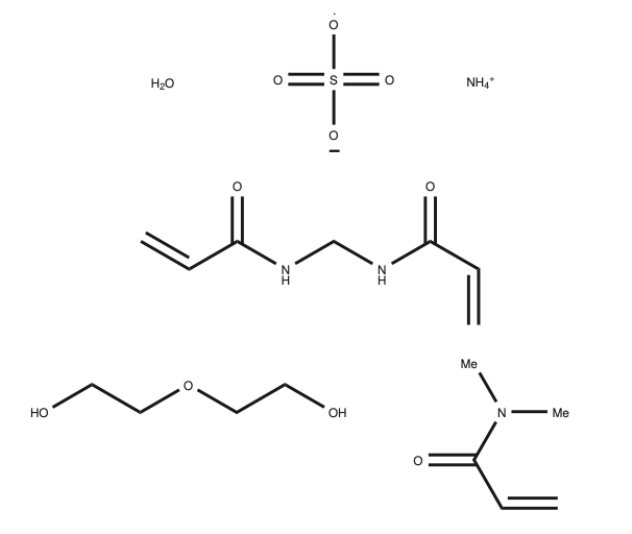}
  \caption{This figure outlines the configuration for the most probable structure run. The experimental conditions involve the presence of water as a conditioning agent, with sulfate radical anion (SO4·-) serving as the primary initiator radical. The reactants in this scenario include polyethylene oxide (PEO), N, N-Dimethylacrylamide as a monomer, and N, N’- Methylenebisacrylamide as the crosslinker. These components collectively contribute to the exploration of the most likely reaction pathways and resultant structures.}
  \label{fig:Supplementary Fig. 46}
\end{figure}

\begin{figure}[!]
  \centering
  \includegraphics[width=\linewidth]{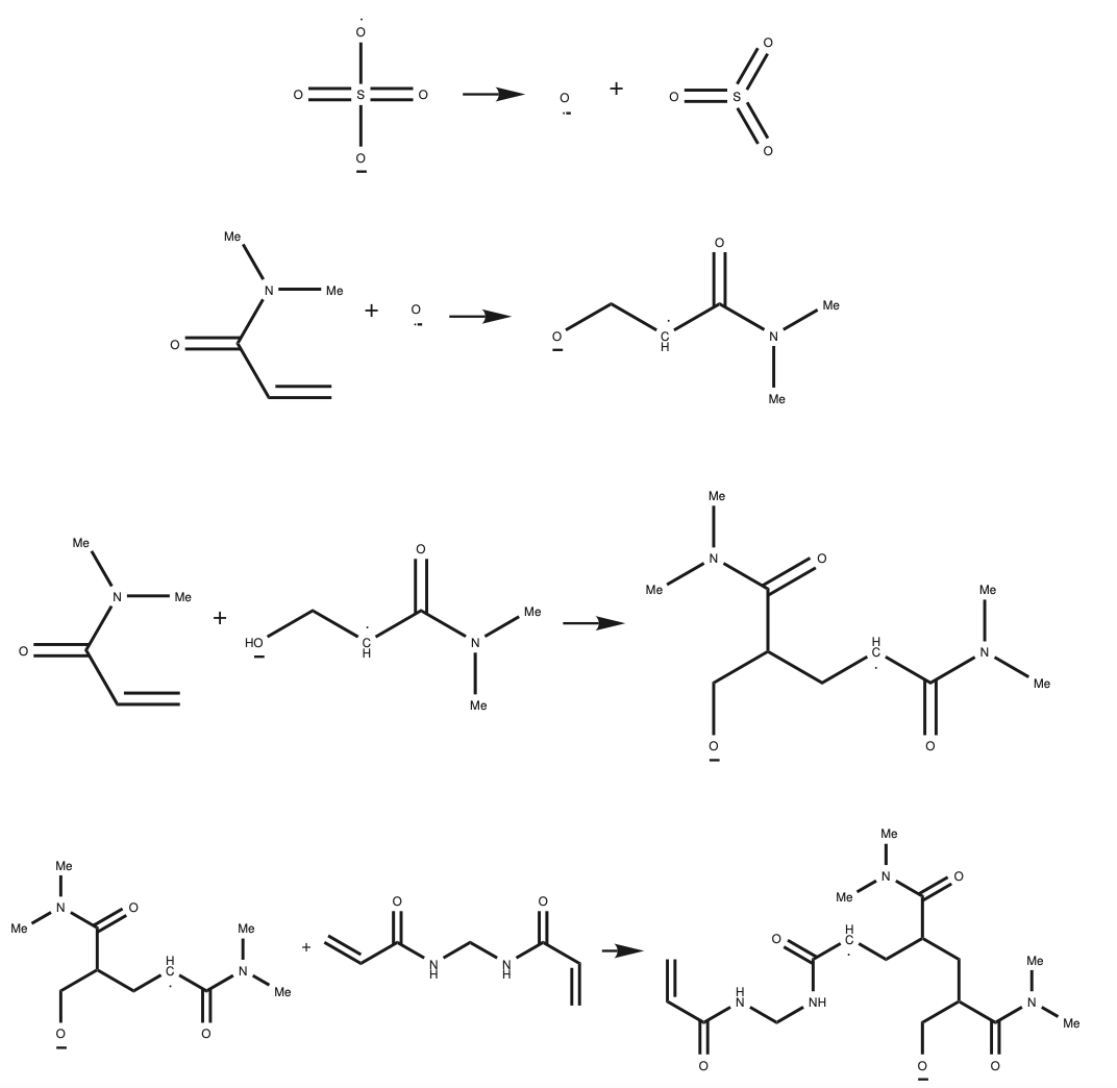}
  \caption{One of the system’s most likely structural formations begins with ammonium persulfate (APS) decomposition, yielding sulfate radical anions (SO4·-). Subsequently, these SO4·- radicals decompose into sulfur trioxide and oxygen anion radicals. The oxygen anion radical initiates the reaction by attacking a monomer, leading to the formation of a radical monomer. This radical monomer subsequently attaches head-to-tail to another monomer, followed by the attachment of a crosslinker to these monomers. As previously demonstrated, radicals within this structure can transfer to the crosslinker’s tail and then engage with other chains. Notably, polymerization growth can occur either on these radicals or via the oxygen anion side group. The structure’s symmetry enables other chains to potentially attach, contributing to the formation of a complex polymerization network.}
  \label{fig:Supplementary Fig. 47}
\end{figure}

\begin{figure}[!]
  \centering
  \includegraphics[width=\linewidth]{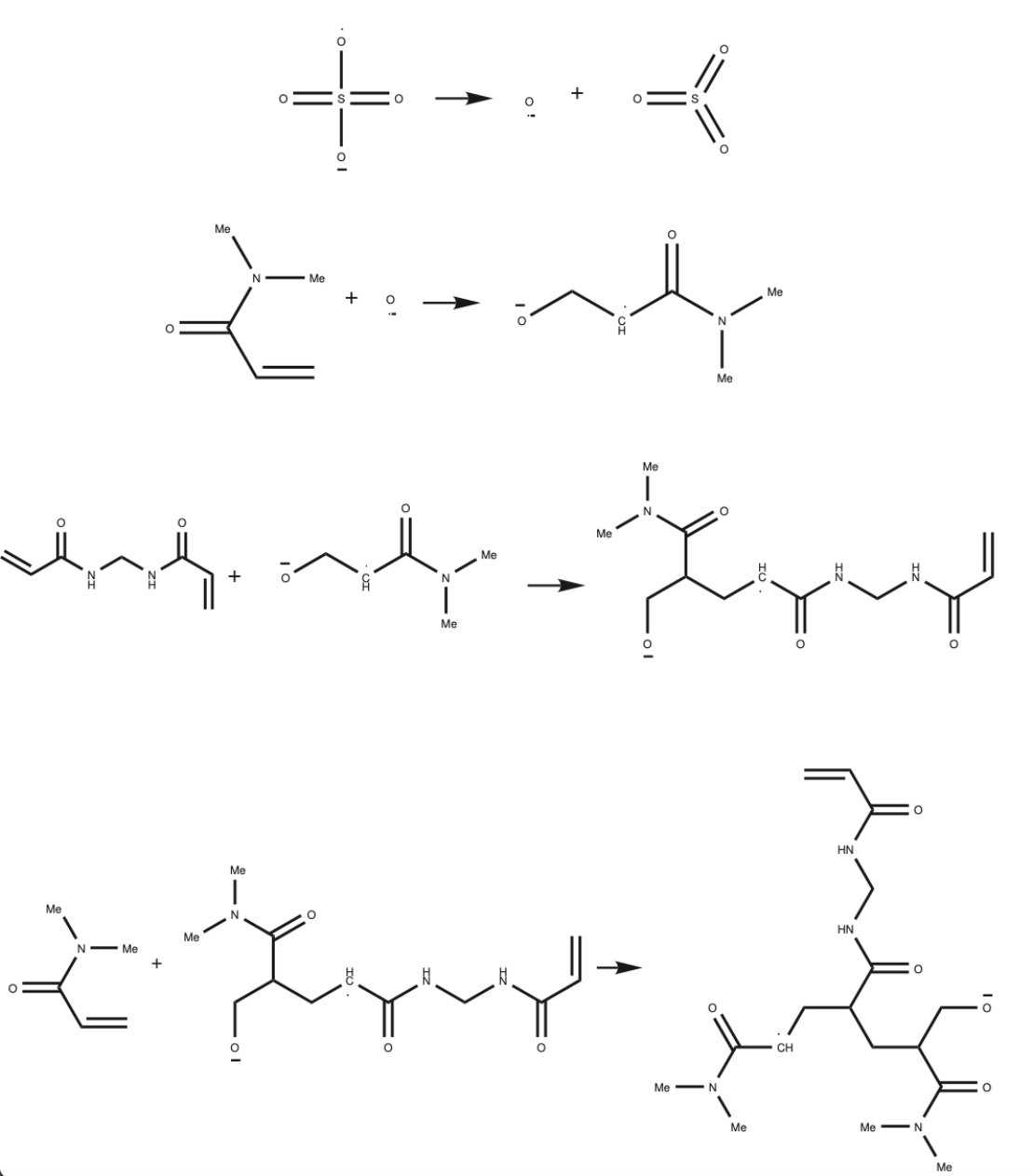}
  \caption{In the second-most likely structure based on the predictor system, the overall setup and conditions mirror those in  Fig. \ref{fig:Supplementary Fig. 47}. However, in this scenario, the radical monomer initiates the reaction by attacking the crosslinker. Subsequently, the reaction progresses analogously to Fig. \ref{fig:Supplementary Fig. 47}, resulting in a similar complex polymerization structure. This figure illustrates the potential for multiple pathways leading to the formation of such intricate polymer networks within the system.}
  \label{fig:Supplementary Fig. 48}
\end{figure}

\begin{figure}[!]
  \centering
  \includegraphics[width=\linewidth]{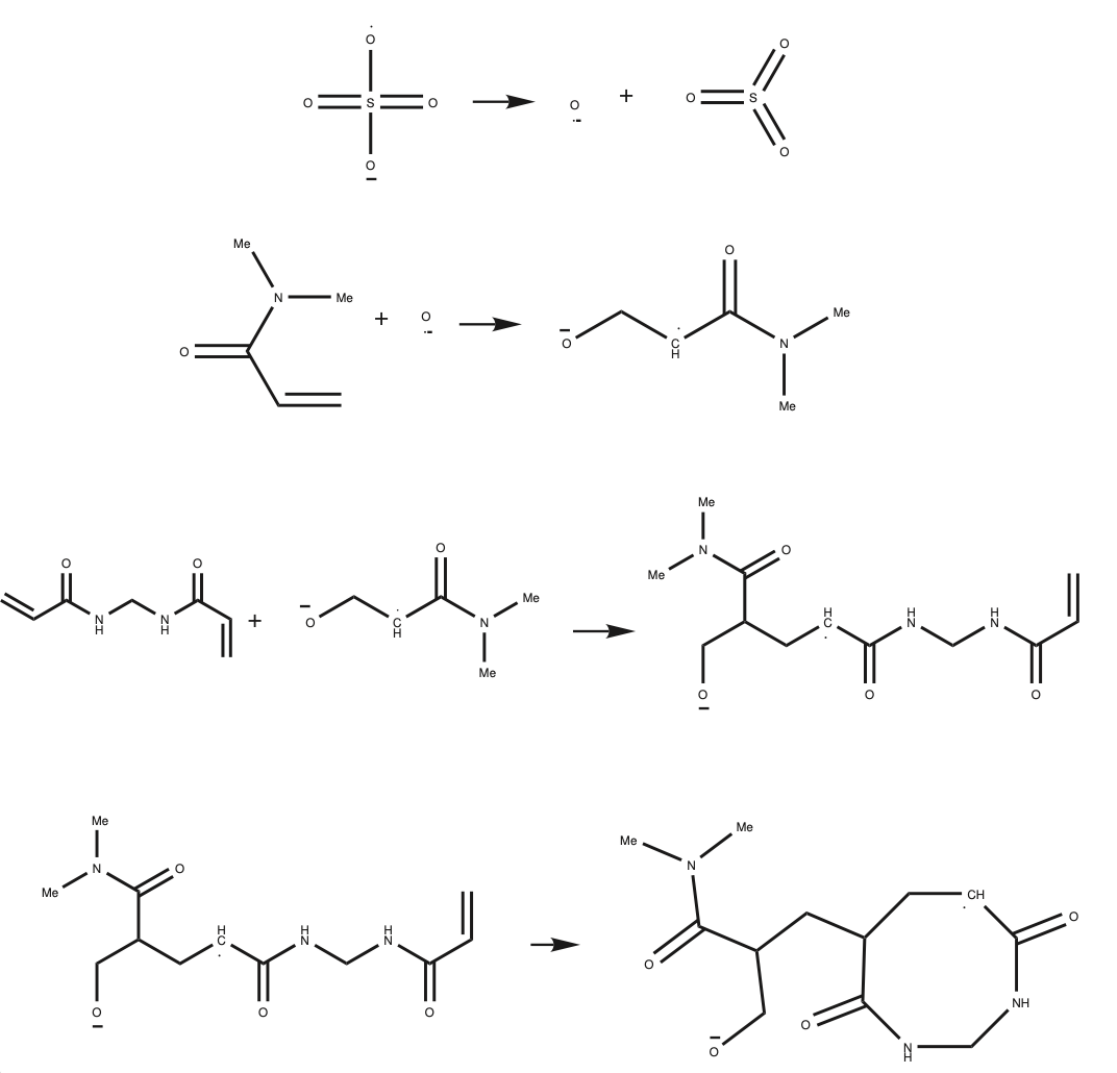}
  \caption{The third plausible mechanism begins by generating a radical monomer, which then attacks the crosslinker. In this scenario, the crosslinker subsequently reverses its attachment and forms a ring structure. This illustration emphasizes that by not advancing to the next step in the reaction, when available monomers are absent, the system has the capability to build rings rather than simply attaching radicals to each chain.}
  \label{fig:Supplementary Fig. 49}
\end{figure}

\begin{figure}[!]
  \centering
  \includegraphics[width=\linewidth]{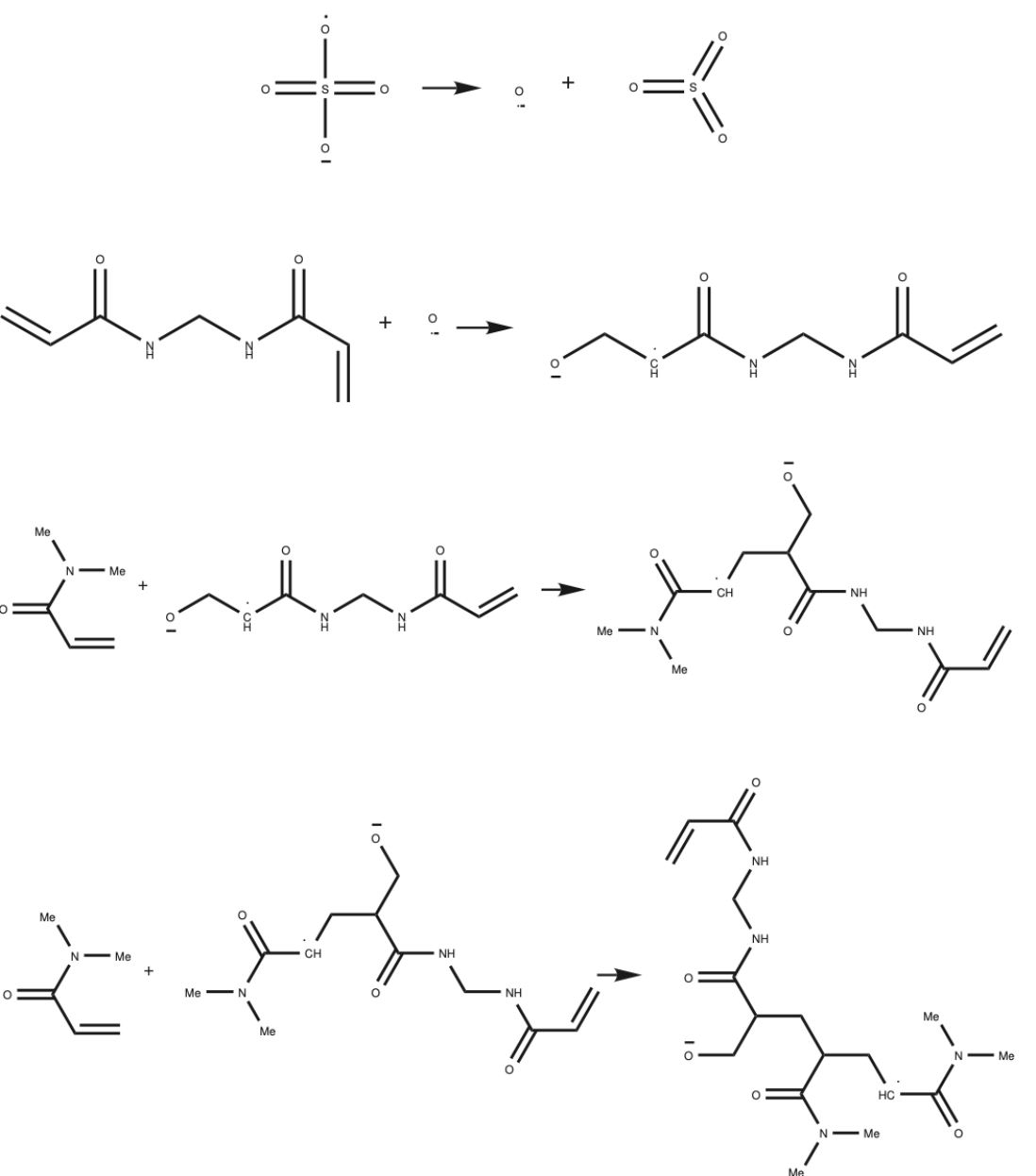}
  \caption{The other most possible final structure shares similarities with the three previous mechanisms. However, the key distinction lies in the initiation step, where the oxygen radical anion directly attacks the crosslinker instead of the monomer. This illustration highlights the system’s dynamic nature, where initiator radicals can simultaneously attack both monomers and crosslinkers. Such parallel reactions within these mechanisms contribute to the formation of a three-dimensional (3D) network, exemplifying the system’s versatility in producing complex structures.}
  \label{fig:Supplementary Fig. 50}
\end{figure}

\begin{figure}[!]
  \centering
  \includegraphics[width=\linewidth]{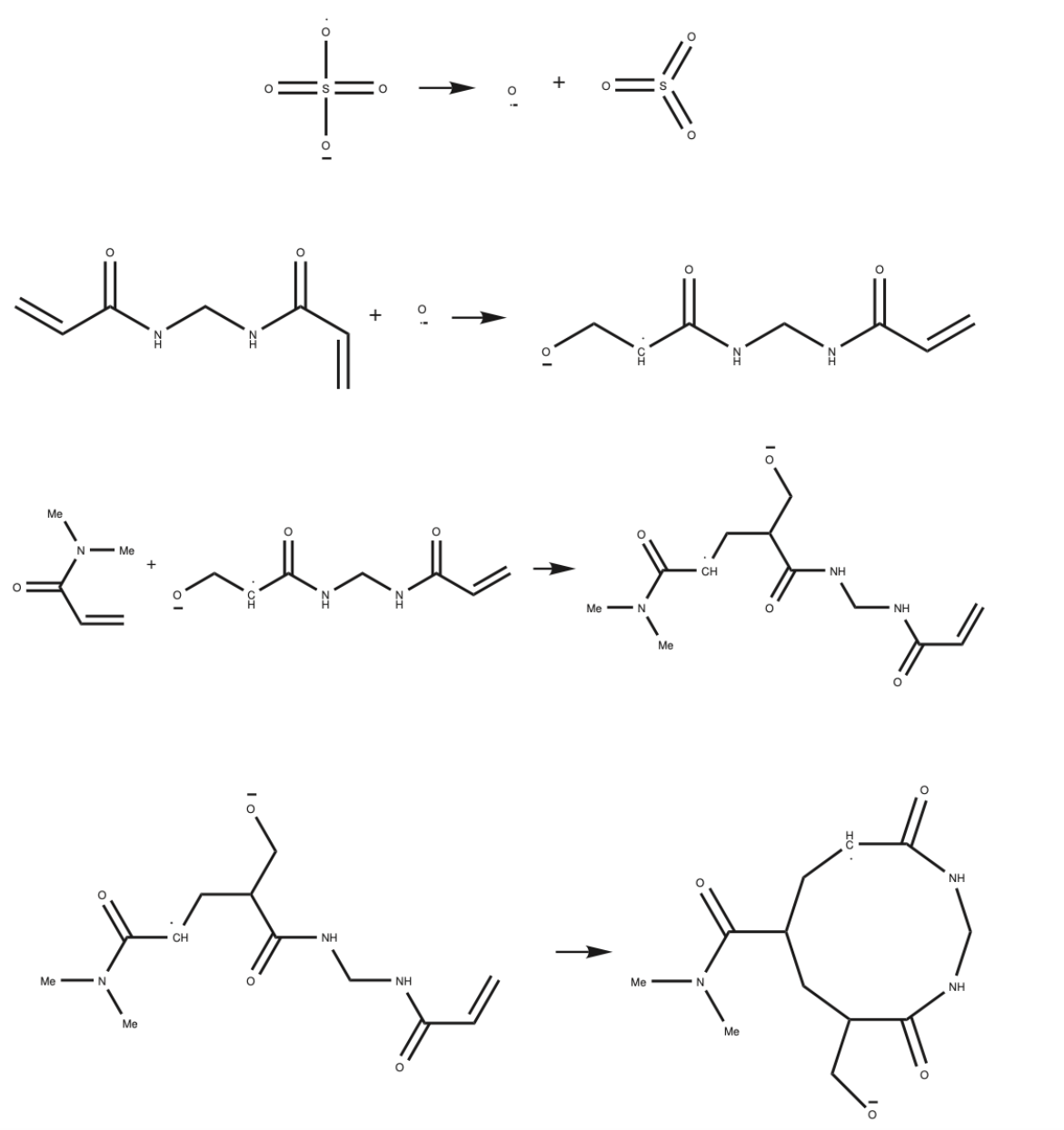}
  \caption{Another most likely mechanism shares similarities with Fig. \ref{fig:Supplementary Fig. 50}. However, in this instance, the final step deviates from the previous mechanisms. Instead of the crosslinker attaching to another monomer, it forms a ring structure. This illustration emphasizes the system’s ability to generate diverse structures based on slight variations in reaction pathways, showcasing its adaptability and complexity in producing intricate arrangements.}
  \label{fig:Supplementary Fig. 51}
\end{figure}

\begin{figure}[!]
  \centering
  \includegraphics[width=0.9\linewidth]{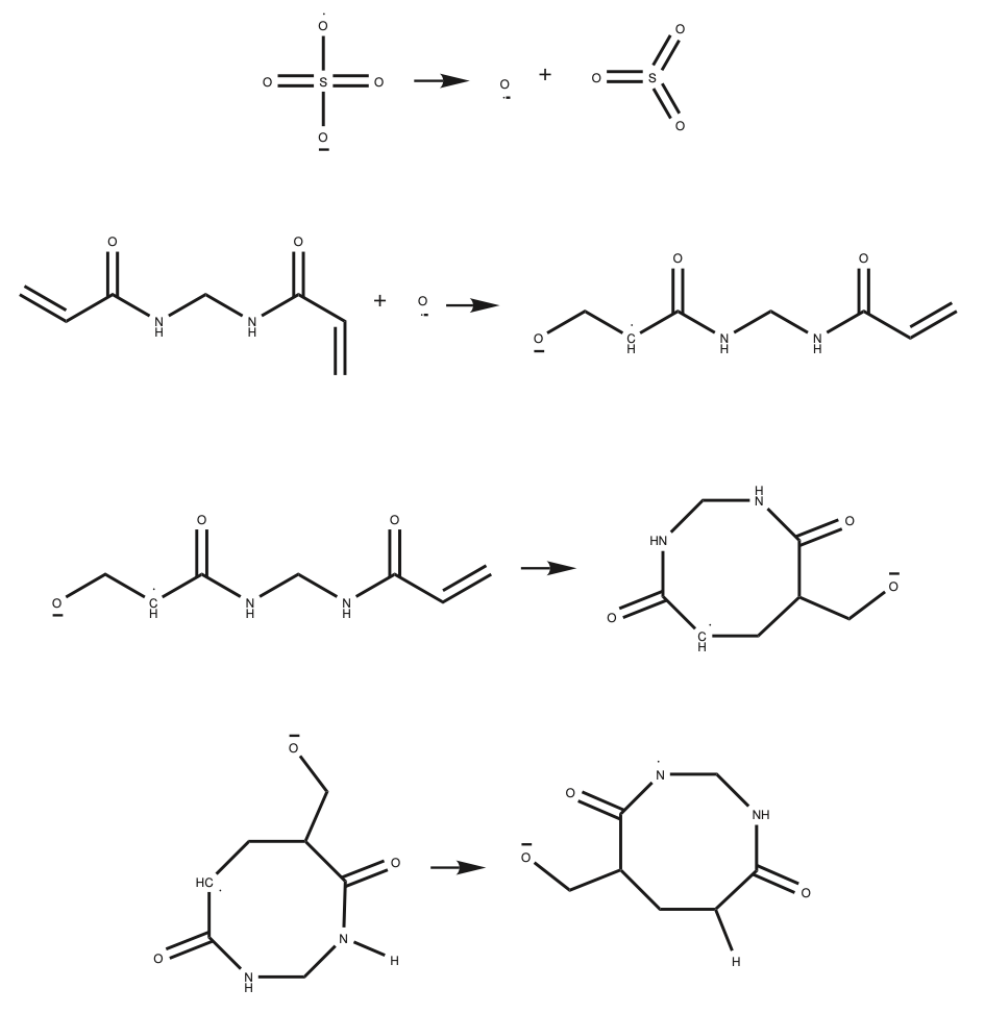}
  \caption{Here, in a distinctive prediction within the system, when the oxygen anion radical initiates by attacking the crosslinkers, the final two steps are primarily dedicated to rearrangement and the transfer of radicals on the crosslinkers. This depiction introduces a unique exception to the typical reaction pathways, highlighting the system’s capacity to exhibit diverse and unexpected behaviors under specific conditions.}
  \label{fig:Supplementary Fig. 52}
\end{figure}

\begin{figure}[!]
  \centering
  \includegraphics[width=0.9\linewidth]{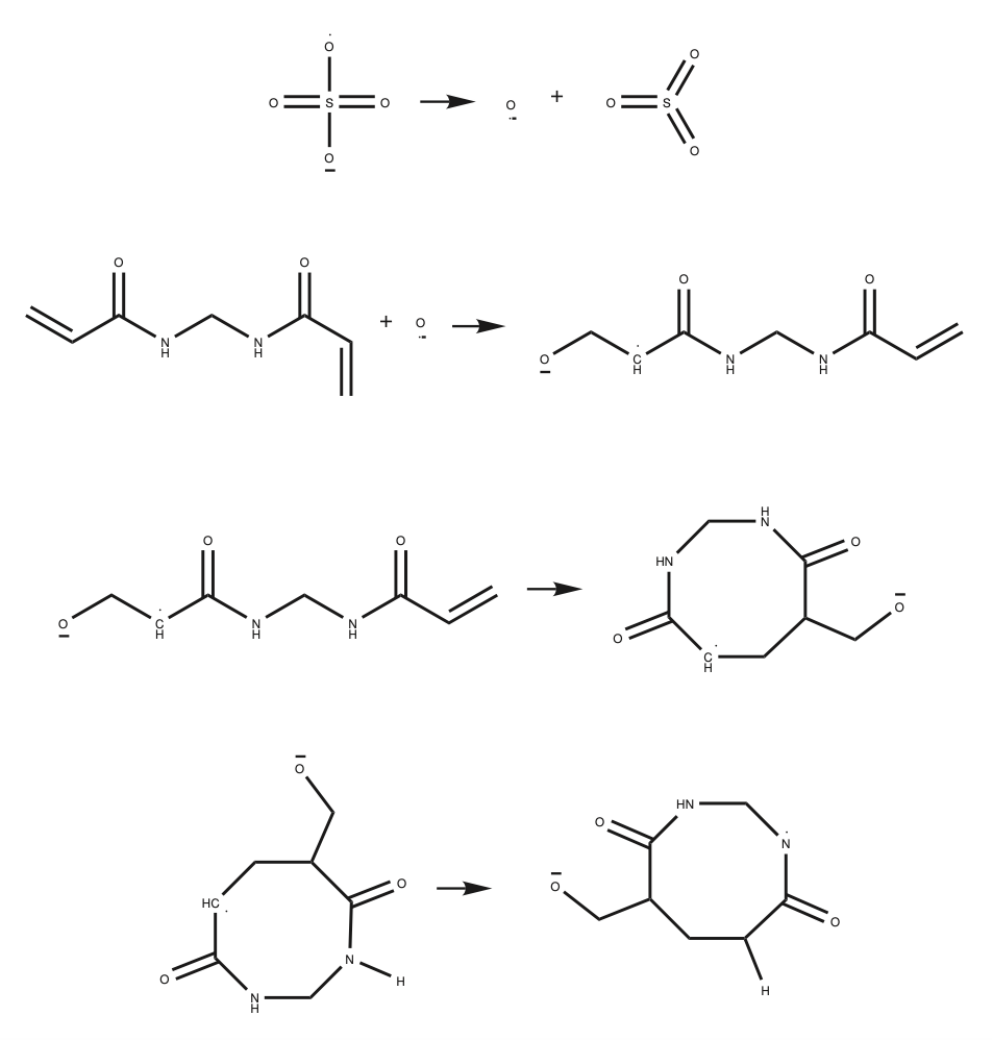}
  \caption{In a unique reaction pathway, radical transfer occurs within the crosslinker, differing from the scenario presented in Fig. \ref{fig:Supplementary Fig. 52}. Specifically, the radical transfers to an alternate nitrogen atom within the crosslinker. This depiction underscores the system’s potential for diverse and unexpected rearrangement patterns, adding complexity to the observed reactions.}
  \label{fig:Supplementary Fig. 53}
\end{figure}

\begin{figure}[!]
  \centering
  \includegraphics[width=\linewidth]{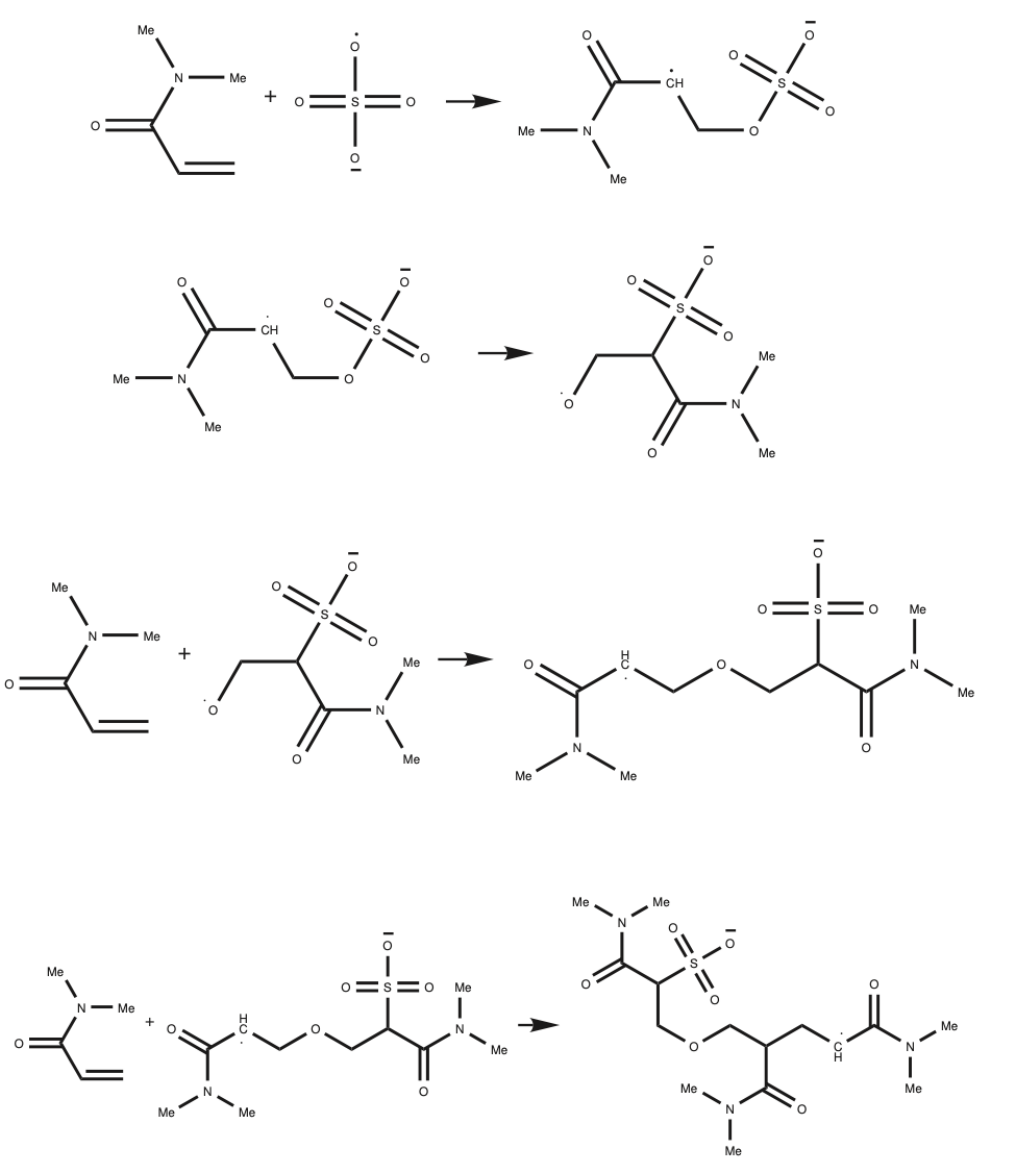}
  \caption{This unconventional reaction pathway is one of five similar cases that are mentioned in the next figures. In this instance, the sulfate radical anion (SO4·-) directly attacks the monomer, activating it in a manner that positions oxygen on the head and SO3 on the tail of the C-C double bond. Subsequently, polymerization occurs in a head-to-head fashion, accompanied by the incorporation of oxygen into the main chain, forming an ether group. This scenario leads to the formation of a trimer, where three monomers are interconnected.}
  \label{fig:Supplementary Fig. 54}
\end{figure}

\begin{figure}[!]
  \centering
  \includegraphics[width=\linewidth]{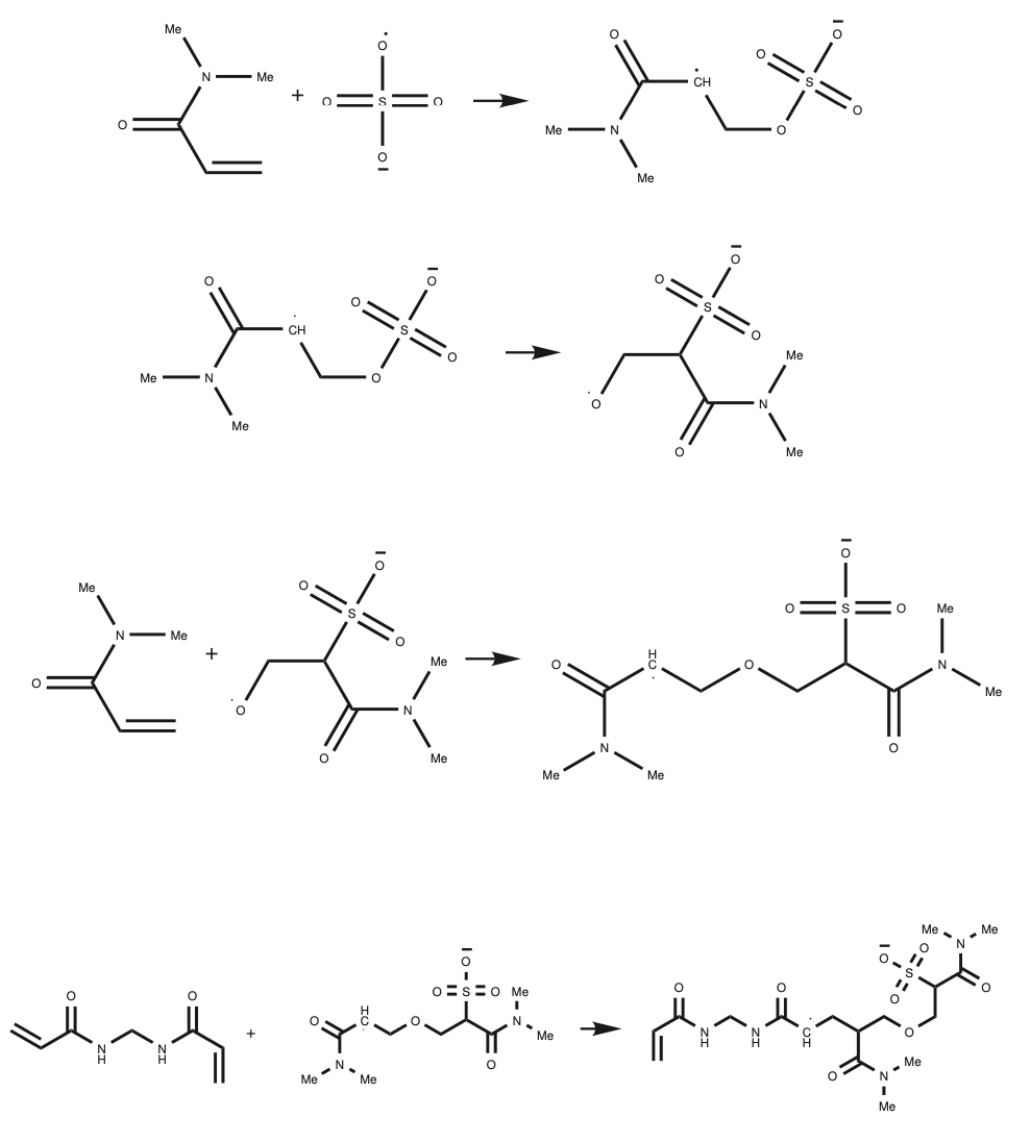}
  \caption{Here, in an atypical reaction mechanism similar to the previous case, the sulfate radical anion (SO4·-) directly attacks the monomer, initiating a chain of reactions. The radical monomer subsequently engages with another monomer, leading to the formation of a dimer, which represents a polymerization event. Importantly, this dimer includes an ether group incorporated into the main chain, although such an outcome is considered rare and may not align with observations from FTIR (Fourier-transform infrared) spectroscopy results. In the final step, the dimer further interacts with crosslinkers, adding complexity to the overall reaction pathway.}
  \label{fig:Supplementary Fig. 55}
\end{figure}

\begin{figure}[!]
  \centering
  \includegraphics[width=\linewidth]{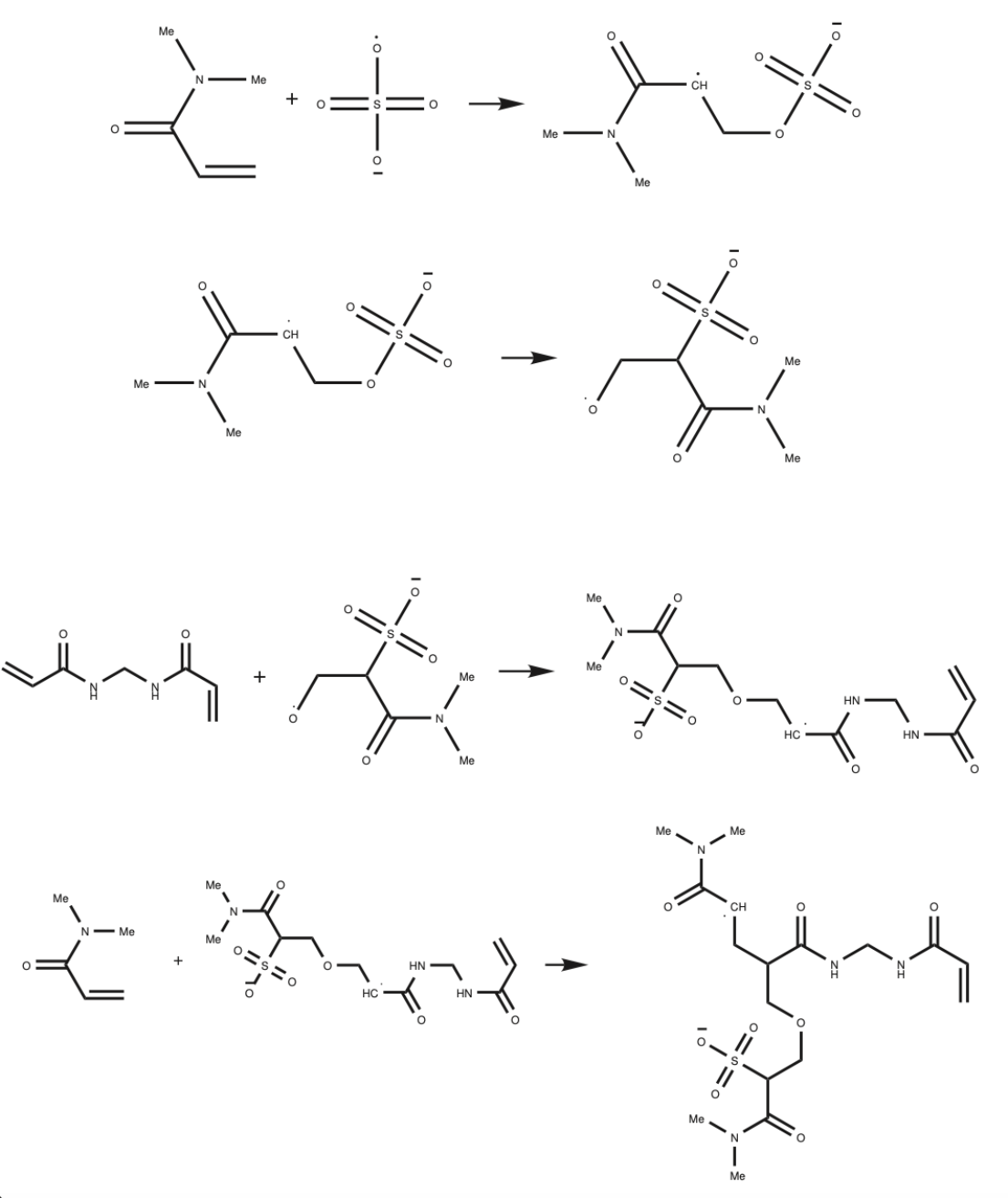}
  \caption{This unconventional mechanism is comparable to the previous cases. In this scenario, the radical monomer initiates the reaction by attacking the crosslinker. Subsequently, this activated crosslinker interacts with a monomer, resulting in a dimer formation. Notably, the connection between the monomer and the crosslinker introduces an ether group into the reaction pathway. This mechanism adds another layer of complexity, where the involvement of the crosslinker, monomer, and ether group shapes the overall reaction outcome.}
  \label{fig:Supplementary Fig. 56}
\end{figure}

\begin{figure}[!]
  \centering
  \includegraphics[width=\linewidth]{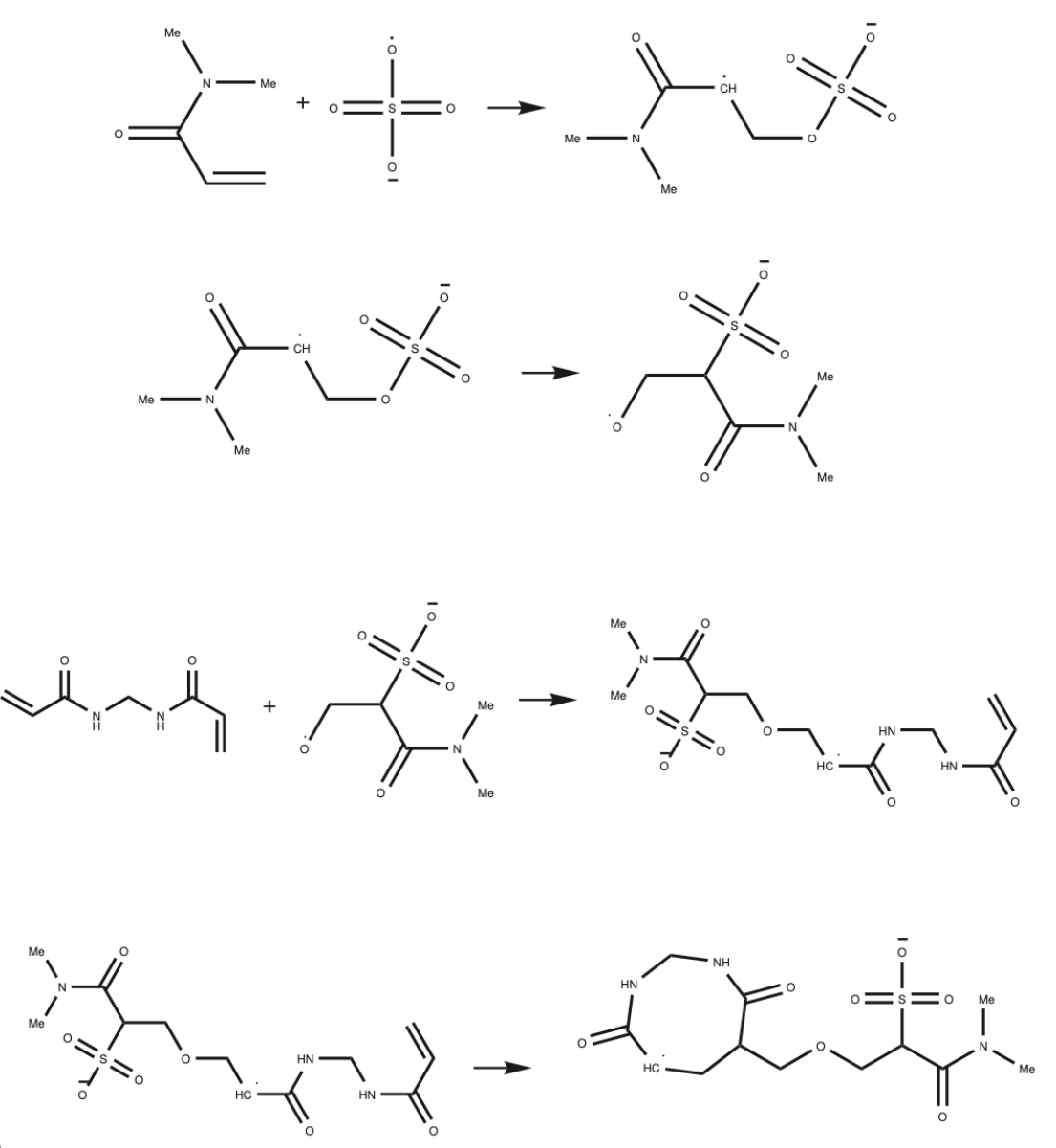}
  \caption{In this unique reaction pathway, following the monomer radical’s attack on the crosslinker, the crosslinker forms a ring structure. Importantly, the presence of an ether group is a notable feature of this mechanism. This depiction highlights the system’s potential to generate complex structures, including ring formations and the incorporation of ether groups, under specific conditions and pathways.}
  \label{fig:Supplementary Fig. 57}
\end{figure}

The four previous groups have primarily demonstrated the validity of mechanisms pertaining to polymerization, encompassing aspects such as the decomposition of APS, radical initiation, monomer radical generation, chain growth, and crosslinking. These investigations were instrumental in analyzing the effects of APS decomposition and validating the polymerization processes. However, our central objective, as previously mentioned, revolves around our second approach, which aims to reduce selectivity and depth to identify the most probable structures within the four identified groups. This endeavor specifically focuses on sulfate radical anion (SO4·-), a species well documented in numerous literature sources and thus regarded as a potential key player\cite{herrera2022role,herrera2023new}. In this section, our experimental system incorporates the reactants as outlined in Fig. \ref{fig:Supplementary Fig. 46}. These runs explored the system down to the fourth depth, reintroducing the monomer up to the final step. Our selectivity was set at 2, and after eliminating redundant structures, we identified a total of 11 structures as the most probable candidate products. These 11 structures are illustrated in Figures \ref{fig:Supplementary Fig. 47}-\ref{fig:Supplementary Fig. 57}. It is important to reiterate that the order in which these 11 mechanisms are presented does not denote any ranking. Instead, the order signifies that all 11 products are considered equally plausible among the entire spectrum of mechanisms that could potentially occur.


\section{Further Investigation and Running}\label{sec4}
In all but one of the mechanisms and runs conducted thus far, we observed that PEO (polyethylene oxide) appeared not to play a substantial role and did not actively participate in the reactions. However, in real-world experiments, we utilized PEO as one of our reactants, detecting its presence through peaks in the FTIR (Fourier-transform infrared) spectra. Furthermore, our experimental data indicated that PEO was indeed incorporated into the final hydrogel structure. This raises the question of how PEO is integrated into the hydrogel network. To address this query, we conducted an in-depth analysis of all initiator radicals and observed that following sulfate radical anion (SO4·-), hydroxyl radicals exhibited the highest likelihood of occurrence, both in accordance with existing literature and as predicted by our modeling system in the decomposition section\cite{liu2020modified}. Notably, some literature sources have attributed a pivotal role to hydroxyl radicals in the activation and initiation steps involving monomers\cite{khan2020effect,bashir2021flexible}. In this context, we sought to elucidate hydroxyl radicals’ potential impact on the polymerization process. This exploration is pivotal as it may lead to a paradigm shift in our understanding of hydrogel formation.

\subsection{The Role and Effect of Hydroxyl Groups}\label{subsec4}
\begin{figure}[!]
  \centering
  \includegraphics[width=0.55\linewidth]{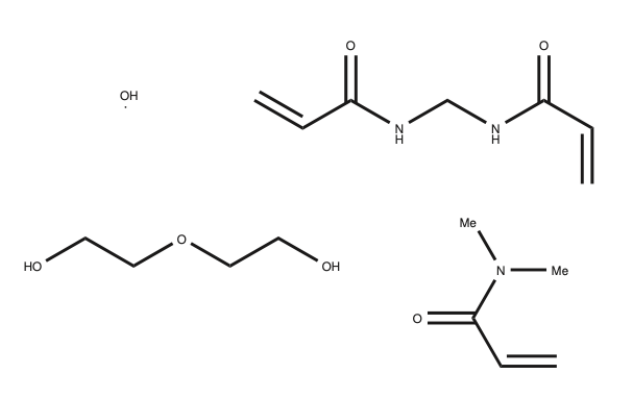}
  \caption{In this illustration of the configuration for Run Data 7, the initiation of reactions is facilitated by hydroxyl radical as the activated center. The reactants involved in this specific run include Polyethylene Oxide (PEO) as polymer, N, N-Dimethylacrylamide as a monomer, and N, N’- Methylenebisacrylamide as the crosslinker. This arrangement serves as the foundation for exploring the unique pathways and structures that arise when hydroxyl radicals initiate the reactions within this system.}
  \label{fig:Supplementary Fig. 58}
\end{figure}

\begin{figure}[!]
  \centering
  \includegraphics[width=\linewidth]{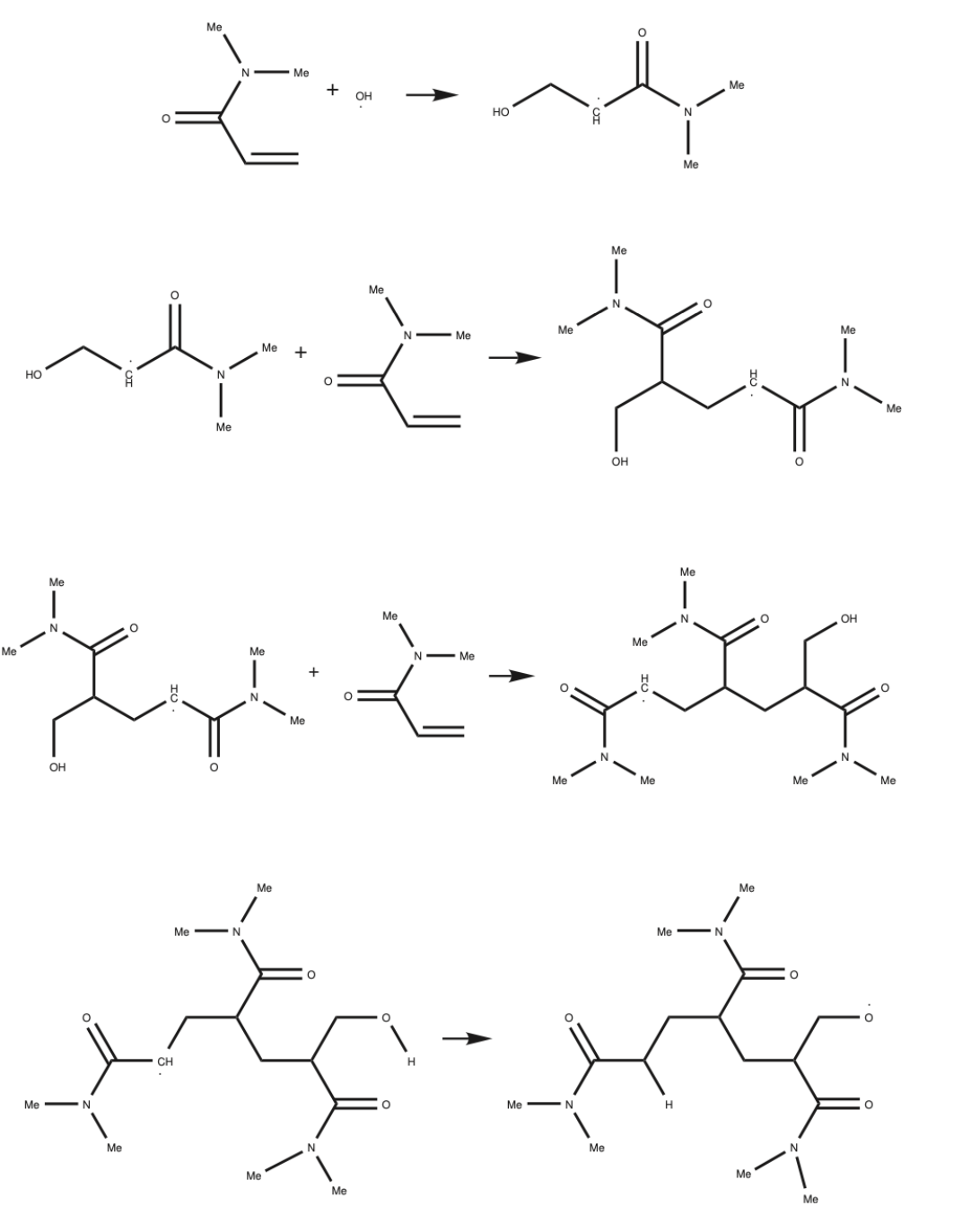}
  \caption{In one case, the hydroxyl radical initiates the reaction by attacking the double bond, leading to its breakage and the generation of a radical monomer. Subsequently, this radical monomer engages in polymerization with other monomers, extending the chain growth, and even progressing to involve three monomers. This illustration underscores the repeatability and reproducibility of polymerization within the system. The final step shows that the radical within the structure may separate hydrogen (H) from the hydroxyl (OH) group, potentially leading to further reactions.}
  \label{fig:Supplementary Fig. 59}
\end{figure}

\begin{figure}[!]
  \centering
  \includegraphics[width=\linewidth]{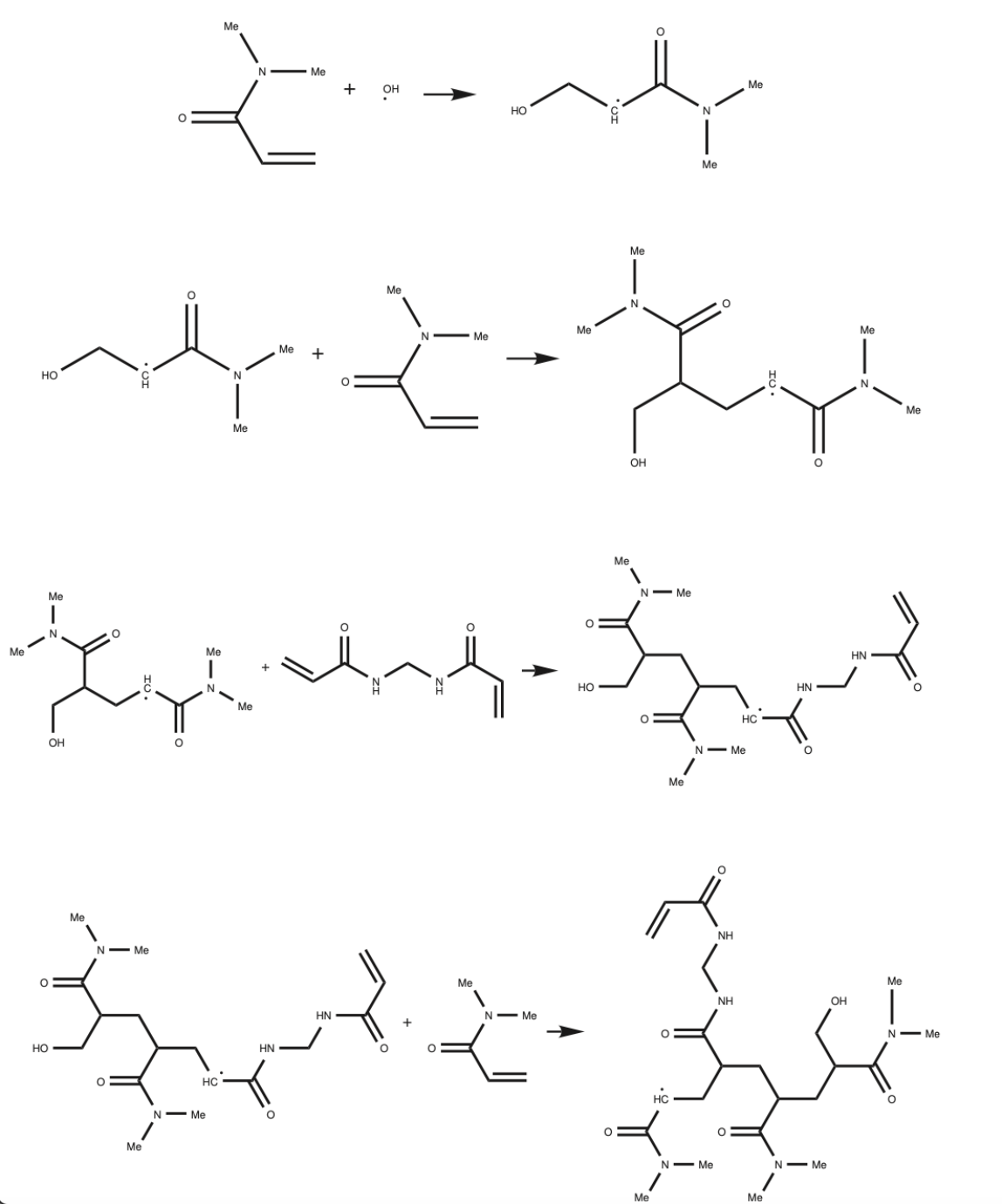}
  \caption{In an alternative mechanism, where the hydroxyl radical initiates the reaction by attacking the double bond of a monomer’s double bond. Subsequently, polymerization occurs, with one monomer attaching head-to-tail with another. Additionally, the crosslinker is introduced into the reaction, connecting to the junction of the dimer. However, with the addition of another monomer, the chain growth continues, showcasing the system’s flexibility of the system. Notably, the symmetry in the crosslinker’s structure and the potential transfer of radicals contribute to the ability to connect to other chains in these plausible mechanisms.}
  \label{fig:Supplementary Fig. 60}
\end{figure}

\begin{figure}[!]
  \centering
  \includegraphics[width=\linewidth]{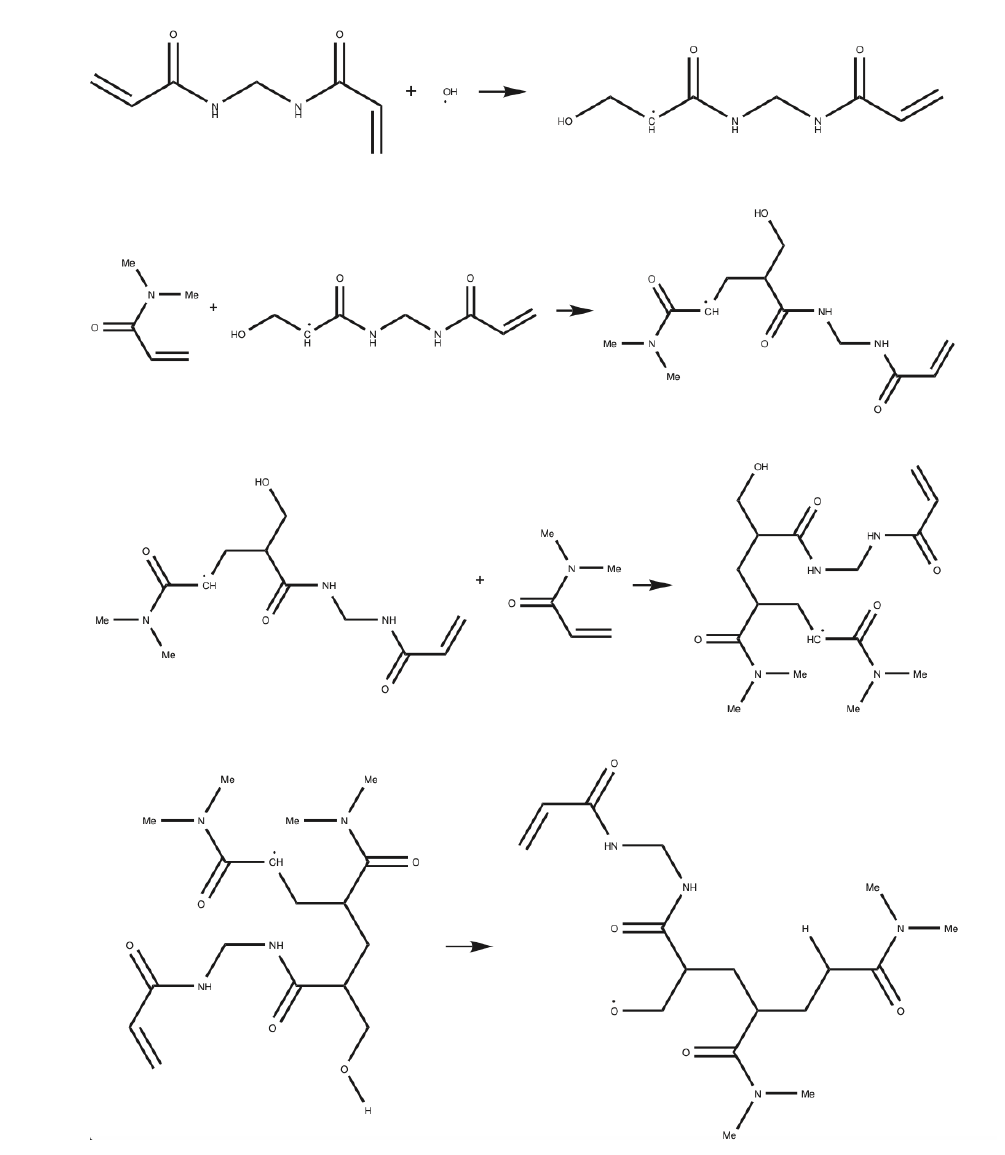}
  \caption{In this illustrative case the hydroxyl radical initiates the reaction by attacking the crosslinker, demonstrating the system’s capacity for simultaneous initiation of reactions involving multiple components. Following the radical crosslinker, two monomers engage and connect, representing a polymerization process facilitated by a single crosslinker. This scenario underscores the system’s versatility in accommodating diverse reaction pathways and mechanisms.}
  \label{fig:Supplementary Fig. 61}
\end{figure}

\begin{figure}[!]
  \centering
  \includegraphics[width=\linewidth]{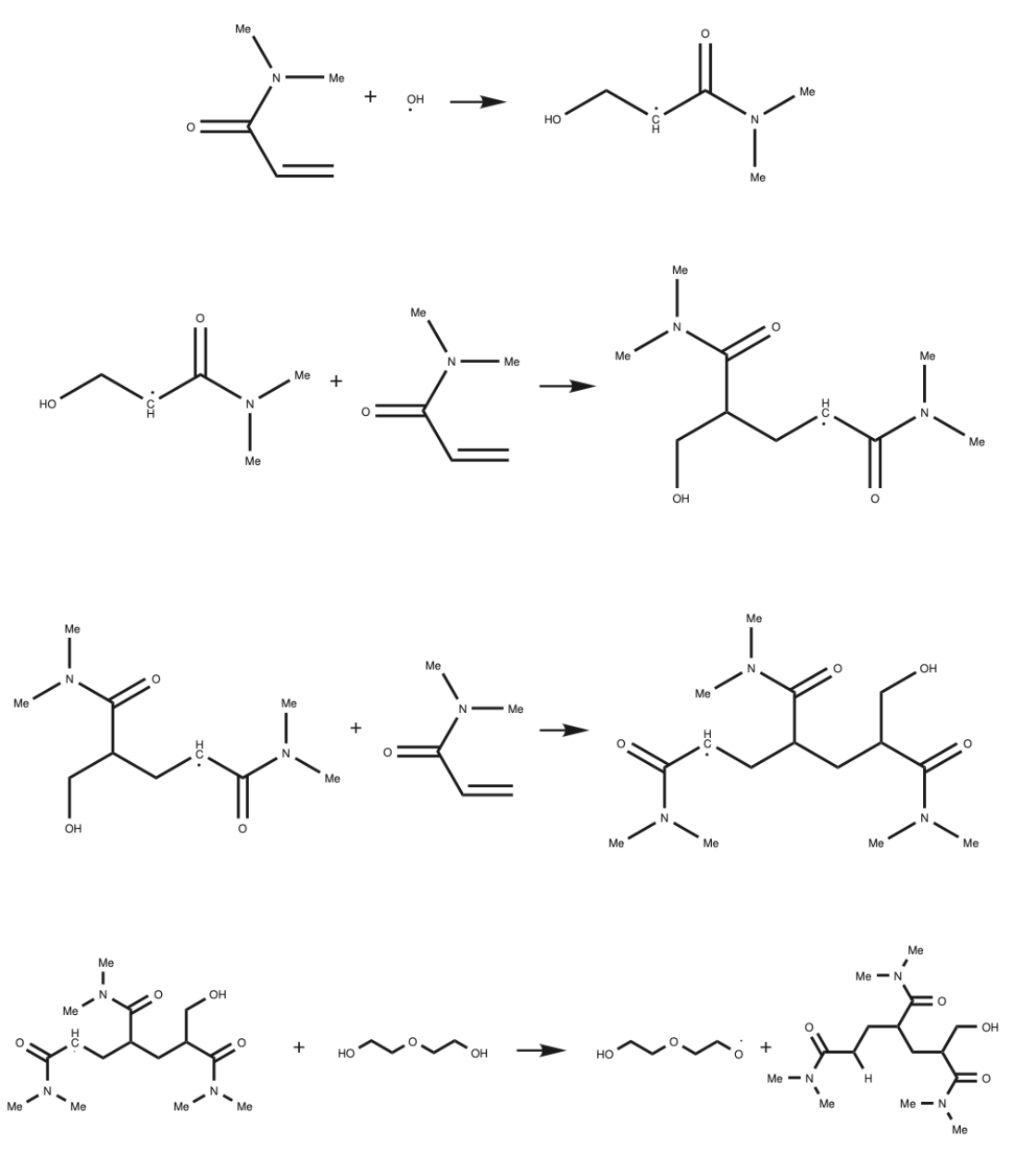}
  \caption{In this unique scenario, following the hydroxyl radical’s attack on the monomer and the subsequent formation of a dimer and trimer, a backbiting mechanism occurs. In this instance, the structure may capture a hydrogen atom (H) from PEO, suggesting a possible pathway for incorporating PEO into the final structures. This illustration raises the possibility that the system’s reaction products include PEO, highlighting its potential role in shaping the final structures.}
  \label{fig:Supplementary Fig. 62}
\end{figure}

In this section, we specifically examined the influence of hydroxyl radicals. For this purpose, we ran experiments utilizing hydroxyl radicals as initiators, along with PEO, N, N-Dimethylacrylamide as the monomer, and N, N’- Methylenebisacrylamide as the crosslinker, as outlined in Fig. \ref{fig:Supplementary Fig. 58}. Within this framework, we conducted supplementary runs, using our monomer as the contextual parameter and reintroducing it at each depth. Our investigations extended to a depth of 4 levels, with 3 selectivity options considered at each depth. Select cases and results are shown in Figures \ref{fig:Supplementary Fig. 59}-\ref{fig:Supplementary Fig. 62}.


\subsection{Effect of PEO Chain Size}\label{subsec4}
\begin{figure}[!]
  \centering
  \includegraphics[width=0.5\linewidth]{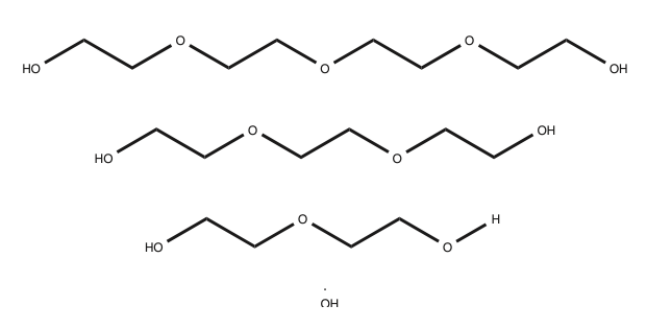}
  \caption{Setup for Run Data 8, where the hydroxyl radical initiates the reactions. This run involves the presence of three Polyethylene Oxide (PEO) chains of different lengths: short, medium, and long. Including various PEO chain lengths allows for the exploration of how different chain lengths may impact the reaction pathways and final structures within the system.}
  \label{fig:Supplementary Fig. 63}
\end{figure}

\begin{figure}[!]
  \centering
  \includegraphics[width=0.9\linewidth]{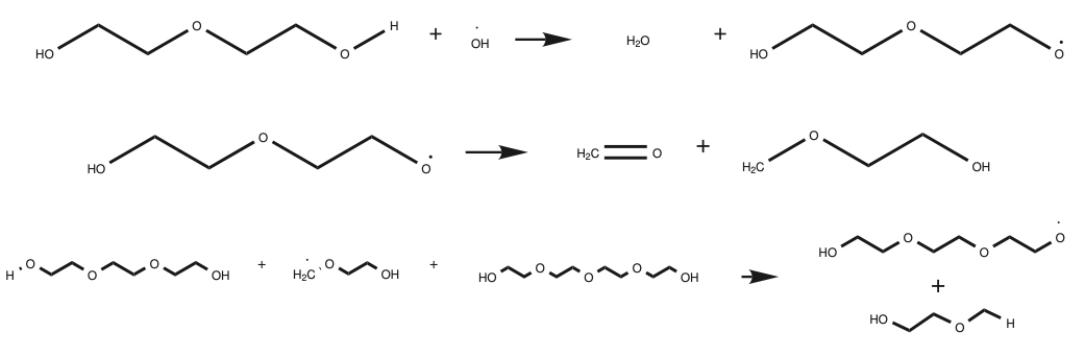}
  \caption{This chain session involves a medium-length Polyethylene Oxide (PEO) chain. During this session, a CH2 group with a double bond and oxygen (O) is formed, suggesting the potential for chain disruption and session occurrence across chains within a given range. This outcome hints at the system’s potential for intricate reactivity and structural diversity.}
  \label{fig:Supplementary Fig. 64}
\end{figure}

\begin{figure}[!]
  \centering
  \includegraphics[width=0.9\linewidth]{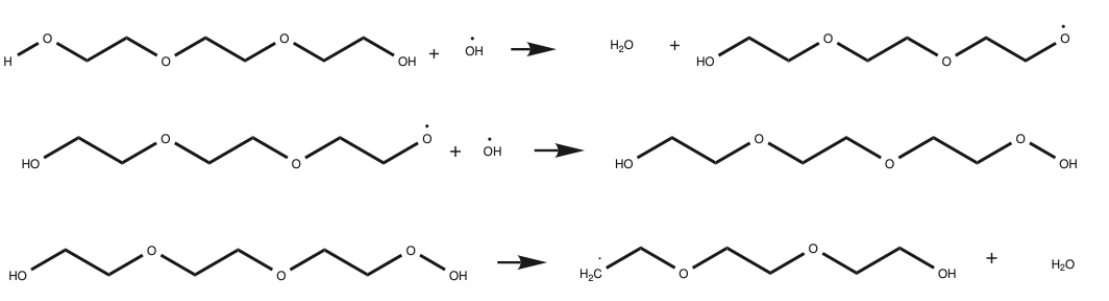}
  \caption{This chain session involves a medium-length polyethylene oxide (PEO) chain, accompanied by the release of water molecules. This illustration highlights the recurrent nature of chain sessions, which occur across various cases within the system.}
  \label{fig:Supplementary Fig. 65}
\end{figure}

\begin{figure}[!]
  \centering
  \includegraphics[width=0.9\linewidth]{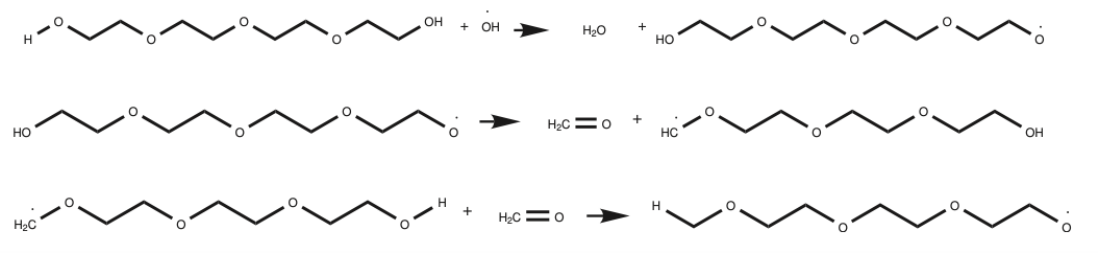}
  \caption{In this scenario, radicals are positioned on carbon atoms rather than oxygen. In a 3-depth simulation, each step of the depicted mechanism can occur individually. However, the noteworthy observation is that, if other components are simultaneously accessible in the system, chain sessions may be triggered, potentially leading to the incorporation of polyethylene oxide (PEO) or other components. This emphasizes the dynamic nature of the system’s reactions and the potential for diverse outcomes.}
  \label{fig:Supplementary Fig. 66}
\end{figure}

We are acutely aware that our use of a restricted number of PEO units responds to a combination of both system constraints and, notably, computational cost considerations. Our strategy consolidates all pertinent factors into a single, unified unit, wherein we conceptualize a dimer—comprising two monomers—as emblematic of the polymer itself\cite{kayik2014stereoselective}. This reduction to a mere pair of numerical values is motivated by our methodology’s inherent intricacies. In delineating the fundamental mechanisms governing PEO, we discern a pair of predominant pathways. Initially, PEO exhibits a propensity to relinquish hydrogen atoms from its terminal groups. Subsequently, due to the presence of radicals along the chain, a propitious confluence occurs, resulting in the amalgamation of these radicals into the ultimate molecular architecture. It is indispensable to acknowledge that in the physical realm,as the chain elongates, adopting a random coil configuration, it restricts terminal end group accessibility and becomes resistant to attachment or incursion by other molecular constituents. Our approach executed three distinct runs, each featuring varying lengths of PEO chains. Specifically, one run featured a short chain consisting of a mere two units, another employed a medium chain comprising three units, and the third incorporated a long chain encompassing five units, as shown in Fig. \ref{fig:Supplementary Fig. 63}.  Intriguingly, all three chains exhibited the attachment of OH radicals, a phenomenon cogently depicted in the first reactions of Figures \ref{fig:Supplementary Fig. 64}-\ref{fig:Supplementary Fig. 66}. This observation underscores the pivotal role played by the sulfate radical anion (SO4·-), along with other components, which manifest significantly higher reactivity relative to hydroxyl radicals, and the literature reports that the first reactions in initiation steps belong to the sulfate radical anion (SO4·-) and that the hydroxyl radical emerges after a long time at the end of reaction\cite{lee2020persulfate}. Consequently, toward the culmination of polymerization, it becomes feasible to incorporate PEO. This elucidates why, in our prior run, PEO’s participation in reactions was seemingly hindered, primarily due to these radicals’ heightened reactivity. Moreover, it becomes evident that as we progress through the runs, each characterized by an increasing computational depth, the potential for PEO to be incorporated into the reaction mix increases. Furthermore, our investigation explores the fascinating dynamics of chain sessions\cite{emami2002peroxide}, as visually expounded in the next steps of these figures. Remarkably, these chain sessions manifest predominantly in the long chain, where they are much more likely to occur. The predictive system anticipates and integrates the effects of chain size, offering invaluable insights into the intricacies of these molecular interactions.


\section{The Final Step}\label{sec5}
\begin{figure}[!]
  \centering
  \includegraphics[width=\linewidth]{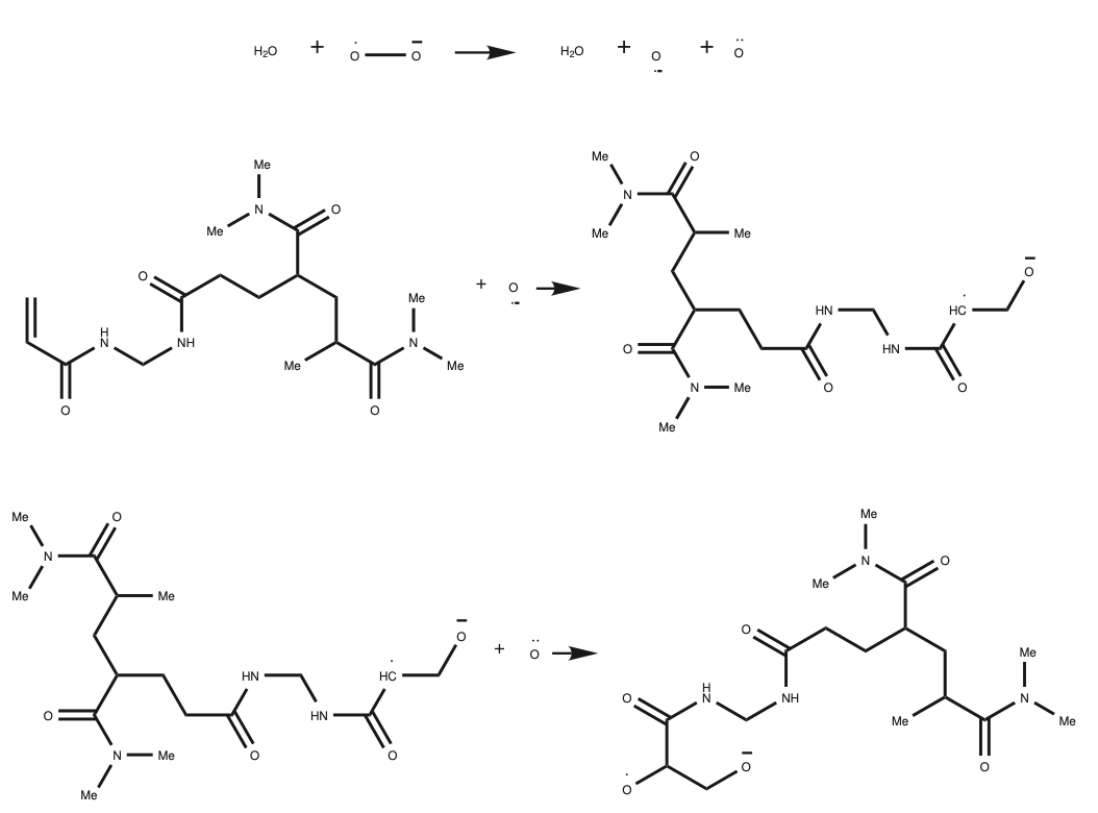}
  \caption{In this scenario a superoxide anion interacts with water, generating free oxygen and an oxygen radical anion. The oxygen radical anion then attacks the head of a monomer within the structure. Notably, in this illustration, other components or monomers were not added, but the oxygen radical anion’s interaction with free oxygen showcases the system’s capacity for reactivating the final structure. This observation underscores the repeatability and reactivation potential of the system’s main chain.}
  \label{fig:Supplementary Fig. 67}
\end{figure}

\begin{figure}[!]
  \centering
  \includegraphics[width=\linewidth]{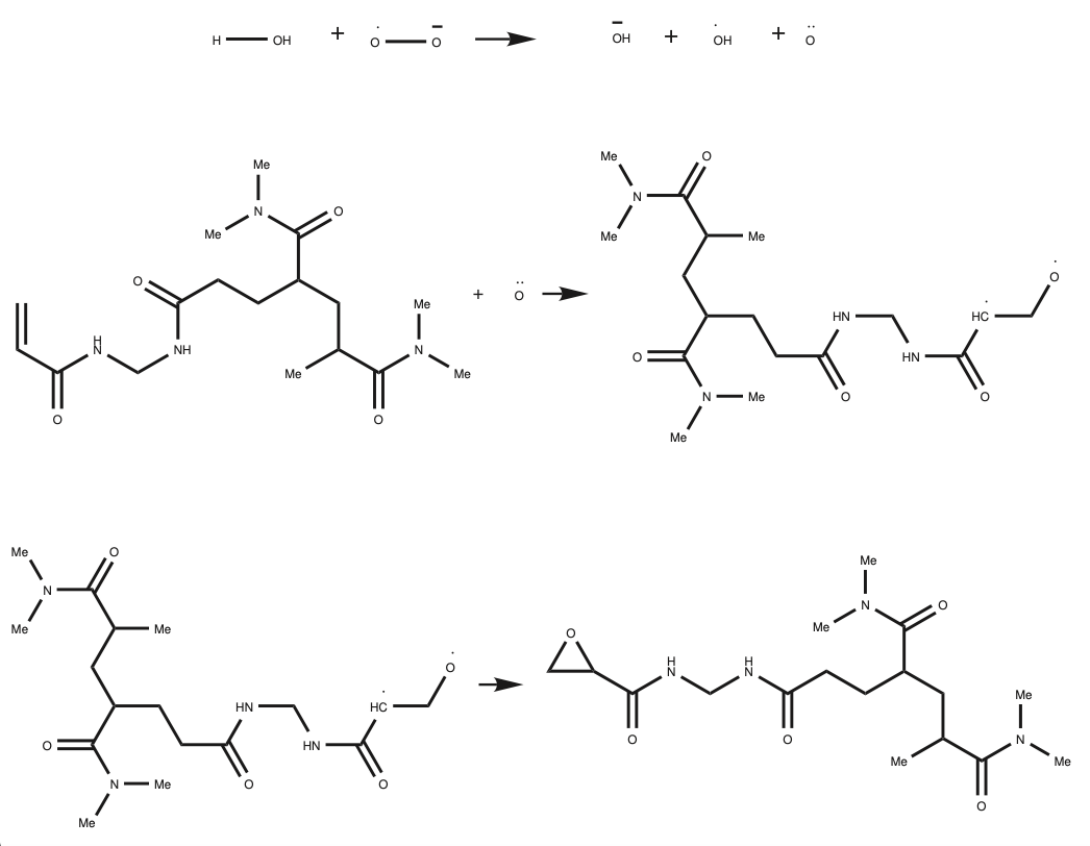}
  \caption{In this scenario free oxygen, rather than a superoxide anion, attacks. In the final step, due to the absence of additional components, a 3-point ring structure forms. This observation emphasizes that when other components, such as monomers and crosslinkers (which are typically present in real-world systems), are available in the system, the chain growth can continue beyond this point.}
  \label{fig:Supplementary Fig. 68}
\end{figure}

\begin{figure}[!]
  \centering
  \includegraphics[width=\linewidth]{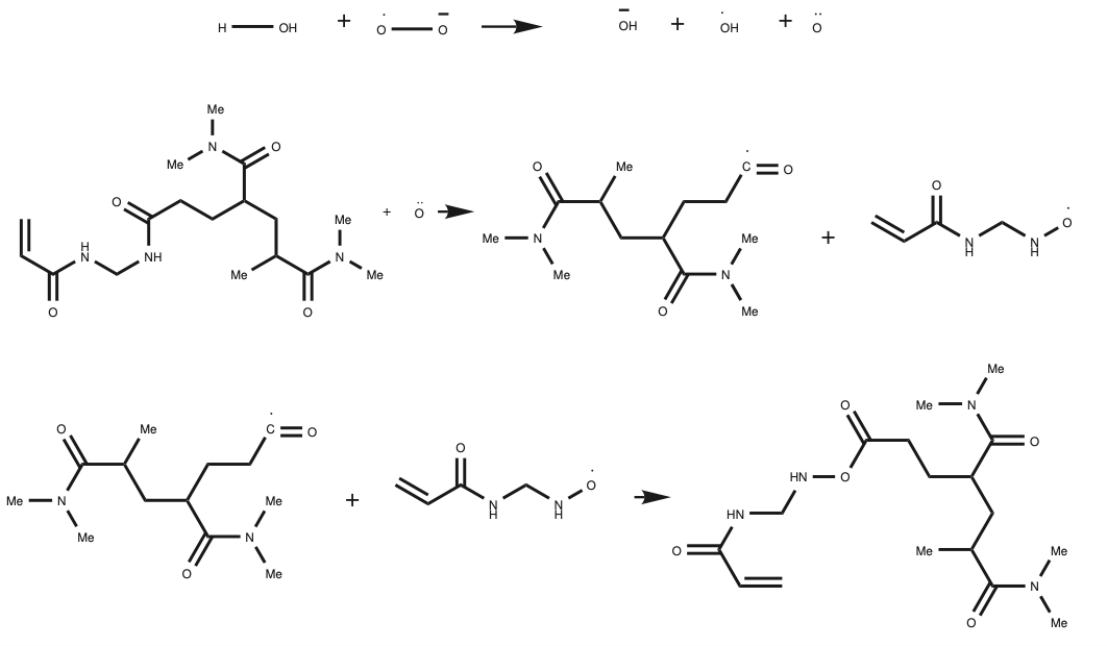}
  \caption{The interaction and reaction between water and a superoxide anion produces a hydroxyl radical, hydroxyl anion, free oxygen, and free oxygen’s interaction with the final structure. This interaction leads to the breaking and rejoining of the crosslinker at the carbon double bond, forming ester groups. This mechanism highlights one of the processes that could potentially contribute to the formation of ester groups within the repeated crosslink structure. The presence of ester groups can be significant for the system’s overall structure and properties and is reported in the FTIR results. }
  \label{fig:Supplementary Fig. 69}
\end{figure}

\begin{figure}[!]
  \centering
  \includegraphics[width=\linewidth]{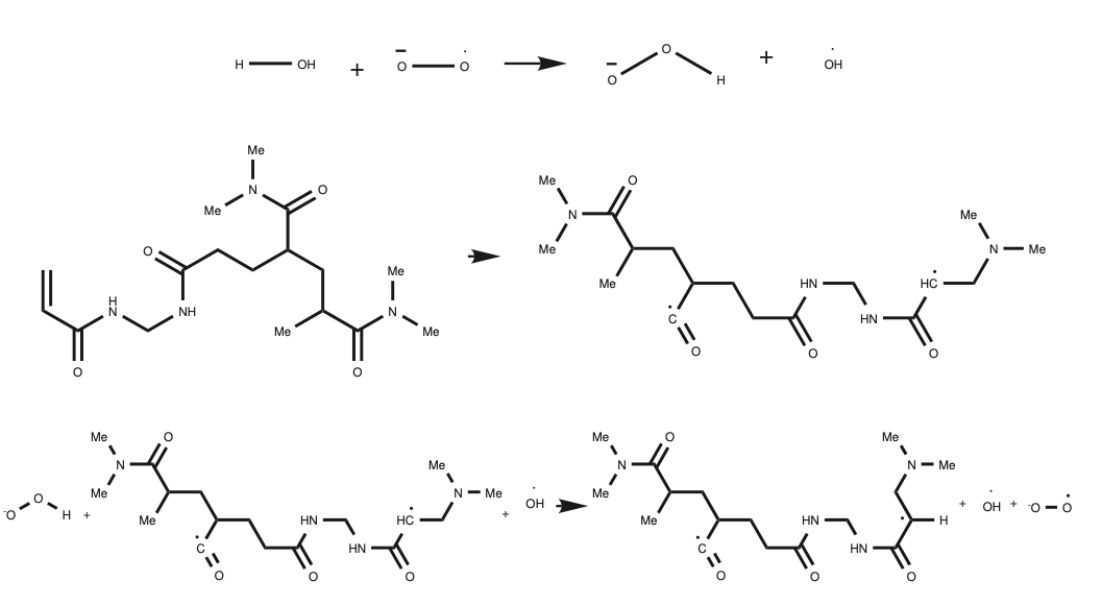}
  \caption{In an expected occurrence, the reaction between a superoxide anion and water generates OOH-. Of particular significance is the final structure’s reaction in the presence of OOH-, which results in a rearrangement leading to the production of a radical c=O. This transformation suggests that the structure becomes poised for subsequent reactions. This observation raises the intriguing possibility that this point serves as a potential site for PEO, which has undergone chain session and is ready for reaction, to interact with the newly formed radical c=O. Such interactions may lead to the formation of ester groups within the structure, highlighting their role in the system’s reactivity and potential for further modifications\cite{gomez2003kinetics}.}
  \label{fig:Supplementary Fig. 70}
\end{figure}

Upon the culmination of our exhaustive series of runs, we can deduce that the initiator disintegrates, yielding various radicals such as the sulfate radical anion (SO4·-) and other species, including hydroxyl radicals. Our postulation centers on the SO4 radical anion’s capacity to initiate polymerization and subsequently engage in gelation with a crosslinking agent. The ultimate structural configuration emerging from this complex interplay is meticulously presented as our final outcome, chosen from among 11 distinct possible structures. Additionally, we scrutinize the superoxide anion, another radical generated through APS decomposition. In the initial stages, we facilitate its interaction with water, and then introduce our aforementioned final structure. This conjuncture is meticulously documented in Figures \ref{fig:Supplementary Fig. 67}-\ref{fig:Supplementary Fig. 70}.

\section{Additional Experimental Test}\label{sec6}

\subsection{Relaxation Time}\label{subsec6}
\begin{figure}[!]
  \centering
  \includegraphics[width=0.8\linewidth]{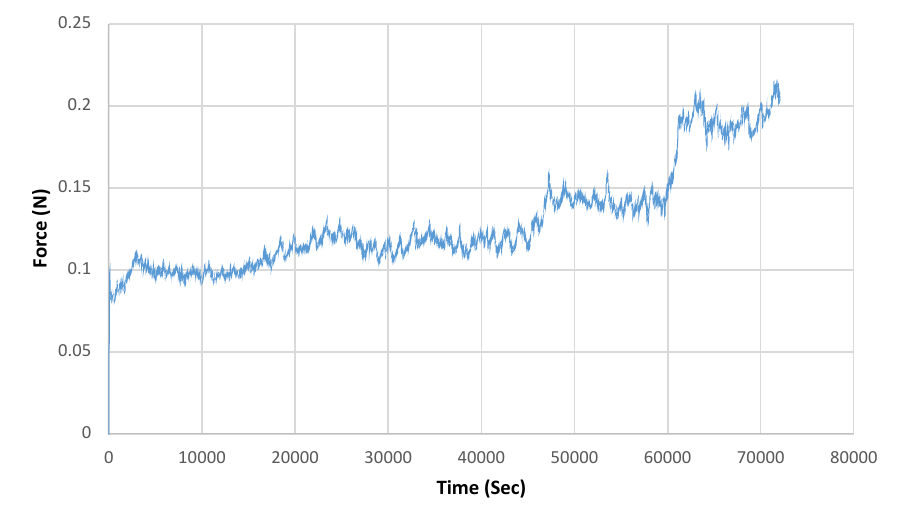}
  \caption{Relaxation experiment overnight for 20 hours at 150 mm extension with a gradual increase in stress under predetermined strain. The specimen is kept in that stretched position for a prolonged time. In typical materials, over time, a smaller amount of load is required to maintain that strain, also known as stress decay. Stress relaxation causes stress decay. In contrast to typical solids, stress flourishes as the sample dries up and more chains align in an axial direction.}
  \label{fig:Supplementary Fig. 71}
\end{figure}

A relaxation test was performed at 150 mm displacement for 20 hours and as seen in Fig. \ref{fig:Supplementary Fig. 71}, Force gradually and steadily increases to 0.2 N and is dissimilar from typical solid materials’ stress relaxation curves. As the time increases, it also decreases in stress-bearing capacity\cite{fitzgerald2015tunable,bauer2017hydrogel,chaudhuri2016hydrogels}. Water removal from PDMA hydrogel enhances mechanical properties. Stepwise stress relaxation revealed 5 steps with high- and low-stress regions. For example, the highest points are (0.11N at 3410 Sec) and (0.210 N at 63480 Sec), and the lowest points are (0.09N at 11550Sec) and (0.171 N at 64210 Sec) for steps 1 and 5, respectively. Moreover, stress-bearing was increased from almost 3.3 KPa at 3410 Sec to 7 KPa at 71880 Sec\cite{drouinviscoelastic}. The reason may be that incremental random coils opened in 20 hours, numerous hydrogen bonds formed between polymer chains placed in close vicinity to each other as a result of the drying effect, and the time sweep experiment suggested stronger force-bearing capacity. This behavior is especially appealing for applications that need a material to stay intact under constant strain. At around 16-17 hours, most of the excess free water was removed and a steep step increase in stress was obvious, indicating an increase in stress at constant strain due to the sudden increase in intermolecular hydrogen bonds.

\subsection{Storage and Loss Modulus (G’ and G’’)}\label{subsec6}
\begin{figure}[!]
  \centering
  \includegraphics[width=0.8\linewidth]{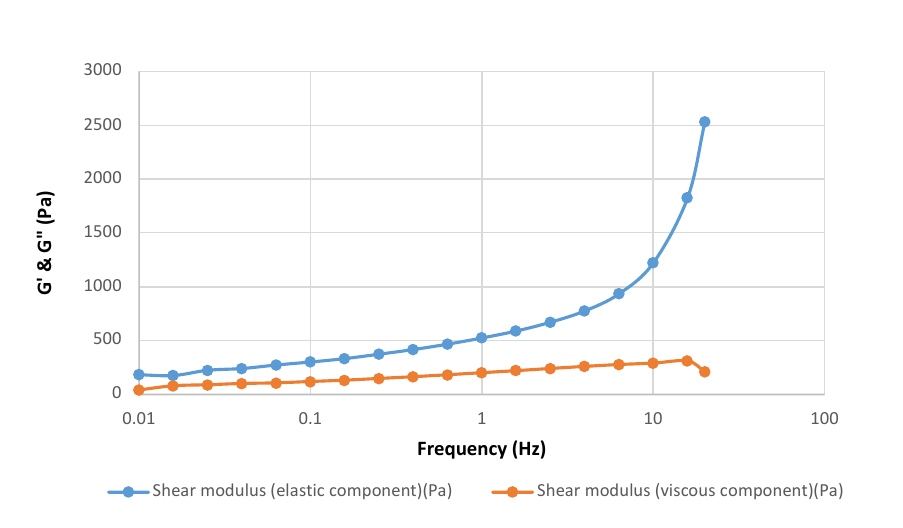}
  \caption{Dynamic mechanical analysis performed in parallel plates. Rheometer in shear mode and frequency sweep.}
  \label{fig:Supplementary Fig. 72}
\end{figure}

Previous works\cite{weng2007rheological,yan2010rheological} have studied hydrogels’ rheological properties. Dynamic mechanical analysis in shear mode reveals an increasing trend in storage modulus or G’, at frequencies above 10 Hz, and Fig. \ref{fig:Supplementary Fig. 72} depicts its strong elastic behavior up to 2500 pa at around 20 Hz. The viscoelastic hydrogel sample at high frequency tends to behave more like an elastic material, and the disruption of physical crosslinks fails to give the slow-responding viscose components time to respond to fast-oscillating shear stress. At higher frequencies, disrupted molecular networks lag behind the physical crosslinks’ elastic behavior and specifically intensify at around 10 Hz. This behavior may imply either the viscose components' large contribution at low frequencies or the extent of physical crosslink formation and microstructure rearrangement at high frequencies. Moreover, dynamic mechanical measurement in shear mode demonstrates that when G’ is larger than G”, the applied force is smaller than the molecular forces. In other words, the material has some capacity to store energy and can return, to some extent, to its initial configuration and behave as an elastic solid, although not an ideal one since some of the mechanical energy is dissipated. However, at higher applied forces, the microstructure collapses and the mechanical energy given to the material is dissipated, so that the material flows, and G” becomes larger than G’. Furthermore, since the imaginary part, G”, represents rearrangements of deformed polymer segments to maximize entropy, this is a viscous response and dissipates some of the deformation energy. A small sinusoidal deformation is not large enough to impose a significant viscose response on PDMA hydrogel and dissipate some of the deformation energy. In contrast, at large deformations such as cyclic experiments, the response may not be in the linear viscoelastic region, so the rearrangement of polymer segments causes energy dumping to intensify.

\subsection{Dynamic Mechanical in Axial Mode (E’ and E’’)}\label{subsec6}
\begin{figure}[!]
  \centering
  \includegraphics[width=0.8\linewidth]{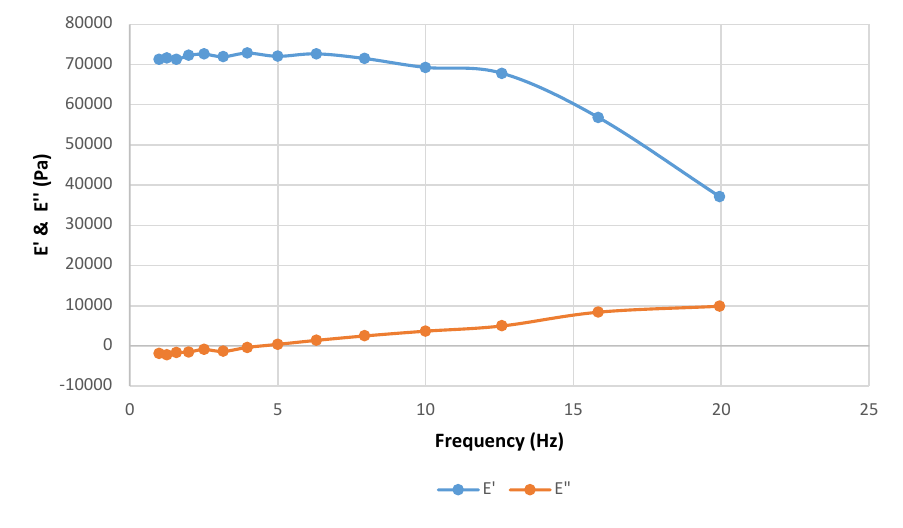}
  \caption{Dynamic mechanical analysis performed in the tensile fixture or axial geometry at frequency sweep mode.}
  \label{fig:Supplementary Fig. 73}
\end{figure}

Prior articles have reviewed dynamic mechanical analysis in axial mode for hydrogels, and Fig. \ref{fig:Supplementary Fig. 73} reveals a contrasting result compared to the shear mode experiment, reducing E’ by increasing frequency\cite{gasik2017viscoelastic,zhu2017high,meyvis2002comparison}. The discrepancy in Figure 8 may well be explained by the unique structure of entangled chains that resist movement in dynamic shear force. In the axial direction, physical crosslinkers play a major role in the static mode’s large extensibility. However, in axial dynamic mode, since PEO chains are physically bonded, they start to loosen drastically in response to a steep reduction in axial storage modulus at frequencies above 12 Hz. The loss modulus in both Figures \ref{fig:Supplementary Fig. 72} and \ref{fig:Supplementary Fig. 73} don't show significant values, but the storage modulus in G’ and E’ suggests polymer molecules’ elastic behavior and storage of some of the deformation energy. This storage at 12 Hz is 2.5 and 59 KPa for G’ and E’, respectively, indicating chain slippage in the shear mode as compared to the axial one.

\end{document}